\documentclass[usenatbib,usegraphicx]{mn2e}

\usepackage{times}
\usepackage{latexsym,amssymb}
\usepackage[fleqn]{amsmath}
\newcommand\aj{AJ}% 
\newcommand\apj{ApJ}% 
\newcommand\apjs{ApJS}% 
\newcommand\mnras{MNRAS}% 
\newcommand\aap{A\&A}% 
\newcommand\aaps{A\&AS}% 
\newcommand\nat{Nature}% 
\newcommand\physrep{Phys.~Rep.}% 
\numberwithin{equation}{section}

\newcommand{\kms}{km\,s$^{-1}$}
\newcommand{\dgr}{$^\circ$}
\newcommand{\Msun}{M$_\odot$}
\newcommand{\Msunpcsq}{M$_\odot$\,pc$^{-2}$}

\newcommand{\Lsunpcsq}{L$_\odot$\,pc$^{-2}$}

\newcommand{\MLsun}{M$_\odot$/L$_\odot$}
\newcommand{\du}{\mathrm{d}}               % upright d in integrals
\newcommand{\sauron}{\texttt{SAURON}}      % IFS SAURON

\title[Recovery orbital structure]{Recovery of the internal orbital
  structure of galaxies} \author[G. van de Ven et al.]  {G. van de
  Ven$^{1,2,3}$\thanks{Hubble Fellow}\thanks{E-mail: glenn@ias.edu},
  P.~T. de Zeeuw$^{1,4}$, R.~C.~E. van den Bosch$^1$\\
  $^1$Leiden Observatory, Leiden University, P.O.\ Box 9513, 2300 RA
  Leiden, The Netherlands\\
  $^2$Department of Astrophysical Sciences, Peyton Hall, Princeton, NJ
  08544, USA\\
  $^3$Institute for Advanced Study, Einstein Drive, Princeton, NJ 08540, USA\\
  $^4$ European Southern Observatory, D-85748 Garching bei M\"unchen}

\date{Accepted 0000 Month 00. Received 0000 Month 00; in original 0000 Month 00}
\pagerange{\pageref{firstpage}--\pageref{lastpage}}
\pubyear{2007}

\begin{document}

%% \date{Draft \today \hfill\fbox{\textbf{\emph{DO NOT DISTRIBUTE}}}}

\label{firstpage}

\maketitle

\begin{abstract}
  We construct axisymmetric and triaxial galaxy models with a
  phase-space distribution function that depends on linear
  combinations of the three exact integrals of motion for a separable
  potential. These Abel models, first introduced by Dejonghe \&
  Laurent and subsequently extended by Mathieu \& Dejonghe, are the
  axisymmetric and triaxial generalisations of the well-known
  spherical Osipkov--Merritt models. We show that the density and
  higher order velocity moments, as well as the line-of-sight velocity
  distribution (LOSVD) of these models can be calculated efficiently
  and that they capture much of the rich internal dynamics of
  early-type galaxies. We build a triaxial and oblate axisymmetric
  galaxy model with projected kinematics that mimic the
  two-dimensional kinematic observations that are obtained with
  integral-field spectrographs such as \sauron. We fit the simulated
  observations with axisymmetric and triaxial dynamical models
  constructed with our numerical implementation of Schwarzschild's
  orbit-superposition method. We find that Schwarzschild's method is
  able to recover the internal dynamics and three-integral
  distribution function of realistic models of early-type galaxies.
\end{abstract}


%============================= section 1 =============================
\section{Introduction}
\label{sec:abelintroduction}
%=====================================================================

The equilibrium state of a collisionless stellar system such as an
elliptical or lenticular galaxy is completely described by its
distribution function (DF) in the six-dimensional phase space of
positions and velocities.  The recovery of the DF from observations is
difficult, as for galaxies other than our own, we can usually only
measure the projected surface brightness and the line-of-sight
velocity distribution (LOSVD) of the integrated light as a function of
position on the plane of the sky.  Moreover, we generally do not know
the intrinsic shape of the galaxy, nor the viewing direction, or the
contribution to the gravitational potential provided by a super
massive central black hole and/or an extended halo of dark matter. By
Jeans (1915\nocite{1915MNRAS..76...70J}) theorem, the DF is a function
of the isolating integrals of motion admitted by the potential, but it
is not evident how to take advantage of this property other than for
the limiting case of spherical systems. Orbits in axisymmetric
geometry have two exact integrals of motion, the energy $E$ and the
angular momentum component $L_z$ parallel to the symmetry $z$-axis,
but the third effective or non-classical integral $I_3$ obeyed by all
regular orbits is generally not known in closed form. In stationary
triaxial geometry $E$ is conserved, but regular orbits now have two
additional effective integrals of motion, $I_2$ and $I_3$, which are
not known explicitly.

Schwarzschild (1979\nocite{1979ApJ...232..236S},
1982\nocite{1982ApJ...263..599S}) devised a numerical method which
sidesteps our ignorance about the non-classical integrals of motion.
It allows for an arbitrary gravitational potential, which may include
contributions from dark components, integrates the equations of motion
for a representative library of orbits, computes the density
distribution of each orbit, and then determines the orbital weights
such that the combined orbital densities reproduce the density of the
system. The best-fitting orbital weights represent the DF (cf.\
Vandervoort 1984\nocite{1984ApJ...287..475V}).  Pfenniger
(1984\nocite{1984A&A...134..373P}) and Richstone \& Tremaine
(1984\nocite{1984ApJ...286...27R}) included kinematic moments in this
method, and Rix et al.\ (1997\nocite{1997ApJ...488..702R}) showed how
to include observed LOSVDs.
A number of groups have developed independent numerical
implementations of Schwarzschild's method for axisymmetric geometry
which fit the projected surface brightness and line-of-sight velocity
distributions of early-type galaxies in detail %
(van der Marel et al.\ 1998\nocite{1998ApJ...493..613V}; %
Cretton et al.\ 1999\nocite{1999ApJS..124..383C}; %
Gebhardt et al.\ 2000\nocite{2000AJ....119.1157G}, %
Valluri, Merritt \& Emsellem 2004\nocite{2004ApJ...602...66V}; %
Thomas et al.\ 2004\nocite{2004MNRAS.353..391T}; %
Cappellari et al.\ 2006\nocite{2006MNRAS.366.1126C}). %
Applications include the determination of central black hole masses
(see also %
van der Marel et al.\ 1997\nocite{1997Natur.385..610V}; %
Cretton \& van den Bosch 1999\nocite{1999ApJ...514..704C}; %
Verolme et al.\ 2002\nocite{2002MNRAS.335..517V}; %
Cappellari et al.\ 2002\nocite{2002ApJ...578..787C}; %
Gebhardt et al.\ 2003\nocite{2003ApJ...583...92G}; %
Copin, Cretton \& Emsellem 2004\nocite{2004A&A...415..889C}), %
accurate global dynamical mass-to-light ratios %
(Cappellari et al.\ 2006\nocite{2006MNRAS.366.1126C}), %
as well as dark matter profiles as a function of radius %
(Cretton, Rix \& de Zeeuw 2000\nocite{2000ApJ...536..319C}; %
Thomas et al.\ 2005\nocite{2005MNRAS.360.1355T}), %
and recovery of the DF %
(Krajnovi\'c et al.\ 2005\nocite{2005MNRAS.357.1113K}). %
Van de Ven et al.\ (2006\nocite{2006A&A...445..513V}) and van den
Bosch et al.\ (2006\nocite{2006ApJ...641..852V}) included proper
motion measurements in order to model nearby globular clusters, and
determine their distance, inclination as well as mass-to-light ratio
as function of radius.
Finally, Verolme et al.\ (2003\nocite{2003LNP...626..279V}) and the
companion paper van den Bosch et al.\ (2007\nocite{2007vvz+},
hereafter vdB07) describe an extension to triaxial geometry that
includes all line-of-sight kinematics.

Although Schwarzschild models have significantly increased our
understanding of the dynamical structure and evolution of early-type
galaxies, questions remain about the uniqueness and the accuracy with
which they are able to recover the global parameters as well as the
internal dynamics of these galaxies. Many tests have been done to
establish how the axisymmetric code recovers known input models, but
these generally have been limited to spherical geometry or to an input
axisymmetric DF that is a function of $E$ and $L_z$ only (van der
Marel et al.\ 1998\nocite{1998ApJ...493..613V}; Cretton et al.\
1999\nocite{1999ApJS..124..383C}; Verolme \& de Zeeuw
2002\nocite{2002MNRAS.331..959V}; Valluri et al.\
2004\nocite{2004ApJ...602...66V}; Cretton \& Emsellem
2004\nocite{2004MNRAS.347L..31C}; Thomas et al.\
2004\nocite{2004MNRAS.353..391T}; Krajnovi\'c et al.\
2005\nocite{2005MNRAS.357.1113K}).

One could construct a numerical galaxy model with Schwarzschild's
method itself, compute the observables, and then use these as input
for the code and determine how well it recovers the input model. This
is useful, but does not provide a fully independent test of the
software. An alternative is to consider the special family of models
with gravitational potential of St\"ackel form, for which all three
integrals of motion are exact and known explicitly. These separable
potentials have a core rather than a central cusp, so the
corresponding models cannot include a central black hole, and are
inadequate for describing galactic nuclei. However, they can be
constructed for a large range of axis ratios (Statler
1987\nocite{1987ApJ...321..113S}), and their observed kinematic
properties are as rich as those seen in the main body of early-type
galaxies (Statler 1991\nocite{1991AJ....102..882S},
1994\nocite{1994ApJ...425..458S}; Arnold, de Zeeuw \& Hunter
1994\nocite{1994MNRAS.271..924A}).

A small number of analytic DFs have been constructed for triaxial
separable models. The `thin-orbit' models (Hunter \& de Zeeuw
1992\nocite{1992ApJ...389...79H}) have the maximum possible streaming
motions, but their DF contains delta functions, and they are therefore
not particularly useful for a test of general-purpose numerical
machinery. Dejonghe \& Laurent (1991\nocite{1991MNRAS.252..606D},
hereafter DL91) constructed separable triaxial models in which the DF
depends on a single parameter $S=E+wI_2+uI_3$, which is a linear
combination of the three exact integrals $E$, $I_2$ and $I_3$ admitted
by these potentials, and is quadratic in the velocity components. For
a given radial density profile, the DF follows by simple inversion of
an Abel integral equation. These so-called Abel models have no net
mean streaming motions, and are the axisymmetric and triaxial
generalisations of the well-known spherical Osipkov--Merritt models
(Osipkov 1979\nocite{1979PAZh....5...77O}; Merritt
1985\nocite{1985AJ.....90.1027M}), for which the observables can be
calculated easily (Carollo, de Zeeuw \& van der Marel
1995\nocite{1995MNRAS.276.1131C}). Mathieu \& Dejonghe
(1999\nocite{1999MNRAS.303..455M}, hereafter MD99) generalised the
results of DL91 by including two families of DF components with net
internal mean motions around the long and the short axis,
respectively, and compared the resulting models with observations of
Centaurus~A. Although the Abel character of the non-rotating
components is no longer conserved, the expressions for the velocity
moments in these more general models can still be evaluated in a
straightforward way. When the entire DF depends on the same single
variable $S$ the famous ellipsoidal hypothesis (e.g., Eddington
1915\nocite{1915MNRAS..76...37E}; Chandrasekhar
1940\nocite{1940ApJ....92..441C}) applies, so that self-consistency is
only possible in the spherical case (Eddington
1916\nocite{1916MNRAS..76..572E}; Camm
1941\nocite{1941MNRAS.101..195C}).  This does not hold for Abel models
with a DF that is a sum of components for which the variable $S$ has
different values of the parameters $w$ and $u$. Such multi-component
Abel models can provide (nearly) self-consistent models with a large
variety of shapes and dynamics.

Here, we show that for Abel models, in addition to the velocity
moments, the full LOSVD can be calculated in a simple way. Next, we
construct axisymmetric and triaxial Abel models to test our numerical
implementation of Schwarzschild's method. We assume a convenient form
for the gravitational potential, and construct the DF that reproduces
a realistic surface brightness distribution. We compute the LOSVDs of
the models and derive two-dimensional maps of the resulting
kinematics. We show that, despite the simple form of the DF, these
models display the large variety of features observed in early-type
galaxies with integral-field spectrographs such as \sauron\ (Emsellem
et al.\ 2004\nocite{2004MNRAS.352..721E}). By fitting axisymmetric and
triaxial three-integral Schwarzschild models to the simulated
observables we find that Schwarzschild's method is able to recover the
internal dynamics and three-integral DF of early-type galaxies. In
this paper we fix the mass-to-light ratio and viewing direction to
those of the Abel models, while in our companion paper vdB07 we
investigate how well these global parameters can be determined by
Schwarzschild's method, along with a full description of our numerical
implementation in triaxial geometry.

This paper is organised as follows. In Section~\ref{sec:triaxabel} we
summarise the properties of the triaxial Abel models of DL91 and MD99
and present the intrinsic velocity moments in a form which facilitates
their numerical implementation. We describe the conversion to
observables in Section~\ref{sec:observables}, including the
computation of the LOSVD. In Section~\ref{sec:triaxgalmodels} we
construct a specific triaxial galaxy model and in
Section~\ref{sec:recoverytriax} we fit the simulated observables with
our triaxial Schwarzschild models to investigate how well the
intrinsic moments and three-integral DF are recovered. In
Section~\ref{sec:axigalmodels} we consider Abel models in the
axisymmetric limit and construct a three-integral oblate galaxy model
to test our axisymmetric implementation of Schwarzschild's method. We
summarise our conclusions in Section~\ref{sec:abeldiscconcl}. In
Appendix~\ref{sec:limitingcases}, we describe the simpler Abel models
for the elliptic disc, large distance and spherical limit, and link
them to the classical Osipkov--Merritt solutions for spheres. Readers
who are mainly interested in the tests of the Schwarzschild method may
skip Sections~\ref{sec:triaxabel}~--~\ref{sec:triaxgalmodels} and
\ref{sec:axivelmom}~--~\ref{sec:axiabelmodel}.

%============================= section 2 =============================
\section{Triaxial Abel models}
\label{sec:triaxabel}
%=====================================================================

The triaxial Abel models introduced by DL91 have gravitational
potentials of St\"ackel form, for which the equations of motion
separate in confocal ellipsoidal coordinates. We briefly describe
these potentials, and refer for further details to de Zeeuw
(1985a\nocite{1985MNRAS.216..273D}). We then make a specific
choice for the DF, for which the velocity moments simplify.

%---------------------------------------------------------------------
\subsection{St\"ackel potentials}
\label{sec:triaxpot}
%---------------------------------------------------------------------

We define confocal ellipsoidal coordinates ($\lambda,\mu,\nu$) as
the three roots for $\tau$ of
\begin{equation}
  \label{eq:car2ell}
  \frac{x^2}{\tau+\alpha} + \frac{y^2}{\tau+\beta} +
  \frac{z^2}{\tau+\gamma} = 1,
\end{equation}
with ($x,y,z$) the usual Cartesian coordinates, and with constants
$\alpha,\beta$ and $\gamma$ such that $-\gamma \leq \nu\leq -\beta
\leq \mu \leq -\alpha \leq \lambda$. From the inverse relations
\begin{equation}
  \label{eq:ell2car}
  x^2 = \frac{(\lambda+\alpha)(\mu+\alpha)(\nu+\alpha)}
  {(\alpha-\beta)(\alpha-\gamma)},
\end{equation}
and similarly for $y^2$ and $z^2$ by cyclic permutation of $\alpha \to
\beta \to \gamma \to \alpha$, it follows that a combination
($\lambda,\mu,\nu$) generally corresponds to eight different points
($\pm x, \pm y, \pm z$). In these coordinates, the St\"ackel
potentials have the following form (Weinacht
1924\nocite{histWeinacht1924})
\begin{multline}
  \label{eq:triaxV_S}
  V_S(\lambda,\mu,\nu) =
  \frac{U(\lambda)}{(\lambda-\mu)(\lambda-\nu)} +
  \frac{U(\mu)}{(\mu-\nu)(\mu-\lambda)} \\ +
  \frac{U(\nu)}{(\nu-\lambda)(\nu-\mu)},
\end{multline}
where $U(\tau)$ is an arbitrary smooth function
$(\tau=\lambda,\mu,\nu)$. The right-hand side of
eq.~\eqref{eq:triaxV_S} can be recognised as the second order
divided difference of $U(\tau)$. Henceforth, we denote it with the
customary expression $U[\lambda,\mu,\nu]$, which is symmetric in its
arguments (see Hunter \& de Zeeuw 1992, eqs 2.1--2.3, 2.13 and
2.14\nocite{1992ApJ...389...79H}). Addition of a linear function of
$\tau$ to $U(\tau)$ does not change $V_S$.

The density $\rho_S$ that corresponds to $V_S$ can be found from
Poisson's equation
\begin{equation}
  \label{eq:triaxrho_S}
    4\pi G \rho_S(\lambda,\mu,\nu) = \nabla^2 V_S(\lambda,\mu,\nu),
\end{equation}
or alternatively by application of Kuzmin's (1973\nocite{1973Kuzmin})
formula (see de Zeeuw 1985b\nocite{1985MNRAS.216..599D}). This
formula shows that, once we have chosen the confocal coordinate system
and the density along the short axis, the mass model is fixed
everywhere by the requirement of separability\footnote{A
  third method for the calculation of the density is to use
  $4\pi G \rho_S = H[\lambda, \lambda, \mu, \mu, \nu, \nu]$, where the
  fifth-order divided difference is of the function $H(\tau) =
  4a(\tau)U'(\tau) - 2a'(\tau)U(\tau)$ with $a(\tau) = (\tau+\alpha)
  (\tau+\beta) (\tau+\gamma)$ and $U(\tau)$ defines the potential as
  in eq.~(\ref{eq:triaxV_S}). This result was obtained by Hunter in
  1989 (priv.\ comm.) and by Mathieu \& Dejonghe
  (1996\nocite{1996A&A...314...25M}).
  Similar expressions exist for
  the related families of potential-density pairs introduced in de
  Zeeuw \& Pfenniger (1988\nocite{1988MNRAS.235..949D}).}. For centrally
concentrated mass models, $V_S$ has the $x$-axis as long-axis and the
$z$-axis as short-axis. In most cases this is also true for the
associated density (de Zeeuw, Peletier \& Franx
1986\nocite{1986MNRAS.221.1001D}).

%---------------------------------------------------------------------
\subsection{Orbital structure}
\label{sec:triaxorbitstructure}
%---------------------------------------------------------------------

The Hamilton-Jacobi equation separates in $(\lambda,\mu,\nu)$ for
the potentials \eqref{eq:triaxV_S}, so that every orbit has three
exact integrals of motion (cf.\ de Zeeuw \& Lynden-Bell
1985\nocite{1985MNRAS.215..713D})
\begin{eqnarray}
  \label{eq:triaxEI2I3}
  E   & = & \textstyle{\frac12} \left( v_x^2 + v_y^2 + v_z^2 \right)
  + U[\lambda,\mu,\nu],
  \nonumber \\
  I_2 & = & \textstyle{\frac12} T L_y^2
  + \textstyle{\frac12} L_z^2
  + \textstyle{\frac12}(\alpha-\beta)v_x^2
  \nonumber \\ & &
  + (\alpha-\beta) x^2 U[\lambda,\mu,\nu,-\alpha],
  \\
  I_3 & = & \textstyle{\frac12} L_x^2
  + \textstyle{\frac12} (1-T) L_y^2
  + \textstyle{\frac12}(\gamma-\beta)v_z^2
  \nonumber \\ & &
  + (\gamma-\beta) z^2 U[\lambda,\mu,\nu,-\gamma],
  \nonumber
\end{eqnarray}
where $v_x$, $v_y$ and $v_z$ are the velocity components in the
Cartesian coordinate system, and $L_x=yv_z-zv_y$, the component of the
angular momentum vector parallel to the $x$-axis. The other two
components, $L_y$ and $L_z$, follow by cyclic permutation of $x\to
y\to z\to x$ and $v_x\to v_y\to v_z\to v_x$. Furthermore, $T$ is a
triaxiality parameter defined as
\begin{equation}
  \label{eq:triaxparT}
  %T = \frac{\beta-\alpha}{\gamma-\alpha},
  T = (\beta-\alpha)/(\gamma-\alpha),
\end{equation}
and $U[\lambda,\mu,\nu,\sigma]$ is the third-order divided difference
of $U(\tau)$. All models for which $U'''(\tau)>0$ have a similar
orbital structure and support four families of regular orbits: boxes
with no net rotation, inner and outer long-axis tubes with net
rotation around the $x$-axis, and short-axis tubes with net rotation
around the $z$-axis (Kuzmin 1973\nocite{1973Kuzmin}; de Zeeuw
1985a\nocite{1985MNRAS.216..273D}; Hunter \& de Zeeuw
1992\nocite{1992ApJ...389...79H}).

According to Jeans (1915\nocite{1915MNRAS..76...70J}) theorem the
phase-space distribution function (DF) is a function $f(E,I_2,I_3)$ of
the isolating integrals of motion (cf.\ Lynden-Bell
1962\nocite{1962MNRAS.124....1L}; Binney
1982\nocite{1982MNRAS.201...15B}). The velocity moments of the DF are
defined as
\begin{equation}
  \label{eq:defvelmom}
  \mu_{lmn}(\lambda,\mu,\nu) = \iiint \! \! v_\lambda^l v_\mu^m v_\nu^n
  \, f(E,I_2,I_3) \,
  \du v_\lambda \, \du v_\mu \, \du v_\nu,
\end{equation}
where $l$, $m$ and $n$ are non-negative integers, and $v_\lambda$,
$v_\mu$ and $v_\nu$ are the velocity components in the confocal
ellipsoidal coordinate system. Many of the velocity moments vanish due
to the symmetry of the orbits in these coordinates. The zeroth-order
velocity moment is the mass density that corresponds to the DF
\begin{equation}
  \label{eq:triaxrho}
    \rho_\star(\lambda,\mu,\nu) = \mu_{000}(\lambda,\mu,\nu).
\end{equation}
In self-consistent models, $\rho_\star$ must equal $\rho_S$ given in
eq.~(\ref{eq:triaxrho_S}), the mass density that is related to the
potential $V_S$ by Poisson's equation.

%---------------------------------------------------------------------
\subsection{Abel distribution function}
\label{sec:abeldf}
%---------------------------------------------------------------------

Following DL91, we choose the DF to be a function of the three
integrals of motion $E$, $I_2$ and $I_3$ as given in
eq.~\eqref{eq:triaxEI2I3} through one variable
\begin{equation}
  \label{eq:dfabel}
  f(E,I_2,I_3) = f(S),
  \quad \mathrm{with} \quad
  S = -E+w\,I_2+u\,I_3,
\end{equation}
and $w$ and $u$ are two parameters\footnote{In contrast with DL91 and
  MD99, we choose $V_S\le0$ and $E\le0$, consistent with e.g.\ de Zeeuw
  (1985a\nocite{1985MNRAS.216..273D}).}.
This choice for the DF is equivalent to the celebrated ellipsoidal
hypothesis (e.g., Eddington 1915\nocite{1915MNRAS..76...37E};
Chandrasekhar 1940\nocite{1940ApJ....92..441C}). Self-consistency
is only possible in the spherical case (Eddington
1916\nocite{1916MNRAS..76..572E}; Camm
1941\nocite{1941MNRAS.101..195C}).  On the other hand, these DFs
can produce realistic (luminous) mass densities $\rho_\star$, which differ
from the (total) mass density $\rho_S$, as in galaxies with dark
matter (see also \S~\ref{sec:combmultcomp} below when we combine DFs of
the form [\ref{eq:dfabel}] with different values for $w$ and $u$.)

DL91 and MD99 divided the DF into three types of components. The
non-rotating (NR) type is made of box orbits and tube orbits with both
senses of rotation populated equally. The two rotating types, LR and
SR, consist of tube orbits, and have net rotation around either the
long axis or the short axis.

%---------------------------------------------------------------------

\subsubsection{Velocity moments}
\label{sec:triaxvelmom}

%%%FIG
\begin{figure}[t]
  \begin{center}
    \includegraphics[width=1.0\columnwidth]{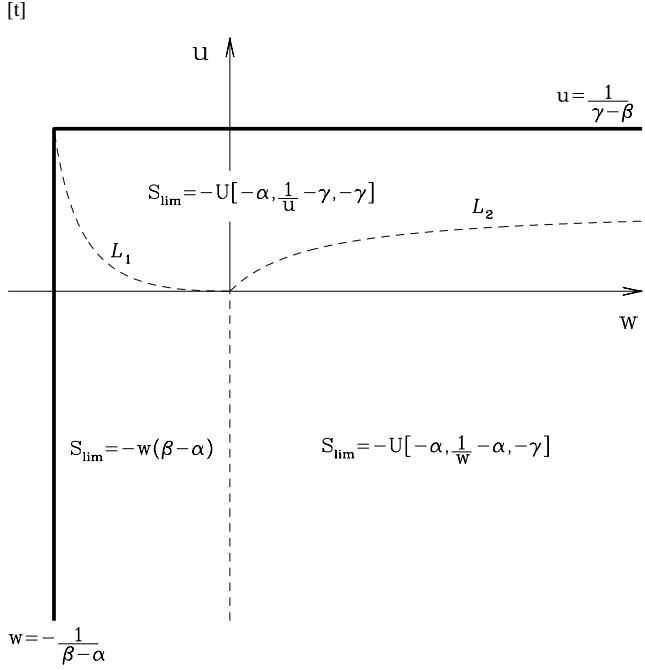}
  \end{center}
  \caption{
    The limiting value $S_\mathrm{lim}$ of the variable
    $S=-E+w\,I_2+u\,I_3$ as function of the parameters $w$ and $u$.
    The physical region is bounded by the relations \eqref{eq:limwu},
    indicated by the thick solid lines. The dashed curves divide this
    region into three parts, each with a different expression for
    $S_\mathrm{lim}$. The relations for the separatrices $L_1$ and
    $L_2$ are given in eq.~\eqref{eq:sepL1andL2}. }
  \label{fig:Slim}
\end{figure}
%%%FIG

%%%FIG
\begin{figure}[t]
  \begin{center}
    \includegraphics[width=1.0\columnwidth]{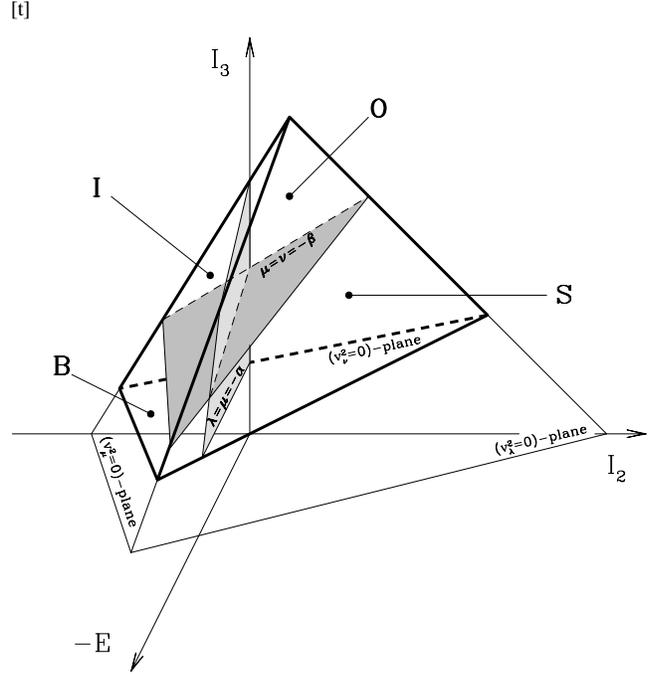}
  \end{center}
  \caption{
    The tetrahedron shows all accessible points in integral space
    $(E,I_2,I_3)$ for a given position $(\lambda,\mu,\nu)$. The
    tetrahedron is bounded by the planes for which $v_\lambda^2=0$,
    $v_\mu^2=0$, $v_\nu^2=0$ and $E=0$, respectively. The two shaded planes, which
    are given by $v_\lambda^2=v_\mu^2=0$ at $\lambda=\mu=-\alpha$ and
    $v_\mu^2=v_\nu^2=0$ at $\mu=\nu=-\beta$, divide the tetrahedron
    into the parts corresponding to the four general orbit families in a
    triaxial separable potential: box (B) orbits, inner (I) and outer
    (O) long-axis tube orbits and short-axis (S) tube orbits.}
  \label{fig:tetrahedron}
\end{figure}
%%%FIG

Due to the choice \eqref{eq:dfabel} of the DF, the general expression
(\ref{eq:defvelmom}) for the velocity moments can be simplified, as
shown by DL91 for the non-rotating components and by MD99 for the
rotating components. We recast their expressions into a different form
to facilitate the numerical implementation. The resulting velocity
moments are given by
\begin{multline}
  \label{eq:mugeneral}
  \mu_{lmn}(\lambda,\mu,\nu) = \sqrt{ \frac{2^{l+m+n+3}}
    {H_{\mu\nu}^{l+1} H_{\nu\lambda}^{m+1} H_{\lambda\mu}^{n+1}}}
    \\ \hspace{-5pt} \times \hspace{-5pt}
    \int\limits_{S_\mathrm{min}}^{S_\mathrm{max}}
      T_{lmn} \, \left[ S_\mathrm{top}(\lambda,\mu,\nu) - S \right]^{(l+m+n+1)/2}
  f(S) \, \du S,
\end{multline}
and set to zero at positions for which $S_\mathrm{max} \le
S_\mathrm{min}$.  The terms $H_{\mu\nu}$, $H_{\nu\lambda}$ and
$H_{\lambda\mu}$ in the square root in front of the integral are
defined as
\begin{equation}
  \label{eq:Hsigtau}
  H_{\sigma\tau} = 1 +
  \frac{(\sigma+\alpha)(\tau+\alpha)}{\gamma-\alpha} \, w +
  \frac{(\sigma+\gamma)(\tau+\gamma)}{\alpha-\gamma} \, u,
\end{equation}
with $\sigma,\tau=\lambda,\mu,\nu$. Orbits are confined to the region
of space for which all three terms are non-negative. In general, this
condition will not be satisfied for all points, so that the
Abel components have finite extent. From the requirement that at least
the origin $(\lambda,\mu,\nu)=(-\alpha,-\beta,-\gamma)$ should be
included, we find the following limits on $w$ and $u$
\begin{equation}
  \label{eq:limwu}
  w \ge -\frac{1}{\beta-\alpha}
  \quad \mathrm{and} \quad
  u \le \frac{1}{\gamma-\beta}.
\end{equation}
The factor $T_{lmn}$ in the integrand as well as the upper limit
$S_\mathrm{max}$ of the integral are different for each of the three
Abel component types NR, LR and SR, and are discussed in
\S\S~\ref{sec:triaxNR}--\ref{sec:triaxSR} below. The lower limit of
the integral $S_\mathrm{min}$ has to be at least as large as the
smallest value possible for the variable $S$. This limiting value
$S_\mathrm{lim}$ depends on the choice of the DF parameters $w$ and
$u$ in \eqref{eq:dfabel}, as is shown in Fig.~\ref{fig:Slim} (cf.\ 
Fig.~7 of DL91). The boundaries follow from (\ref{eq:limwu}) and the
separatrices $L_1$ and $L_2$ are given by
\begin{eqnarray}
  \label{eq:sepL1andL2}
  L_1 & : & w =
  \frac{U[-\alpha,\frac1u-\gamma,-\gamma]}{(\beta-\alpha)}
  \nonumber\\[-5pt]\\[-5pt]\nonumber
  L_2 & : & w = \frac{u}{1-(\gamma-\alpha)\,u}.
\end{eqnarray}
At a given position $(\lambda,\mu,\nu)$, orbits with different values
of the integrals of motion $E$, $I_2$ and $I_3$, and hence different
values of $S$, can contribute to the integral \eqref{eq:mugeneral}.
The restriction to bound orbits ($E\le0$) together with the
requirement that $v_\lambda^2$, $v_\mu^2$ and $v_\nu^2$ all three have
to be non-negative determines the part of the integral space that is
accessible by orbits that go through $(\lambda,\mu,\nu)$. An example
of the resulting tetrahedron in the $(E,I_2,I_3)$-space is shown in
Fig.~\ref{fig:tetrahedron} (cf.\ Fig.~1 of MD99). The largest possible
value of $S$ is given by the top of this tetrahedron
\begin{eqnarray}
  \label{eq:S_top}
  S_\mathrm{top}(\lambda,\mu,\nu) 
  \hspace{-5pt}&=&\hspace{-5pt} 
  - U[\lambda,\mu,\nu] \nonumber \\
  &&\hspace{-30pt} - w\, 
  \frac{(\lambda+\alpha)(\mu+\alpha)(\nu+\alpha)}{\gamma-\alpha} \; 
  U[\lambda,\mu,\nu,-\alpha] \nonumber \\
  &&\hspace{-30pt} - u\, 
  \frac{(\lambda+\gamma)(\mu+\gamma)(\nu+\gamma)}{\alpha-\gamma} \; 
  U[\lambda,\mu,\nu,-\gamma],
\end{eqnarray}
which is thus a function of the position $(\lambda,\mu,\nu)$. At the
origin $S_\mathrm{top}(-\alpha,-\beta,-\gamma) =
U[-\alpha,-\beta,-\gamma]$, which is the central value of the
potential $V_S$. In what follows, we normalise $V_S$ by setting
$U[-\alpha,-\beta,-\gamma]=-1$, so that $0 \le S_\mathrm{top} \le 1$.

%---------------------------------------------------------------------

\subsubsection{Non-rotating components (NR)}
\label{sec:triaxNR}

Since the non-rotating component type can exist everywhere in the
accessible integral space (the tetrahedron in Fig.~\ref{fig:tetrahedron}), we simply have that
$S_\mathrm{max} = S_\mathrm{top}(\lambda,\mu,\nu)$. Spatially the
NR components are thus bounded by the surface
$S_\mathrm{top}(\lambda,\mu,\nu)=S_\mathrm{min}$.

The factor $T_{lmn}$ follows from the cross section of the
$S$-plane within the tetrahedron and can be written in compact
form as
\begin{equation}
  \label{eq:Tlmn_NR}
  T^\mathrm{NR}_{lmn} =
  B(\textstyle{\frac{l+1}{2}},\textstyle{\frac{m+1}{2}},\textstyle{\frac{n+1}{2}}),
\end{equation}
where $B$ is the beta function of three variables\footnote{The beta
  function of $k$ variables in terms of the complete gamma
  function $\Gamma$ is defined as $B(\beta_1,\dots,\beta_k) = \Gamma(\beta_1)
  \cdots \Gamma(\beta_k) / \Gamma(\beta_1 + \cdots + \beta_k)$.}.
Since $T^\mathrm{NR}_{lmn}$ is independent of $S$ it can be taken out
of the integral (cf.\ eq.\ [3.10] of DL91), which then becomes of Abel
form.  Unfortunately, the inversion of eq.~(\ref{eq:mugeneral}) for
any chosen moment $\mu_{lmn}(\lambda, \mu, \nu)$, including the case
$l=m=n=0$, is generally impossible, as the left-hand side is a
function of three variables, while the DF depends on only one
variable, $S$. The density $\rho_\star$ specified along any given
curve will define a different $f(S)$. A case of particular interest is
to choose the density along the short axis to be $\rho_\star(0,0,z) =
\rho_S(0,0,z)$.  This defines a unique $f(S)$, and hence gives
$\rho_\star$ everywhere. Kuzmin's formula applied to $\rho_S(0,0,z)$
similarly defines the density $\rho_S$ everywhere. For single Abel DF
components these will not be the same, except in the spherical limit
(see Appendix~\ref{sec:limspherical}).

Since the orbits have no net rotation, the velocity moments
$\mu^\mathrm{NR}_{lmn}$ are only non-zero when $l$, $m$ and $n$
are all three even, and vanish in all other cases.

%---------------------------------------------------------------------

\subsubsection{Long-axis rotating components (LR)}
\label{sec:triaxLR}

The long-axis rotating component type only exists in the part of the
integral space that is accessible by the (inner and outer) long-axis
tube orbits. Within the tetrahedron for all orbits this is the region
for which $v_\nu^2\ge0$ at $\nu=-\beta$. It follows that $S_\mathrm{max}
= S_\mathrm{top}(\lambda,\mu,-\beta) \le
S_\mathrm{top}(\lambda,\mu,\nu)$.
%, so that the spatial extent of the LR components is generally smaller
%than the NR components.

The term $T_{lmn}$ follows from the cross section of the $S$-plane
within the tetrahedron \emph{and} with the above boundary plane
$v_\nu^2=0$ at $\nu=-\beta$. Without any further constraint this
results in zero net rotation, because each orbit with positive
rotation around the long axis with $v_\nu>0$, is balanced by an orbit
with opposite direction of rotation with $v_\nu<0$. Therefore, we
restrict to orbits with $v_\nu~\ge~0$, resulting in maximum streaming
around the long axis for each LR component. This reduces the
accessible integral space, and thus also the term $T_{lmn}$, by a
factor of two, so that the latter becomes
\begin{equation}
  \label{eq:Tlmn_LR}
  T^\mathrm{LR}_{lmn} =
  \frac{2\,(-2)^{(l+m)/2}\sqrt{a_0^{l+1} \, b_0^{m+1}}
  \, \mathcal{M}_0^\mathrm{LR}}{(s+1)(s-1)\dots(s+1-(l+m))},
\end{equation}
with $s=l+m+n$, the parameters $a_0$ and $b_0$ defined as
\begin{eqnarray}
  \label{eq:defa0b0}
  a_0 & = & \frac{(\lambda+\beta)\,H_{\mu\nu}\,
    \left[ S_\mathrm{top}(\lambda,\mu,-\beta) - S \right]}
  {(\lambda-\nu)\,H_{\mu(-\beta)}\,
    \left[ S_\mathrm{top}(\lambda,\mu,\nu) - S \right]},
  \nonumber\\[-5pt]\\[-5pt]\nonumber
  b_0 & = & \frac{(\mu+\beta)\,H_{\nu\lambda}\,
    \left[ S_\mathrm{top}(\lambda,\mu,-\beta) - S \right]}
  {(\mu-\nu)\,H_{(-\beta)\lambda}\,
    \left[ S_\mathrm{top}(\lambda,\mu,\nu) - S \right]},
\end{eqnarray}
which for $S \le S_\mathrm{max} = S_\mathrm{top}(\lambda,\mu,-\beta)$
are non-negative, and
\begin{equation}
  \label{eq:C_LR}
  \mathcal{M}_0^\mathrm{LR} =
  \begin{cases}
    \mathcal{M}(s,\frac{l}{2},\frac{m}{2};a_0,b_0,\frac{\pi}{2}),
    & \text{$a_0 \le b_0$}, \\
    \mathcal{M}(s,\frac{m}{2},\frac{l}{2};b_0,a_0,\frac{\pi}{2}),
    & \text{$a_0 > b_0$}.
  \end{cases}
\end{equation}
The function $\mathcal{M}(s,i,j;a,b,\phi)$ is defined in
Appendix~\ref{sec:funcM}, where we evaluate it in terms of elementary
functions (odd $s$) and elliptic integrals (even $s$).

The LR components have maximum streaming around the long axis, but the
motion parallel to the intermediate axis and short axis cancels. As a
result, the velocity moments $\mu^\mathrm{LR}_{lmn}$ vanish when $l$
or $m$ are odd\footnote{Since $l+m$ is even, the factor
  $(-2)^{(l+m)/2}$ in eq.~\eqref{eq:Tlmn_LR} is always real.}.
Multiplying $\mu^\mathrm{LR}_{lmn}$ with $(-1)^n$ results in maximum
streaming in the opposite direction. By choosing different weights for
both senses of rotation, we can control the direction and the amount
of long-axis streaming motion for each LR component.

%---------------------------------------------------------------------

\subsubsection{Short-axis rotating components (SR)}
\label{sec:triaxSR}

The short-axis component type reaches the part of integral space
accessible by the short-axis tube orbits. Within the tetrahedron for
all orbits this is the region for which $v_\mu^2\ge0$ both at
$\mu=-\beta$ and $\mu=-\alpha$ (Fig.~\ref{fig:tetrahedron}). The
latter requirement is equivalent to $I_2\ge0$. In this case,
$S_\mathrm{max} = S_\mathrm{top}(\lambda,-\alpha,\nu) \le
S_\mathrm{top}(\lambda,\mu,\nu)$.
%, and the spatial extent of the SR components is generally smaller
%than the NR components.

The form of the term $T_{lmn}$ depends on the cross section of the
$S$-plane within the tetrahedron \textit{and} with the above two
boundary planes. In case each SR component has maximum streaming
around the short axis ($v_\mu\ge0$), it is given by
\begin{equation}
  \label{eq:Tlmn_SR}
  T^\mathrm{SR}_{lmn} = \frac{2\,(-2)^{(l+n)/2}
    \sum_{i=1}^2 \sqrt{a_i^{l+1} \, c_i^{n+1}} \, \mathcal{M}_i^\mathrm{SR}}
  {(s+1)(s-1)\dots(s+1-(l+n))}.
\end{equation}
The parameters $a_1$ and $c_1$ follow from $a_0$ and $b_0$ defined in
(\ref{eq:defa0b0}) by interchanging $\mu \leftrightarrow \nu$, and in
turn $a_2$ and $c_2$ follow from $a_1$ and $c_1$ by interchanging
$\alpha \leftrightarrow \beta$. For the terms
$\mathcal{M}_i^\mathrm{SR}$ we have two possibilities, I and II,
\begin{eqnarray}
  \label{eq:C_SR_I}
  \mathcal{M}_\mathrm{I}^\mathrm{SR} & = &
  \begin{cases}
    \mathcal{M}(s,\frac{l}{2},\frac{n}{2};a_\mathrm{I},c_\mathrm{I},\theta_\mathrm{I}),
    & \text{$a_\mathrm{I} \le c_\mathrm{I}$}, \\
    \mathcal{M}(s,\frac{n}{2},\frac{l}{2};c_\mathrm{I},a_\mathrm{I},\frac{\pi}{2}) \\
    \quad - \mathcal{M}(s,\frac{n}{2},\frac{l}{2};c_\mathrm{I},a_\mathrm{I},
    \frac{\pi}{2}\!-\!\theta_\mathrm{I}),
    & \text{$a_\mathrm{I} > c_\mathrm{I}$},
  \end{cases} \\
  \label{eq:C_SR_II}
  \mathcal{M}_\mathrm{II}^\mathrm{SR} & = &
  \begin{cases}
    \mathcal{M}(s,\frac{l}{2},\frac{n}{2};a_\mathrm{II},c_\mathrm{II},\frac{\pi}{2}) \\
    \quad - \mathcal{M}(s,\frac{l}{2},\frac{n}{2};a_\mathrm{II},c_\mathrm{II},\theta_\mathrm{II}),
    & \text{$a_\mathrm{II} \le c_\mathrm{II}$}, \\
    \mathcal{M}(s,\frac{n}{2},\frac{l}{2};c_\mathrm{II},a_\mathrm{II},
    \frac{\pi}{2}\!-\!\theta_\mathrm{II}),
    & \text{$a_\mathrm{II} > c_\mathrm{II}$},
  \end{cases}
\end{eqnarray}
where $\mathcal{M}$ is given in Appendix~\ref{sec:funcM}, and $\theta_\mathrm{I}$ and $\theta_\mathrm{II}$
follow from
\begin{equation}
  \label{eq:deftheta}
  \tan^2\theta_\mathrm{I} =
  \frac{c_\mathrm{II}\,(a_\mathrm{I}-a_\mathrm{II})}
    {a_\mathrm{II}\,(c_\mathrm{II}-c_\mathrm{I})}, \quad
  \tan^2\theta_\mathrm{II} =
  \frac{c_\mathrm{I}\,(a_\mathrm{II}-a_\mathrm{I})}
  {a_\mathrm{I}\,(c_\mathrm{I}-c_\mathrm{II})}.
\end{equation}
For the assignment of the labels $I$ and $II$, we discriminate between
four cases
\begin{eqnarray}
  \label{eq:assignCi_SR}
  a_1 \le a_2, \quad c_1 \ge c_2 & : &
  \mathrm{I} \to 1, \quad \mathrm{II} \to 2 ,
  \nonumber \\
  a_1 \ge a_2, \quad c_1 \le c_2 & : &
  \mathrm{I} \to 2, \quad \mathrm{II} \to 1,
  \nonumber\\[-8pt]\\[-8pt]%
  a_1 \le a_2, \quad c_1 \le c_2 & : &
  \mathrm{I} \to 1, \, \theta_\mathrm{I}=\pi/2,
  \quad C_2^\mathrm{SR}=0,
  \nonumber \\
  a_1 \ge a_2, \quad c_1 \ge c_2 & : &
  \mathrm{I} \to 2, \, \theta_\mathrm{I}=\pi/2,
  \quad C_1^\mathrm{SR}=0.
  \nonumber
\end{eqnarray}
The SR components have maximum streaming around the short axis, so
that the velocity moments $\mu^\mathrm{SR}_{lmn}$ vanish when $l$ or
$n$ are odd. Multiplying $\mu^\mathrm{SR}_{lmn}$ with $(-1)^m$ results
in SR components with maximum streaming around the short axis in the
opposite direction.

%---------------------------------------------------------------------
\subsection{Combination of multiple DF components}
\label{sec:combmultcomp}
%---------------------------------------------------------------------

Until now, we have chosen the Abel DF to be a function of a single
variable $S=-E+wI_2+uI_3$, and we have separated it in three component
types, NR, LR and SR, but we have not made any assumption about the
form of the DF (apart from the obvious requirement that it has to be
non-negative everywhere and that it decreases to zero at large radii).
Following MD99, we choose the DF to be a linear combination of
basis functions of the form
\begin{equation}
  \label{eq:simpleDF}
  f_\delta(S) =
  \left(\frac{S-S_\mathrm{min}}{1-S_\mathrm{min}}\right)^\delta,
\end{equation}
which, like the velocity moments \eqref{eq:mugeneral}, are
non-vanishing as long as $S_\mathrm{lim} \le S_\mathrm{min} \le S \le
S_\mathrm{max} \le S_\mathrm{top} \le 1$. The exponent $\delta$ is a
(non-negative) constant.

Once the St\"ackel potential \eqref{eq:triaxV_S} is known by defining
the function $U(\tau)$, we can use the above relations
(\S~\ref{sec:abeldf}) together with the expressions in
Appendix~\ref{sec:funcM}, to compute for a given basis function
$f_\delta(S)$ the velocity moments \eqref{eq:mugeneral} for the NR, SR
and LR components in an efficient way, where at most the integral over
$S$ has to be evaluated numerically.  For the NR components this
integral can even be evaluated explicitly, resulting in
\begin{multline}
  \label{eq:simpleDFmulmnNR}
  \mu^\mathrm{NR}_{lmn,\delta}(\lambda,\mu,\nu) =
  \sqrt{ \frac{[2(S_\mathrm{max}-S_\mathrm{min})]^{l+m+n+3}}
    {H_{\mu\nu}^{l+1} H_{\nu\lambda}^{m+1} H_{\lambda\mu}^{n+1}}}
  \\ \times
  \left(\frac{S_\mathrm{max}-S_\mathrm{min}}{1-S_\mathrm{min}}\right)^\delta
  B(\textstyle{\frac{l+1}{2}},%
    \textstyle{\frac{m+1}{2}},%
    \textstyle{\frac{n+1}{2}},%
    \delta\!+\!1),
\end{multline}
where $S_\mathrm{max}=S_\mathrm{top}(\lambda,\mu,\nu)$ (cf.\
eq.~\ref{eq:S_top}).

Each DF component and corresponding velocity moments thus depend on
the choice of the DF parameters $w$, $u$ and $\delta$, the type of
component, and for the rotating components (LR and SR), they also
depend on the sense of rotation around the axis of symmetry. By
summing a series of DF basis functions over $w$, $u$ and $\delta$, one
might even expect to cover a large fraction of all physical DFs. Due
to the different values of $w$ and $u$, such a sum of DF components is
no longer a function of the same, single variable $S$, so that the
ellipsoidal hypothesis does not apply.  Consequently, it becomes
possible to construct (nearly) self-consistent dynamical models, with
the (combined) luminous mass density $\rho_\star$ equal (or close) to
the mass density $\rho_S$ associated to the potential.

%============================= section 3 =============================
\section{Observables}
\label{sec:observables}
%=====================================================================

We describe how the intrinsic velocity moments can be converted to
projected velocity moments on the plane of the sky. Alternatively,
these line-of-sight velocity moments follow as moments of the LOSVD,
which we show can be calculated in a straightforward way for Abel
models. Parameterising the LOSVD as a Gauss-Hermite series, we obtain
the observable quantities: the surface brightness, the mean
line-of-sight velocity $V$, velocity dispersion $\sigma$, and
higher-order Gauss-Hermite moments $h_3$, $h_4$, \dots
%, all as function of position on the sky plane.

%---------------------------------------------------------------------
\subsection{From intrinsic to observer's coordinate system}
\label{sec:int2obscoordsystem}
%---------------------------------------------------------------------

In order to calculate line-of-sight velocity moments, we introduce a
new Cartesian coordinate system $(x'',y'',z'')$, with $x''$ and $y''$
in the plane of the sky and the $z''$-axis along the line-of-sight.
Choosing the $x''$-axis in the $(x,y)$-plane of the intrinsic
coordinate system (cf.\ de Zeeuw \& Franx
1989\nocite{1989ApJ...343..617D} and their Fig.~2), the transformation
between both coordinate systems is known once two viewing angles, the
polar angle $\vartheta$ and azimuthal angle $\varphi$, are specified.
The intrinsic $z$-axis projects onto the $y''$-axis, which for an
axisymmetric galaxy model aligns with the short axis of the projected
surface density $\Sigma$. However, for a triaxial galaxy model the
$y''$-axis in general lies at an angle $\psi$ with respect to the
short axis of $\Sigma$. This misalignment $\psi$ can be expressed in
terms of the viewing angles $\vartheta$ and $\varphi$ and the
triaxiality parameter $T$ (defined in eq.~\ref{eq:triaxparT}) as
follows (cf.\ eq.~B9 of Franx 1988)
\begin{equation}
  \label{eq:misalignment_psi}
    \tan2\psi = - \frac{T\sin2\varphi\cos\vartheta}
    {\sin^2\vartheta
      -T\left(\cos^2\varphi-\sin^2\varphi\cos^2\vartheta\right)}
\end{equation}
with $\sin2\psi\sin2\varphi\cos\vartheta\le0$ and
$-\pi/2\le\psi\le \pi/2$. A rotation over $\psi$ transforms the
coordinate system $(x'',y'',z'')$ to $(x',y',z')$, with the
$x'$-axis and $y'$-axis aligned with respectively the major and
minor axis of $\Sigma$, whereas $z'=z''$ is along the
line-of-sight.

The expressions in \S~\ref{sec:abeldf} involve the velocity components
in the confocal ellipsoidal coordinate system $(\lambda,\mu,\nu)$. The
conversion to line-of-sight quantities can be done by four successive
matrix transformations. First, we obtain the velocity components in
the first octant of the intrinsic Cartesian coordinate system
$(x,y,z)$ via $\mathbf{Q}$, of which the first element is given by
(cf.\ DL91)
\begin{equation}
  \label{eq:convN11}
  Q_{11} = \mathrm{sign}(\lambda+\alpha)\,
  \sqrt{\frac{(\mu+\alpha)(\nu+\alpha)(\lambda+\beta)(\lambda+\gamma)}
  {(\alpha-\beta)(\alpha-\gamma)(\lambda-\mu)(\lambda-\nu)}},
\end{equation}
and the other elements follow horizontally by cyclic permutation of
$\lambda \to \mu \to \nu \to \lambda$ and vertically by cyclic
permutation of $\alpha \to \beta \to \gamma \to \alpha$. The second
matrix uses the symmetries of the orbits to compute the appropriate
signs of the intrinsic Cartesian velocities in the other octants. The
result depends on whether or not the orbit has a definite sense of
rotation in one of the confocal coordinates.  For the three types of
Abel components this results in the following matrices
\begin{eqnarray}%
\label{eq:signmatS}%
\text{NR} &:& \mathbf{S} =
\mathrm{diag}[\mathrm{sgn}(x),\mathrm{sgn}(y),\mathrm{sgn}(z)]
\nonumber\\%
\text{LR} &:& \mathbf{S} =
\mathrm{diag}[\mathrm{sgn}(xyz),\mathrm{sgn}(z),\mathrm{sgn}(y)]
\\%
\text{SR} &:& \mathbf{S} =
\mathrm{diag}[\mathrm{sgn}(y),\mathrm{sgn}(x),\mathrm{sgn}(xyz)]
\nonumber
\end{eqnarray}
Finally, the conversion from the intrinsic to the observer's
Cartesian velocities involves the same projection and rotation as
for the coordinates. We represent these two coordinate
transformations respectively by the projection matrix
\begin{equation}\label{eq:projmatP}
  \mathbf{P} =
  \begin{pmatrix}
    -\sin\varphi           & \cos\varphi            & 0 \\
    -\cos\vartheta\cos\varphi & -\cos\vartheta\sin\varphi & \sin\vartheta \\
     \sin\vartheta\cos\varphi & \sin\vartheta\sin\varphi  & \cos\vartheta
  \end{pmatrix},
\end{equation}
and the rotation matrix
\begin{equation}\label{eq:rotmatR}
  \mathbf{R} =
  \begin{pmatrix}
     \cos\psi & -\sin\psi & 0 \\
     \sin\psi &  \cos\psi & 0 \\
     0        &  0        & 1
  \end{pmatrix}.
\end{equation}
In this way, we arrive at the following relation
\begin{equation}
  \label{eq:vel_ell2carprime}
  \begin{pmatrix}
    v_{x'} \\
    v_{y'} \\
    v_{z'}
  \end{pmatrix}
  = \mathbf{M}
  \begin{pmatrix}
    v_\lambda \\
    v_\mu \\
    v_\nu
  \end{pmatrix},
  \quad \mathrm{with} \quad
  \mathbf{M}\equiv\mathbf{R}\mathbf{P}\mathbf{S}\mathbf{Q},
%  \mathbf{M}\equiv\mathbf{S}\mathbf{P}\mathbf{Q}\mathbf{R},
% "Senatus Populusque Romanus" or "Sono Pazzi Questi Romani" by Obelix
\end{equation}
where the full transformation matrix $\mathbf{M}$ is thus a function
of $(\lambda,\mu,\nu)$, the constants $(\alpha,\beta,\gamma)$ and the
viewing angles $(\vartheta,\varphi,\psi)$.

\subsection{Line-of-sight velocity moments}
\label{sec:losvelmom}

We can now write each velocity moment in the observer's Cartesian
coordinate system $(x',y',z')$ as a linear combination of the
velocity moments in the confocal ellipsoidal coordinate system
\begin{equation}
  \mu_{ijk}(x',y',z') =
  \sum_{l,m,n} c_{l,m,n}(s) \, \mu_{lmn}(\lambda,\mu,\nu),
\end{equation}
with $s=i+j+k=l+m+n$. The coefficients $c_{l,m,n}(s)$ are products of
elements of the transformation matrix $\mathbf{M}$ in
eq.~\eqref{eq:vel_ell2carprime}.  They can be obtained with the
following recursive algorithm
\begin{equation}
  \label{eq:coeff_cslmn}
  c_{l,m,n}(s) = 
  \begin{cases}
    c_{1,0,0}(s) \, c_{l-1,m,n}(s-1),
    & \text{if $l>0$}, \\
    c_{0,1,0}(s) \, c_{0,m-1,n}(s-1),
    & \text{if $m>0$}, \\
    c_{0,0,1}(s) \, c_{0,0,n-1}(s-1),
    & \text{if $n>0$},
  \end{cases}
\end{equation}
with the first order expressions given by
\begin{equation}\label{eq:coeff_cs=1}
  c_{1,0,0}(s) = M_{e_s1}, \quad
  c_{0,1,0}(s) = M_{e_s2}, \quad
  c_{0,0,1}(s) = M_{e_s3},
\end{equation}
and $c_{0,0,0} = 1$. The index $e_s$ is the $s$\,th element of the
vector $\mathbf{e} = [3,..,3,2,..,2,1,..,1]$, where the number of
integers 3 ($\#3$) is equal to the value of the velocity moment index
$k$, and similarly $\#2=j$ and $\#1=i$.

The line-of-sight velocity moments now follow from (numerical)
integration of $\mu_{00k}$ along the line-of-sight
\begin{equation}%
  \label{eq:deflosvdmom1}%
  \mu_k(x',y') = \int_{-\infty}^\infty  \mu_{00k}(x',y',z') \, \du z',
\end{equation}
which are thus functions of position on the sky plane.

%---------------------------------------------------------------------
\subsection{Line-of-sight velocity distribution}
\label{sec:losveldistr}
%---------------------------------------------------------------------

Using the definition of the intrinsic velocity moments of the DF
(eq.~\ref{eq:defvelmom}) and rearranging the sequence of integration,
we rewrite eq.~\eqref{eq:deflosvdmom1} for the line-of-sight
velocity moments as
\begin{equation}%
  \label{eq:deflosvdmom2}%
  \mu_k(x',y') = \int_{-\infty}^\infty v_{z'}^k
  \mathcal{L}(x',y',v_{z'}) \, \du v_{z'},
\end{equation}
where we have introduced the LOSVD
\begin{equation}
  \label{eq:defLOSVD}
  \mathcal{L}(x',y',v_{z'}) = \iiint \!\! f(E,I_2,I_3) \,
  \du v_{x'} \, \du v_{y'} \, \du z'.
\end{equation}
Although the integral over $z'$ in general can only be evaluated
numerically, we show that for the choice~\eqref{eq:dfabel} of the DF,
the double integral over the velocities can be simplified
significantly. 

Our analysis generalises the results for the well-known spherical
Osipkov--Merritt models. We describe the spherical limit together with
the elliptic disc and large distance limit in Appendix~A, while we
present axisymmetric Abel models in \S~\ref{sec:axivelmom}.

%---------------------------------------------------------------------

\subsubsection{Abel LOSVD}
\label{sec:triaxabelLOSVD}

Substituting the expressions \eqref{eq:triaxEI2I3} for the integrals
of motion in $S=-E+w I_2+u I_3$, we obtain
\begin{equation}
  \label{eq:triaxexprS}
  S = S_\mathrm{top}(\lambda,\mu,\nu) - \textstyle{\frac12}
  \left(H_{\mu\nu}\,v_\lambda^2 + H_{\nu\lambda}\,v_\mu^2 + H_{\lambda\mu}\,v_\nu^2\right),
\end{equation}
where the expression for $S_\mathrm{top}(\lambda,\mu,\nu)$ is given by
eq.~\eqref{eq:S_top} and the terms $H_{\mu\nu}$, $H_{\nu\lambda}$ and
$H_{\lambda\mu}$ are defined in eq.~\eqref{eq:Hsigtau}. Defining
\begin{equation}
  \label{eq:triaxdefXYZ}
  X^2 = \frac{ H_{\mu\nu} }
  {2\left[ S_\mathrm{top}(\lambda,\mu,\nu) - S \right]} \, v_\lambda^2,
%  X = \sqrt{\frac{ H_{\mu\nu} }
%  {2\left[ S_\mathrm{top}(\lambda,\mu,\nu) - S \right]}} \; v_\lambda,
\end{equation}
and similarly $Y$ and $Z$ by cyclic permutation of
$\lambda\to\mu\to\nu\to\lambda$, we can write the expression for $S$
as
\begin{equation}
  \label{eq:triaxsphereXYZ}
  X^2 + Y^2 + Z^2 = 1.
\end{equation}
For a given position $(\lambda,\mu,\nu)$, each value of $S$ thus
defines the surface of the unit sphere in the variables $(X,Y,Z)$. In
these variables, we can write the integral of the DF over velocities,
i.e., the stellar mass density, as
\begin{eqnarray}
  \label{eq:triaxrhoSXYZ}
  \rho_\star
  \hspace{-5pt} & = & \hspace{-5pt}
  \iiint \!\! f(S) \, \du v_{x'} \, \du v_{y'} \, \du v_{z'},
  \nonumber \\
  \hspace{-5pt} & = & \hspace{-5pt}
  \int\limits_{S_\mathrm{min}}^{S_\mathrm{max}} \!\!
  \sqrt{\frac{2\left[S_\mathrm{top} - S\right]}
    {H_{\mu\nu} H_{\nu\lambda} H_{\lambda\mu}}} \; f(S)
  \left[ \underset{X^2+Y^2+Z^2=1}{\iiint\!\!\du X\du Y\du Z} \right]
  \du S.
\end{eqnarray}
This is the same expression as for the zeroth-order velocity moment of
the DF, $\mu_{000}$, in eq.~\eqref{eq:mugeneral}, where $2\,T_{000}$
is equal to the integral between square brackets.

The matrix $\mathbf{M}$ in eq.~\eqref{eq:vel_ell2carprime} provides
the conversion from the velocity components in the confocal
ellipsoidal coordinate system, $(v_\lambda,v_\mu,v_\nu)$, to those in
the observer's Cartesian coordinate system, $(v_{x'},v_{y'},v_{z'})$.
Hence, for a given line-of-sight velocity $v_{z'}$, we find
\begin{equation}
  \label{eq:triaxplaneXYZ}
  e_1 \, X + e_2 \, Y + e_3 \, Z = v_{z'} \, / \, g(S).
\end{equation}
The coefficients $e_1$, $e_2$ and $e_3$ are defined as
\begin{eqnarray}
  \label{eq:defe1e2e3}
  h\,e_1 & = & \sqrt{H_{\nu\lambda}H_{\lambda\mu}} \, M_{31},
  \nonumber \\
  h\,e_2 & = & \sqrt{H_{\lambda\mu}H_{\mu\nu}} \, M_{32},
  \\ \nonumber
  h\,e_3 & = & \sqrt{H_{\mu\nu}H_{\nu\lambda}} \, M_{33},
\end{eqnarray}
and normalised with respect to $h$ given by
\begin{equation}
  \label{eq:defnormh}
  h^2 = H_{\nu\lambda}H_{\lambda\mu} \, M_{31}^2
  + H_{\lambda\mu}H_{\mu\nu} \, M_{32}^2
  + H_{\mu\nu}H_{\nu\lambda} \, M_{33}^2.
\end{equation}
These coefficients are functions of the position $(\lambda,\mu,\nu)$,
the constants $(\alpha,\beta,\gamma)$ and the viewing angles
$(\vartheta,\varphi,\psi)$ through the components of the matrix
$\mathbf{M}$, and also depend on the DF parameters $w$ and $u$ through
the terms $H_{\lambda\mu}$, $H_{\mu\nu}$ and $H_{\nu\lambda}$. It
follows that
\begin{equation}
  \label{eq:deffuncg}
  g(S) =  h \, \sqrt{\frac{2\left[ S_\mathrm{top} - S \right]}
    { H_{\lambda\mu} H_{\mu\nu} H_{\nu\lambda} }},
\end{equation}
which is a function of the variable $S$.

We thus find that each combination of values of $S$ and $v_{z'}$
results in the cross section of the surface of the unit sphere in
eq.~\eqref{eq:triaxsphereXYZ} with the plane in
eq.~\eqref{eq:triaxplaneXYZ}, i.e., a circle, in the variables
$(X,Y,Z)$. We rotate the latter coordinate system such that the normal
vector $(e_1,e_2,e_3)$ of the plane of the circle coincides with the
$Z'$-axis of the system given by
\begin{equation}
  \label{eq:triaxXYZtoXYZprime2}
  \begin{pmatrix}
    X \\
    Y \\
    Z
  \end{pmatrix}
  \!\! = \!\!
  \begin{pmatrix}
    \cos\Phi  & \sin\Phi\cos\Theta & \sin\Phi\sin\Theta \\
    -\sin\Phi & \cos\Phi\cos\Theta & \cos\Phi\sin\Theta \\
    0         &        -\sin\Theta &         \cos\Theta
  \end{pmatrix}
  \begin{pmatrix}
    X' \\
    Y' \\
    Z'
  \end{pmatrix}.
\end{equation}
where the rotation angles $\Phi$ and $\Theta$ follow from
\begin{equation}
  \label{eq:defPhiTheta}
  \tan\Phi = \frac{e_1}{e_2}, \qquad \tan\Theta = \frac{\sqrt{e_1^2+e_2^2}}{e_3}.
\end{equation}
In these coordinates the circle is conveniently parameterised as
\begin{equation}
  \label{eq:triaxparcircle}
  X' = \sqrt{1-Z'^2}\,\cos\xi', \quad
  Y' = \sqrt{1-Z'^2}\,\sin\xi',
\end{equation}
where $Z'=v_{z'}/g(S)$. We can now rewrite the integral between square brackets in
eq.~\eqref{eq:triaxrhoSXYZ} as
\begin{equation}
  \label{eq:triaxintgsphsurf}
  \iint \left| \frac{\partial \mathbf{R}}{\partial \xi'} \wedge
  \frac{\partial \mathbf{R}}{\partial Z'} \right| \, \du \xi' \, \du Z' =
  \frac{1}{g(S)} \iint \du \xi' \, \du v_{z'},
\end{equation}
where the vector $\mathbf{R}=(X,Y,Z)$ and $\wedge$ indicates the cross
product. The integral over $\xi'$ is the length of the part of the
circle, $\Delta\xi'$, for which the corresponding integral space is
accessible by orbits, and hence is in general a function of $S$ and
$v_{z'}$ and differs for the different types of Abel components as we
show below.

Inserting eq.~\eqref{eq:triaxintgsphsurf} in
eq.~\eqref{eq:triaxrhoSXYZ}, we obtain
\begin{eqnarray}
  \label{eq:triaxrhoSvzpxip}
  \rho_\star
  \hspace{-5pt} & = & \hspace{-5pt}
  \frac1h
  \int\limits_{S_\mathrm{min}}^{S_\mathrm{max}}
  \int\limits_{-g(S)}^{g(S)} \!\!
  f(S) \, \Delta\xi'(v_{z'},S) \, \du v_{z'} \, \du S,
  \nonumber \\
  \hspace{-5pt} & = & \hspace{-5pt}
  \frac1h
  \int\limits_{-g(S_\mathrm{min})}^{g(S_\mathrm{min})}
  \int\limits_{S_\mathrm{min}}^{S_\mathrm{up}(v_{z'})} \!\!
  f(S) \, \Delta\xi'(v_{z'},S) \, \du S \; \du v_{z'},
\end{eqnarray}
where after changing the order of integration in the last step, the
upper limit of $S$ is given by $S_\mathrm{up} =
\min[G(v_{z'}),S_\mathrm{max}]$, with
\begin{equation}
  \label{eq:triaxdefGvzp}
  G(v_{z'}) = S_\mathrm{top}(\lambda,\mu,\nu) -
  H_{\mu\nu} H_{\nu\lambda} H_{\lambda\mu} \frac{v_{z'}^2}{2h^2}.
\end{equation}
Comparing the first line of eq.~\eqref{eq:triaxrhoSXYZ} with the
second line of eq.~\eqref{eq:triaxrhoSvzpxip}, we see that the choice
of the Abel DF, $f(E,I_2,I_3)=f(S)$, indeed reduces the triple
integration \eqref{eq:defLOSVD} for the LOSVD to a double integral:
\begin{equation}
  \label{eq:triaxAbelLOSVD}
  \mathcal{L}(x',y',v_{z'}) =
  \int\limits_{-\infty}^{\infty} \!\!
  \frac1h \!\!
  \int\limits_{S_\mathrm{min}}^{S_\mathrm{up}(v_{z'})} \!\!
  f(S)\,\Delta\xi'(v_{z'},S)
  \, \du S \, \du z',
\end{equation}
and vanishes when $|v_{z'}|$ exceeds the 'terminal velocity' $v_t =
g(S_\mathrm{min})$. The expressions for $h$ and $S_\mathrm{up}$ follow
from eqs \eqref{eq:defnormh} and \eqref{eq:triaxdefGvzp}, whereas
$S_\mathrm{max}$ and $\Delta\xi'$ are different for each of the three
Abel component types and are considered next.

%---------------------------------------------------------------------

\subsubsection{Non-rotating components (NR)}
\label{sec:triaxLOSVD_NR}

As for the intrinsic moments in \S~\ref{sec:triaxNR}, we have for the
non-rotating component type that $S_\mathrm{max} =
S_\mathrm{top}(\lambda,\mu,\nu)$, and, since the full integral space is
accessible, $\Delta\xi'_\mathrm{NR} = 2\pi$, independent of $S$ and
$v_{z'}$.

In the case of a basis function $f_\delta(S)$ as defined in
eq.~\eqref{eq:simpleDF}, the integral over $S$ can be evaluated
explicitly resulting in
\begin{equation}
  \label{eq:triaxLOSVD_NRsimpleDf}
  \mathcal{L}^\mathrm{NR}_\delta =
  \frac{2\pi}{(\delta\!+\!1)(1\!-\!S_\mathrm{min})^\delta}
  \int\limits_{-\infty}^{\infty} \!\!
  \frac1h
  \left[ G(v_{z'}) \!-\! S_\mathrm{min} \right]^{\delta+1} \, \du z'.
\end{equation}
%

%%%FIG
\begin{figure*}
  \begin{center}
    \includegraphics[width=0.45\textwidth]{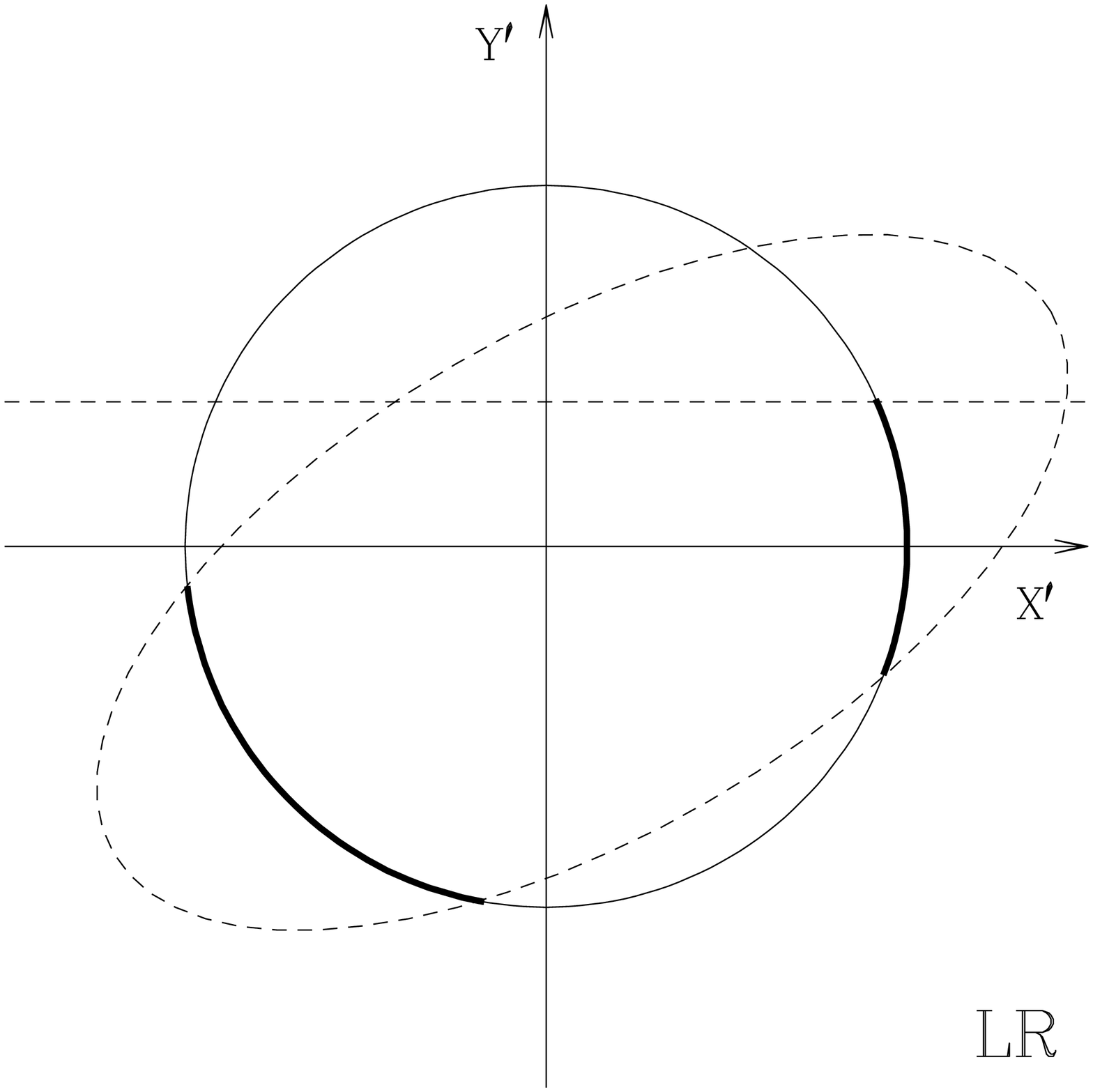}
    \hfill
    \includegraphics[width=0.45\textwidth]{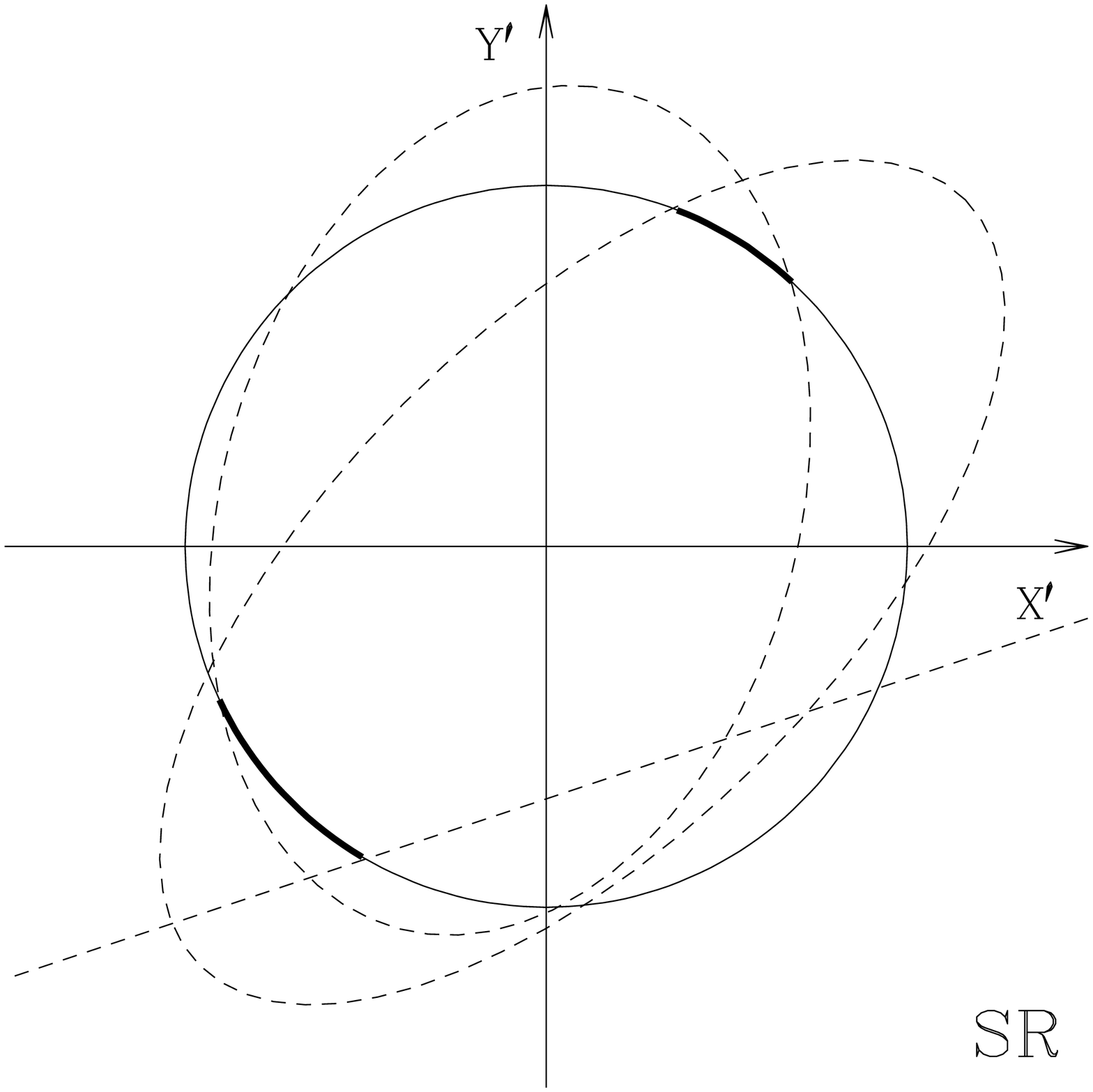}
  \end{center}
  \caption{The part of integral space accessible by long-axis (left
    panel) and short-axis (right panel) tube orbits, for given values
    of the Abel DF variable $S$ and the line-of-sight velocity
    $v_{z'}$. This corresponds to the thick part of the circle, which
    is restricted to be within the dashed ellipse(s) and below [above]
    the dashed line for long-axis [short-axis] rotating components
    with maximum streaming (see text for details). The length of the
    thick part of the circle equals $\Delta\xi'$ in the expression
    \eqref{eq:triaxAbelLOSVD} of the LOSVD.}
  \label{fig:partintspace}
\end{figure*}
%%%FIG

%---------------------------------------------------------------------

\subsubsection{Long-axis rotating components (LR)}
\label{sec:triaxLOSVD_LR}

The integral space accessible by the (inner and outer) long-axis tube
orbits is given by $v_\nu^2\ge0$ at $\nu=-\beta$, so that immediately
$S_\mathrm{max} = S_\mathrm{top}(\lambda,\mu,-\beta)$, whereas the
calculation of $\Delta\xi'_\mathrm{LR}$ is more complex.

Since eq.~\eqref{eq:triaxsphereXYZ} must also hold at $\nu=-\beta$, we
find that for LR components, within the unit sphere in the variables
$(X,Y,Z)$, the space is restricted to that within the elliptic
cylinder given by
\begin{equation}
  \label{eq:triaxellcyl_LR}
  \frac{X^2}{a_0} + \frac{Y^2}{b_0} = 1,
\end{equation}
where $a_0$ and $b_0$ are defined in eq.~\eqref{eq:defa0b0}. In the
rotated coordinate system $(X',Y',Z')$ defined in
eq.~\eqref{eq:triaxXYZtoXYZprime2}, at height $Z'=v_{z'}/g(S)$, the
elliptic cylinder results in an ellipse given by
\begin{equation}
  \label{eq:triaxellxip_LR}
  d_1 X'^2 + d_2 Y'^2 + d_3 X'Y' + d_4 X' + d_5 Y' + d_6 = 0
\end{equation}
for $0\le\xi'\le2\pi$ and with coefficients
\begin{eqnarray}
  \label{eq:triaxellcoeff_LR}
  d_1 & = & (b_0 \, e_2^2 + a_0 \, e_1^2), \nonumber\\
  d_2 & = & e_3^2 \, (b_0 \, e_1^2 + a_0 \, e_2^2), \nonumber\\
  d_3 & = & 2 \, e_1 e_2 e_3 \, (b_0 - a_0),
  \nonumber\\[-7pt]\\[-7pt]
  d_4 & = & 2 \, e_1 e_2 \, (e_1^2 + e_2^2)^\frac12 \,
  (b_0 - a_0) \, (v_{z'}/g), \nonumber\\
  d_5 & = & 2 \, e_3 \, (e_1^2 + e_2^2)^\frac12  \,
  (b_0 \, e_1^2 + a_0 \, e_2^2) \, (v_{z'}/g), \nonumber\\
  d_6 & = & (e_1^2 + e_2^2) \,
  \left[ (b_0 \, e_1^2 + a_0 \, e_2^2) \, (v_{z'}/g)^2 - a_0 b_0 \right]. \nonumber
\end{eqnarray}
Because all the above relations only involve the squared values of
$X$, $Y$ and $Z$, they are independent of the sign of the
corresponding velocities $v_\lambda$, $v_\mu$ and $v_\nu$ (cf.\ 
eq.~\ref{eq:triaxdefXYZ}), which results in zero net rotation. For the
LR components, to obtain net rotation around the long axis, we simply
limit the range of $v_\nu$ values, e.g., requiring $Z\ge0$, results in
maximum streaming motion around the long axis. This restricts the
space in $(X',Y',Z')$, at given height $Z'=v_{z'}/g(S)$, to one side
of the line
\begin{equation}
  \label{eq:triaxstreaming_LR}
  -(e_1^2 + e_2^2)^\frac12 \, Y' + e_3 \, (v_{z'}/g) = 0.
\end{equation}
The restriction to the opposite side of the line inverts the rotation
around the long axis. By choosing different weights for both senses of
rotation, we can control the direction and the amount of long-axis
streaming motion.

For given values of $S$ and $v_{z'}$, the integral space covered by
the LR components is thus the part of the circle in
eq.~\eqref{eq:triaxparcircle} that falls within the ellipse in
eq.~\eqref{eq:triaxellxip_LR} and that is on the correct side of the
line in eq.~\eqref{eq:triaxstreaming_LR} (see also
Fig.~\ref{fig:partintspace}). The length $\Delta\xi'_\mathrm{LR}$ of
this part thus ranges from zero to a maximum of $2\pi$ when the circle
is completely inside the ellipse and on the correct side of the line.
To compute this length, we determine the points where the circle
(possibly) intersects the ellipse and the line. Substituting the
circle parameterisation of eq.~\eqref{eq:triaxparcircle} in the
expression for the ellipse in eq.~\eqref{eq:triaxellxip_LR}, we find
that the intersections with the ellipse are the (real) zero points of
the following fourth order polynomial in $u\equiv\tan(\xi'/2)$
\begin{multline}
  \label{eq:triaxpoly4thorderu}
  (d_6 - d_4 R' + d_1 R'^2) \, u^4
  + 2 R' (d_5 - d_3 R') \, u^3
  \\
  + 2 \left[d_6 + (2 d_2 - d_1) R'^2\right] \, u^2
  + 2 R' (d_5 + d_3 R') \, u
  \\
  + (d_6 + d_4 R' + d_1 R'^2) = 0,
\end{multline}
where we have introduced $R'\equiv\sqrt{1-(v_{z'}/g)^2}$. The
intersections with the line result in the following two solutions
\begin{equation}
  \label{eq:triaxline2angles_LR}
  u_\pm = \frac{(e_1^2+e_2^2)^\frac12 R' \pm 
    \left[1-e_3^2-(v_{z'}/g)^2\right]^\frac12}
  {e_3\,(v_{z'}/g)},
\end{equation}
for $|v_{z'}| \le g(S) (1-e_3^2)^\frac12$, otherwise the line is
outside the circle.

We thus (numerically) find up to six real zero points $u_i$ and
corresponding angles $\xi'_i = 2\arctan(u_i)$, sorted from low to
high. For the set $\{-\pi,\xi'_1,\xi'_2,\dots,\pi\}$, we compute the
lengths of the sequential intervals on the circle for which the
corresponding values fall within the ellipse and on the correct side
of the line. This can be checked by inserting a value from the
corresponding interval, e.g.\ the central value, in
eq.~\eqref{eq:triaxparcircle} and substituting the resulting $X'$ and
$Y'$ into eqs \eqref{eq:triaxellxip_LR} and
\eqref{eq:triaxstreaming_LR}. If the left-hand side is negative
(positive), the interval is inside (outside) the ellipse, and (for
$Z\ge0$) on the wrong (correct) side of the line. Finally, the sum of
the resulting interval lengths provides $\Delta\xi'_\mathrm{LR}$.

%---------------------------------------------------------------------

\subsubsection{Short-axis rotating components (SR)}
\label{sec:triaxLOSVD_SR}

The short-axis tube orbits are restricted to the region of integral
space for which $v_\mu^2\ge0$ both at $\mu=-\beta$ and $\mu=-\alpha$,
and hence $S_\mathrm{max} = S_\mathrm{top}(\lambda,-\alpha,\nu)$. For
the calculation of $\Delta\xi'_\mathrm{SR}$ we have that the space
within the unit sphere in $(X,Y,Z)$ is now restricted to the part that
falls within \textit{both} the elliptic cylinders
\begin{equation}
  \label{eq:triaxellcyl_SR}
  \frac{X^2}{a_1} + \frac{Z^2}{c_1} = 1
  \qquad \mathrm{and} \qquad
  \frac{X^2}{a_2} + \frac{Z^2}{c_2} = 1.
\end{equation}
As in \S~\ref{sec:triaxSR} for the intrinsic moments, $a_1$ and $c_1$
follow from $a_0$ and $b_0$ defined in (\ref{eq:defa0b0}) by
interchanging $\mu \leftrightarrow \nu$, and in turn $a_2$ and $c_2$
follow from $a_1$ and $c_1$ by interchanging $\alpha \leftrightarrow
\beta$.

Both elliptic cylinders result in ellipses in the $Z'$-plane, as in
eq.~\eqref{eq:triaxellxip_LR} for the LR components, but now with
coefficients
\begin{eqnarray}
  \label{eq:triaxellcoeff_SR}
  d_1 & = & c_i \, e_2^2, \nonumber\\
  d_2 & = & c_i \, e_1^2  e_3^2 + a_i \, (e_1^2 + e_2^2)^2, \nonumber\\
  d_3 & = & 2 \, c_i \, e_1 e_2 e_3
  \nonumber\\[-7pt]\\[-7pt]
  d_4 & = & 2 \, c_i \, e_1 e_2 \, (e_1^2 + e_2^2)^\frac12 \, (v_{z'}/g), \nonumber\\
  d_5 & = & 2 \, e_3 \, (e_1^2 + e_2^2)^\frac12
  \left[ c_i \, e_1^2 - a_i \, (e_1^2 + e_2^2) \right] (v_{z'}/g), \nonumber\\
  d_6 & = & (e_1^2 + e_2^2)
  \left[ (c_i \, e_1^2 + a_i \, e_3^2) \, (v_{z'}/g)^2 - a_i c_i \right]. \nonumber
\end{eqnarray}
for $i=1,2$ respectively. The zero points of the corresponding fourth
order polynomials in eq.~\eqref{eq:triaxpoly4thorderu} are again the
intersections with the circle in eq.~\eqref{eq:triaxparcircle}. 

Net rotation around the short axis follows by limiting the range of
$v_\mu$ values, e.g., $Y\ge0$ yields maximum streaming, which
restricts the accessible integral space to one side of the line
\begin{equation}
  \label{eq:triaxstreaming_SR}
  -e_1 \, X' + e_2 e_3 \, Y' + e_2 \, (e_1^2 + e_2^2)^\frac12 \, (v_{z'}/g) = 0.
\end{equation}
The two solutions of the intersection with the circle are
\begin{equation}
  \label{eq:triaxline2angles_SR}
  u_\pm = \frac{-e_2 e_3 R' \pm 
    (e_1^2+e_2^2)^\frac12 \left[1-e_2^2-(v_{z'}/g)^2\right]^\frac12}
  {e_1 R' + (e_1^2+e_2^2)^\frac12 e_2\,(v_{z'}/g)},
\end{equation}
for $|v_{z'}| \le g(S) (1-e_2^2)^\frac12$, otherwise the line is
outside the circle.

The combination of all the (real) zero points provides the (ordered)
set $\{-\pi,\xi'_1,\xi'_2,\dots,\pi\}$, with at most ten intersections
$\xi'_i$ with the circle given in eq.~\eqref{eq:triaxparcircle}.  We
compute the lengths of the circle intervals for which the enclosed
values fall within both ellipses and on the correct side of the line.
This means, for which the corresponding $X'$ and $Y'$ values
substituted in eq.~\eqref{eq:triaxellxip_LR} result in a negative
left-hand side for both pairs of $a_i$ and $b_i$, and (for $Y\ge0$) in
a positive left-hand side of eq.~\eqref{eq:triaxstreaming_SR}.
Finally, $\Delta\xi'_\mathrm{SR}$ is the sum of the resulting interval
lengths.

%---------------------------------------------------------------------

\subsubsection{Other type of components}
\label{sec:triaxLOSVD_other}

When considering the LR type of components we make no distinction
between inner and outer long-axis tube orbits because they have
similar dynamical properties. Similarly, the non-rotating box orbits
are part of the NR type of components and are not considered
separately. Nevertheless, if we are interested in the specific
contribution of these orbit families to the LOSVD, this can be
achieved by a straightforward extension of the above analysis.

As can be seen from Fig.~\ref{fig:tetrahedron}, the inner and outer
long-axis tube orbits are separated by the plane $I_2=0$, or
equivalently the regions for which $v_\lambda^2\ge0$ at
$\lambda=-\alpha$ and $v_\mu^2\ge0$ at $\mu=-\alpha$, respectively.
This is in addition to the restriction $v_\nu^2\ge0$ at $\nu=-\beta$
for both long-axis tube orbits. For the \textit{inner} long-axis tube
orbits this implies that $S_\mathrm{max} =
S_\mathrm{top}(\lambda,\mu,-\beta)$. The space within the unit sphere
in $(X,Y,Z)$ is now restricted to the part that falls within the
intersection of the elliptic cylinders in
eq.~\eqref{eq:triaxellcyl_LR} and
\begin{equation}
  \label{eq:triaxellcyl_Itube}
  \frac{Y^2}{b_3} + \frac{Z^2}{c_3} = 1,
\end{equation}
where $b_3$ and $c_3$ follow from $a_0$ and $b_0$ defined in
(\ref{eq:defa0b0}) by interchanging $\nu \leftrightarrow \lambda$ and
$\beta \leftrightarrow \alpha$. In the $Z'$-plane, these two elliptic
cylinders result in ellipses as in eq.~\eqref{eq:triaxellxip_LR}, with
coefficients respectively given in eq.~\eqref{eq:triaxellcoeff_LR} and
\begin{eqnarray}
  \label{eq:triaxellcoeff_Itube}
  d_1 & = & c_3 \, e_1^2, \nonumber\\
  d_2 & = & c_3 \, e_2^2  e_3^2 + b_3 \, (e_1^2 + e_2^2)^2, \nonumber\\
  d_3 & = & - 2 \, c_3 \, e_1 e_2 e_3
  \nonumber\\[-7pt]\\[-7pt]
  d_4 & = & - 2 \, c_3 \, e_1 e_2 \, (e_1^2 + e_2^2)^\frac12 \, (v_{z'}/g), \nonumber\\
  d_5 & = & 2 \, e_3 \, (e_1^2 + e_2^2)^\frac12
  \left[ c_3 \, e_1^2 - b_3 \, (e_1^2 + e_2^2) \right] (v_{z'}/g), \nonumber\\
  d_6 & = & (e_1^2 + e_2^2)
  \left[ (c_3 \, e_2^2 + b_3 \, e_3^2) \, (v_{z'}/g)^2 - b_3 c_3 \right]. \nonumber
\end{eqnarray}
As before, $\Delta\xi'$ follows from the combination of the real zero
points of the corresponding fourth order polynomials in
eq.~\eqref{eq:triaxpoly4thorderu}, and of
eq.~\eqref{eq:triaxstreaming_LR} in the case of maximum streaming
around the long axis. For the \textit{outer} long-axis tube orbits,
$S_\mathrm{max} = \min[ S_\mathrm{top}(\lambda,\mu,-\beta) ,
S_\mathrm{top}(\lambda,-\alpha,\nu) ]$. The two elliptic cylinders are
the one in eq.~\eqref{eq:triaxellcyl_LR} and the second in
eq.~\eqref{eq:triaxellcyl_SR}, with the coefficients of the
corresponding ellipses in the $Z'$-plane are given in
eq.~\eqref{eq:triaxellcoeff_LR} and eq.~\eqref{eq:triaxellcoeff_SR}
($i=2$), respectively.

The part of integral space accessible by box orbits is the region for
which both $v_\mu^2\ge0$ at $\mu=-\beta$ and $v_\lambda^2\ge0$ at
$\lambda=-\alpha$ (Fig.~\ref{fig:tetrahedron}). Therefore,
$S_\mathrm{max} = S_\mathrm{top}(\lambda,\mu,\nu)$, and the two
elliptic cylinders are the first in eq.~\eqref{eq:triaxellcyl_SR} and
the one in eq.~\eqref{eq:triaxellcyl_Itube}. The coefficients of the
corresponding ellipses in the $Z'$-plane are respectively those in
eq.~\eqref{eq:triaxellcoeff_SR} ($i=1$) and in
eq.~\eqref{eq:triaxellcoeff_Itube}.

%---------------------------------------------------------------------
\subsection{Gauss-Hermite moments}
\label{sec:GHmom}
%---------------------------------------------------------------------

We have seen that the line-of-sight velocity moments $\mu_k(x',y')$
can be derived either via line-of-sight integration of the intrinsic
velocity moments (eq.~\ref{eq:deflosvdmom1}) or as moments of the
LOSVD (eq.~\ref{eq:deflosvdmom2}). The lowest order line-of-sight
velocity moments $\mu_0$, $\mu_1$ and $\mu_2$ provide the surface mass
density $\Sigma$, the mean line-of-sight velocity $V$ and dispersion
$\sigma$ by
\begin{equation}%
  \label{eq:def_velmomlos}%
  \Sigma = \mu_0, \quad
  V = \frac{\mu_1}{\mu_0}, \quad \mathrm{and} \quad
  \sigma^2 = \frac{\mu_0\,\mu_2-\mu_1^2}{\mu_0^2},
\end{equation}
all three as a function of $(x',y')$. Whereas $\Sigma$, $V$ and
$\sigma$ can be measured routinely, determinations of the higher order
moments ($\mu_3$, $\mu_4$, \dots) are more complicated.  Spectroscopic
observations of the integrated light of galaxies provide the LOSVD as
function of position on the sky plane. Unfortunately, the wings of the
LOSVD become quickly dominated by the noise in the observations, and
since the higher order moments significantly depend on the wings,
their measurements can become very uncertain. Instead of these true
higher-order moments, one often uses the Gauss-Hermite moments ($h_3$,
$h_4$, \dots), which are much less sensitive to the wings of the LOSVD
(van der Marel \& Franx 1993\nocite{1993ApJ...407..525V}; Gerhard
1993\nocite{1993MNRAS.265..213G}).

There is no simple (analytic) relation between the true moments and
the Gauss-Hermite moments, including the lower order moments
$\Sigma_\mathrm{GH}$, $V_\mathrm{GH}$ and $\sigma_\mathrm{GH}$ (but
see eq.\ 18 of van der Marel \& Franx 1993\nocite{1993ApJ...407..525V}
for approximate relations to lowest order in $h_3$ and $h_4$).
Nevertheless, we have shown that for Abel models the full LOSVD can be
computed in a efficient way from eq.~\eqref{eq:triaxAbelLOSVD}, so
that by fitting a Gauss-Hermite series to the resulting LOSVD, we can
derive the Gauss-Hermite moments accurately, all as function of
$(x',y')$.

Still, the calculation of the line-of-sight velocity moments through
the intrinsic moments is useful, e.g., in case of investigating a
range of viewing directions. The intrinsic moments have to be computed
once, after which only a (numerical) integration along the
line-of-sight is needed for each viewing direction. This is (much)
faster than calculating the LOSVD separately at each direction. The
higher order true moments can even be used to (numerically) determine
the Gauss-Hermite moments. One way is to find the Gauss-Hermite LOSVD
of which the true moments best-fit those from the Abel model. However,
in practise this direct fitting of the true moments has several
(numerical) problems. Because it is a non-linear minimisation problem,
the convergence can take long and may result in a local instead of the
global best-fit solution, possibly resulting in Gauss-Hermite moments
that are significantly different from their true values.  If, instead,
we first (re)construct the LOSVD from the true moments by means of an
Edgeworth expansion (see Appendix~\ref{sec:edgeworth}) and then fit a
Gauss-Hermite series, the Gauss-Hermite moments can be calculated
accurately and efficiently. Evidently, once the viewing direction is
known, it is more straightforward to compute the full LOSVD to derive
the (higher-order) Gauss-Hermite moments.

When we construct a galaxy model consisting of multiple Abel DF
components (\S~\ref{sec:combmultcomp}), we cannot simply combine the
corresponding Gauss-Hermite moments in a linear way, because they are
non-linear functions of the DF. Instead, we first add together the
LOSVDs of the different DF components\footnote{Or, in case the LOSVD is
  not readily accessible, the true line-of-sight velocity moments,
  which are also linear functions of the DF.}, each multiplied with a
constant weight, and then parameterise the resulting combined LOSVD as
a Gauss-Hermite series.  Because the mass included in each DF
component is different, in order to obtain the mass fractions per DF
component, we multiply the latter weights with the mass of the
corresponding DF component divided by the total (luminous) mass. To
change the sense of rotation of a rotating DF component (LR or SR),
the corresponding observables do not have to be recomputed, as a
change in the sign of the odd velocity moments is sufficient.

%---------------------------------------------------------------------
\subsection{Surface brightness}
\label{sec:surfbrightness}
%---------------------------------------------------------------------

The surface brightness follows upon integration of the luminosity
density along the line-of-sight. The luminosity density in turn is
related to the mass density $\rho_\star$ via the stellar mass-to-light
ratio $M_\star/L$. With $\rho_\star$ the zeroth-order velocity moment
of the DF (eq.~\ref{eq:triaxrho}), the surface brightness follows as
\begin{equation}
  \label{eq:sbviadensstar}
  \mathrm{SB}(x',y') = \int_{-\infty}^\infty
  (M_\star/L)^{-1} \mu_{000}(x',y',z') \, \du z'.
\end{equation}
In the special case when $(M_\star/L)$ does not change (e.g., due to
variation in the underlying stellar populations) with position, we can
take it out of the integral and $\mathrm{SB}=\Sigma/(M_\star/L)$,
where $\Sigma$ is the surface mass density defined in
eq.~\eqref{eq:def_velmomlos}.

In addition to the luminous matter, a galaxy may also contain dark
matter. While in the outer parts of late-type galaxies the presence of
dark matter, as predicted by the cold dark matter paradigm for galaxy
formation (e.g., Kauffmann \& van den Bosch
2002\nocite{2002SciAm.286f..36K}), was demonstrated convincingly
already more than two decades ago (e.g., van Albada et al.\
1985\nocite{1985ApJ...295..305V}), the proof in the outer parts of
early-type galaxies remains uncertain, mainly due to a lack of
kinematic constraints. As a consequence, in the outer parts of
galaxies, commonly a simple functional form for the dark matter
distribution is assumed, often the universal profile from the CDM
paradigm (Navarro, Frenk \& White 1997\nocite{1997ApJ...490..493N}).

The dark matter distribution in the inner parts of galaxies is
probably even more poorly understood (e.g., Primack
2004\nocite{2004IAUS..220...53P}). For this reason, in current
dynamical studies of the central parts of early-type galaxies, it is
commonly assumed that both $(M_\star/L)$ and the dark matter fraction
are constant, i.e., mass follows light. In this case the surface
brightness also follows from $\mathrm{SB}=\Sigma_S/(M/L)$, where
$(M/L)$ is the (constant) total mass-to-light ratio and $\Sigma_S$ the
surface mass density, which after deprojection yields $\rho_S$, the
mass density related to the potential $V_S$ via Poisson's equation
\eqref{eq:triaxrho_S}. In case of a St\"ackel potential
\eqref{eq:triaxV_S}, $\Sigma_S$ (and hence the surface brightness) has
concentric isodensity contours that show no twist (e.g., Franx
1988\nocite{1988MNRAS.231..285F}).

%============================= section 4 =============================
\section{Triaxial three-integral galaxy models}
\label{sec:triaxgalmodels}
%=====================================================================

After choosing a St\"ackel potential, we investigate the shape of the
density generated by the Abel DF components, and use these components
to construct a triaxial galaxy model with three integrals of motion.

%---------------------------------------------------------------------
\subsection{Isochrone potential}
\label{sec:isopotdens}
%---------------------------------------------------------------------

There are various choices for the potential that provide useful test
models for comparison with the kinematics of triaxial elliptical
galaxies (e.g., Arnold et al.\ 1994\nocite{1994MNRAS.271..924A}). One
option is to consider the so-called perfect ellipsoid, for which
Statler (1987\nocite{1987ApJ...321..113S}) already computed numerical
Schwarzschild models and Hunter \& de Zeeuw
(1992\nocite{1992ApJ...389...79H}) investigated the maximum streaming
thin orbit models. It has a density distribution stratified on similar
concentric ellipsoids, but the potential function $U(\tau)$ contains
elliptic integrals, which slows down numerical calculations.  An
alternative is to consider the set of models introduced by de Zeeuw \&
Pfenniger (1988\nocite{1988MNRAS.235..949D}), which have nearly
ellipsoidal density figures, and have a potential and density that are
evaluated easily and swiftly. They are defined by the choice:
\begin{equation}
  \label{eq:isoUtau}
  U(\tau) = -GM\sqrt{\tau}(\tau+\beta),
\end{equation}
so that the triaxial St\"ackel potential has the elegant form
\begin{equation}
  \label{eq:isoV_S}
  V_S(\lambda,\mu,\nu) =
  \frac{-GM \left(
  \sqrt{\lambda\mu}+\sqrt{\mu\nu}+\sqrt{\nu\lambda}-\beta \right)}
  {(\sqrt{\lambda}+\sqrt{\mu}) (\sqrt{\mu}+\sqrt{\nu})
  (\sqrt{\nu}+\sqrt{\lambda})},
\end{equation}
where we set $GM=\sqrt{-\gamma}+\sqrt{-\alpha}$ so that
$V_S(-\alpha,-\beta,-\gamma)=-1$ in the centre. In the oblate
axisymmetric limit this potential is that of the Kuzmin-Kutuzov
(1962\nocite{1962KK}) models of Dejonghe \& de Zeeuw
(1988\nocite{1988ApJ...333...90D}), and in the spherical limit it
reduces to H\'enon's (1959\nocite{1959AnAp...22..126H}) isochrone.
For all these models, $V_S=U[-\alpha,-\beta,\tau]$ along the short
$z$-axis is identical to the isochrone potential
$-GM/(\sqrt{\tau}+\sqrt{-\alpha})$. We therefore refer to models with
$U(\tau)$ of the form \eqref{eq:isoUtau} as isochrone models. Since
the potential falls of as $1/r$ at large radii, all these models have
finite total mass.

The expressions for the integrals of motion are given in
\eqref{eq:triaxEI2I3}, where $U[\lambda,\mu,\nu]=V_S$ and the third
order divided difference $U[\lambda,\mu,\nu,\sigma]$ is given by the
symmetric expression\footnote{Substituting eq.~\eqref{eq:isoV_S} shows
  that $U[\lambda,\mu,\nu,\sigma]$ is in fact fully symmetric:
  $\frac{U[\lambda,\mu,\nu,\sigma]}{GM} = \frac{\sqrt{\lambda\mu\nu}
    + \sqrt{\mu\nu\sigma} + \sqrt{\nu\sigma\lambda} +
    \sqrt{\sigma\lambda\mu}
    -\beta(\sqrt{\lambda}+\sqrt{\mu}+\sqrt{\nu}+\sqrt{\sigma})} {
    (\sqrt{\lambda}\!+\!\sqrt{\mu}) (\sqrt{\mu}\!+\!\sqrt{\nu})
    (\sqrt{\nu}\!+\!\sqrt{\lambda}) (\sqrt{\lambda}\!+\!\sqrt{\sigma})
    (\sqrt{\mu}\!+\!\sqrt{\sigma}) (\sqrt{\nu}\!+\!\sqrt{\sigma})}$}
\begin{equation}
  \label{eq:isoUlmns}
  U[\lambda,\mu,\nu,\sigma] =
  \frac{-GM -V_S (\sqrt{\lambda}+\sqrt{\mu}+\sqrt{\nu}+\sqrt{\sigma})}
  {(\sqrt{\lambda}+\sqrt{\sigma}) (\sqrt{\mu}+\sqrt{\sigma})
  (\sqrt{\nu}+\sqrt{\sigma})}.
\end{equation}
These triaxial isochrone models have the convenient property that the
expressions for the potential and the integrals of motion contain only
elementary functions of the (confocal ellipsoidal) coordinates and
have no singularities.

The same is true for the associated mass density $\rho_S$, of which
the expression is given in Appendix~C of de Zeeuw \& Pfenniger
(1988\nocite{1988MNRAS.235..949D}), and a contour plot of $\rho_S$
in the $(x,z)$-plane is shown in their Fig.~2. These authors also
derive the axis ratios of $\rho_S$ in the centre (their eq.~C7) and at
large radii (their eq.~C11), in terms of the axis ratios $\zeta$ and
$\xi$ of the confocal ellipsoidal coordinate system, defined as
\begin{equation}%
  \label{eq:defzetaxi}%
  \zeta^2 = (-\beta)/(-\alpha), \quad
  \xi^2 = (-\gamma)/(-\alpha).
\end{equation}
Although $\rho_S$ becomes slightly rounder at larger radii, its axis
ratios remain smaller than unity (for $\xi<\zeta<1$) because at large
radii $\rho_S\sim 1/r^4$ in all directions. Characteristic values for
the axis ratios can be obtained from the (normalised) moments of
inertia along the principal axes of the density,
\begin{equation}%
  \label{eq:momentsofinertia}%
  a^2 = \frac{\int x^2\rho(x,0,0)\,\du x}{\int \rho(x,0,0)\,\du x},
\end{equation}
where the intermediate and short semi-axis length, $b$ and $c$, of the
inertia ellipsoid follow from the long semi-axis length $a$ by
replacing $x$ with $y$ and $z$, and at the same time $\rho(x,0,0)$
with $\rho(0,y,0)$ and $\rho(0,0,z)$, respectively. Taking for example
$\zeta=0.8$ and $\xi=0.64$, the semi-axis lengths of the inertia
ellipsoid result in the characteristic axis ratios $b_S/a_S=0.88$ and
$c_S/a_S=0.80$ for the density $\rho_S$. The contours of the projected
density are nearly elliptic with slowly varying axis ratios.

For triaxial mass models with a St\"ackel potential $V_S$, de Zeeuw,
Peletier \& Franx (1986\nocite{1986MNRAS.221.1001D}) have shown that
the corresponding intrinsic mass density $\rho_S$ cannot fall off more
rapidly than $1/r^4$, except along the short $z$-axis. All models in
which $\rho_S$ falls off less rapidly than $1/r^4$ become round at
large radii. When $\rho_S\sim 1/r^4$, as is the case for, e.g., the
above isochrone potential and the perfect ellipsoid (e.g., de Zeeuw
1985a\nocite{1985MNRAS.216..273D}), the model remains triaxial at
large radii. Moreover, mass models containing a linear combination of
different St\"ackel potentials are possible as long as the associated
confocal ellipsoidal coordinate systems share the same foci (e.g., de
Zeeuw \& Pfenniger 1988\nocite{1988MNRAS.235..949D}; Batsleer \&
Dejonghe 1994\nocite{1994A&A...287...43B}). This shows that, although
we choose here a (single-component) isochrone potential, our method is
capable of providing Abel models for a large range of St\"ackel
potentials, with a similarly large range of shapes of the
corresponding mass model. The same holds true for the luminous mass
density, which we consider next.

%---------------------------------------------------------------------
\subsection{The shape of the luminous mass density}
\label{sec:shapedens}
%---------------------------------------------------------------------

%%%FIG
\begin{figure*}
  \begin{center}
    \includegraphics[width=1.0\textwidth]{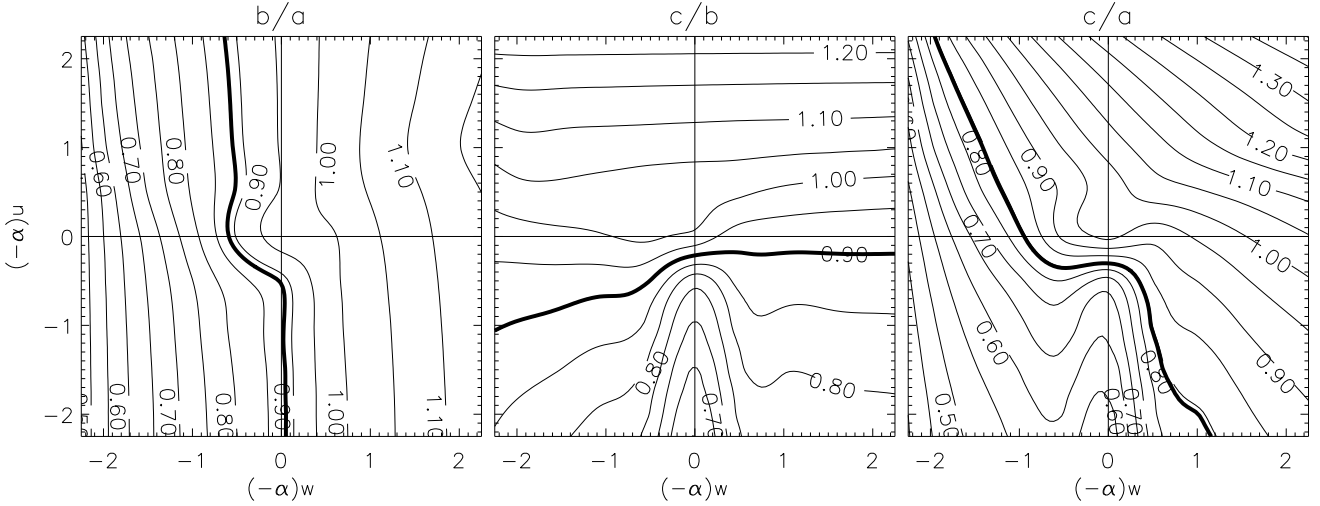}
  \end{center}
  \caption{The characteristic axis ratios $b/a$, $c/b$ and $c/a$ of
    the luminous mass density for a non-rotating Abel component, as
    function of the DF parameters $w$ and $u$, while $\delta=1$. The
    axis ratios of the confocal ellipsoidal coordinate system are
    $\zeta=0.8$ and $\xi=0.64$, so that cf.~\eqref{eq:limwu}
    $(-\alpha) w \ge -25/9 \approx -2.78$ and $(-\alpha) u \le
    625/144 \approx 4.34$. The thick contours are drawn at the
    levels that correspond to the characteristic axis ratios of the
    total mass density $\rho_S$, associated with the underlying
    isochrone St\"ackel potential \eqref{eq:isoV_S}, respectively
    $b_S/a_S=0.88$, $c_S/b_S=0.90$ and $c_S/a_S=0.80$.  The
    intermediate-over-long axis ratio $b/a$ depends mainly on $w$, the
    short-over-intermediate axis ratio $c/b$ depends mainly on $u$,
    and $c/a$ is the product of the previous two.}
  \label{fig:flattening_wu}
\end{figure*}
%%%FIG

%%%FIG
\begin{figure*}
  \begin{center}
    \includegraphics[width=1.0\textwidth]{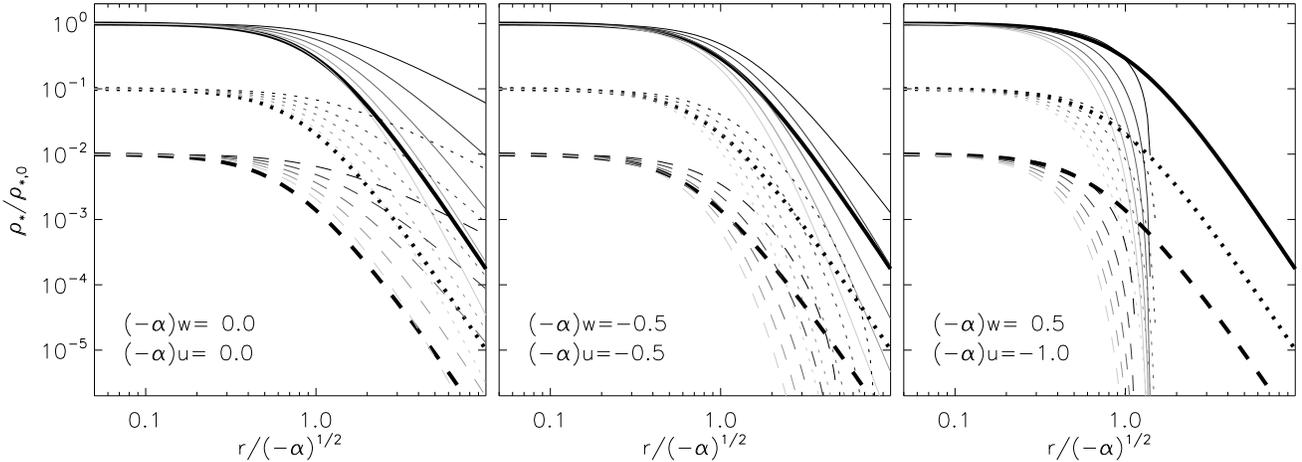}
  \end{center}
  \caption{ Principal axes profiles of the luminous mass density
    $\rho_\star$ for a non-rotating Abel component, normalised to the
    central value $\rho_{\star,0}$. Each panel is for a different
    combination of the DF parameters $w$ and $u$, while the grey scale
    indicates variation in $\delta$ from zero (darkest curve) to four
    (lightest curve), in unity steps. The profiles along the $y$-axis
    (dotted curves) and along the $z$-axis (dashed curves) are
    arbitrarily offset vertically with respect to the profile along
    the $x$-axis (solid curves) to enhance visualisation. The thick
    black curves show the profiles for the (total) mass density
    $\rho_S$, associated with the underlying isochrone St\"ackel
    potential \eqref{eq:isoV_S}, with $\zeta=0.8$ and $\xi=0.64$. When
    the value of either $w$ or $u$ is positive (right panel), the
    profiles show a break at $r\sim\sqrt{-\alpha}$, so that these
    compact components may be used to represent kinematically
    decoupled components.}
  \label{fig:profile_delta}
\end{figure*}
%%%FIG

Whereas the shape of the (total) mass density $\rho_S$ is fixed by the
choice of the potential $V_S$, and $\zeta$ and $\xi$
(eq.~\ref{eq:defzetaxi}), the shape of the (luminous) mass density
$\rho_\star$, which is the zeroth order velocity moment of the DF
(eq.~\ref{eq:triaxrho}), also depends on the DF parameters $w$, $u$
and $\delta$, and the type of component. For $\zeta=0.8$ and
$\xi=0.64$, we show in Fig.~\ref{fig:flattening_wu} for non-rotating
DF components the characteristic (eq.~\ref{eq:momentsofinertia}) axis
ratios of the corresponding density, as function of $w$ and $u$.  We
have set $\delta=1$, but the axis ratios depend only weakly on it,
with $\rho_\star$ becoming slightly flatter for increasing $\delta$.
The thick contours are drawn at the levels that correspond to the
values of the characteristic axis ratios of $\rho_S$, respectively
$b_S/a_S=0.88$, $c_S/b_S=0.90$ and $c_S/a_S=0.80$. These values are
independent of $w$ and $u$ (as well as the other DF parameters).
%Self-consistent: b/a,c/b,c/a:      0.88305050      0.90063854      0.79530932

While the intermediate-over-long axis ratio $b/a$ increases with
increasing $w$, its value is only weakly dependent of $u$. By
contrast, the short-over-intermediate axis ratio $c/b$ mainly
increases with increasing $u$. The short-over-long axis ratio $c/a$ is
the product of the previous two axis ratios and thus depends on both
$w$ and $u$.  When both $w$ and $u$ are negative, the density
$\rho_\star$ has its long-axis along the $x$-axis and its short-axis
along the $z$-axis, in the same way as the potential $V_S$ and the
associated density $\rho_S$. Above certain positive values of either
$w$ or $u$, the axis ratios become larger than unity, which means that
$\rho_\star$ is no longer aligned with the underlying coordinate
system in the same way as $V_S$ and $\rho_S$. For example, when
$(-\alpha)w=-0.5$ and $(-\alpha)u=0.5$, $b/a<1$ but $c/b>1$, so that in this case
$\rho_\star$ has its short axis along the $y$-axis.

A change in the sign of $w$ and $u$ has a strong effect on the
radial slope of $\rho_\star$. In Fig.~\ref{fig:profile_delta}, the
radial profiles of $\rho_\star$ along the principal axes are shown for
three combinations of $w$ and $u$.
The density is normalised to the central value $\rho_0$. The profiles
along the $y$-axis (dotted curves) and along the $z$-axis (dashed
curves) are arbitrarily offset vertically with respect to the profile
along the $x$-axis (solid curves) to enhance visualisation. The thin
curves are the profiles of the (luminous) mass density $\rho_\star$
for varying $\delta$, from $\delta=0$ (darkest curve) to $\delta=4$
(lightest curve), in unit steps. The thick black curves show the
profiles for the (total) mass density $\rho_S$, which is independent
of $w$, $u$ and $\delta$.

The profiles of $\rho_\star$ steepen for increasing $\delta$ and for
increasing absolute values of $w$ and $u$. In particular, when either
$w$ or $u$ becomes positive (right panel), the profiles suddenly
become much steeper and drop to zero already at relatively small radii
$r\sim\sqrt{-\alpha}$. The resulting Abel components are thus compact
and, as we saw above, can be different in shape and orientation from
the main body of the galaxy model. Therefore, they can be used to
represent kinematically decoupled components.  When both $w\le0$ and
$u\le0$ (left and middle panel), $\rho_\star$ falls off much more
gently and the Abel components cover a larger region. When $w=u=0$
(left panel), so that the DF only depends on energy, the profiles as
well as the shape (Fig.~\ref{fig:flattening_wu}) of $\rho_\star$ can
even be flatter than those of $\rho_S$. However, already for small
non-zero values of $w$ and $u$, generally $\rho_\star\le\rho_S$
everywhere in the galaxy model, and $\rho_\star<\rho_S$ in the outer
parts. Although self-consistency $\rho_\star=\rho_S$ is only possible
in the spherical case (for fixed values of $w$ and $u$, see
\S~\ref{sec:abeldf}), one can choose the parameters $w$, $u$ and
$\delta$ so that $\rho_\star\sim\rho_S$.  At the same time, having
$\rho_\star<\rho_S$ in the outer parts of the galaxy model, allows for
a possible dark halo contribution.

The shape of $\rho_\star$ can furthermore change due to the additional
contribution from long-axis rotating and short-axis rotating
components. Although these components have no density along their
rotation axis, the behaviour of their overall shape as function of $w$,
$u$ and $\delta$ is similar as for the corresponding non-rotating
components.

The above analysis shows that, given the triaxial isochrone potential
\eqref{eq:isoV_S}, we can use Abel components to construct a galaxy
model with a realistic density. Depending on the choice of $w$, $u$
and $\delta$, the galaxy model can contain compact (kinematically
decoupled) components and account for possible dark matter (in the
outer parts). Furthermore, we show below that even with a small number
DF components, enough kinematic variation is possible to mimic the
two-dimensional kinematic maps of early-type galaxies provided by
observations with current integral-field spectrographs. This means
that we can construct simple but realistic galaxy models to test our
Schwarzschild software (\S~\ref{sec:recoverytriax} and
\ref{sec:recoveryaxi}).

%---------------------------------------------------------------------
\subsection{A triaxial Abel model}
\label{sec:triaxabelmodel}
%---------------------------------------------------------------------

%%%FIG
\begin{figure*}
  \begin{center}
    \includegraphics[width=1.0\textwidth]{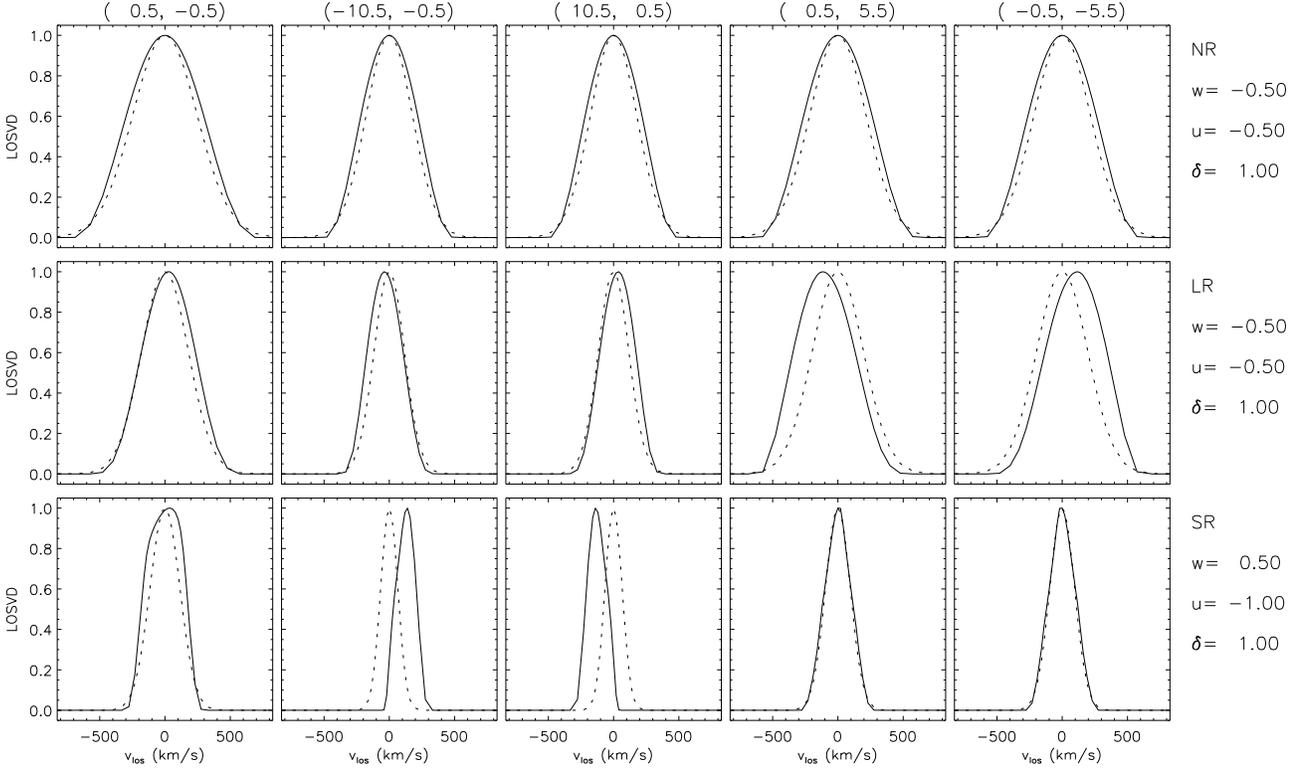}
  \end{center}
  \caption{Line-of-sight velocity distribution (LOSVD) at five
    different positions $(x',y')$ on sky-plane (in arcsec
    at the top of each column) of three different Abel components. The
    isochrone St\"ackel potential \eqref{eq:isoV_S} is used, with
    $\zeta=0.8$ and $\xi=0.64$ ($T=0.61$), and scale length
    $\sqrt{-\alpha}=10$\arcsec. The model is placed at a distance of
    $D=20$ Mpc and the adopted viewing angles are $\vartheta=70$\dgr
    and $\varphi=30$\dgr. From top to bottom the LOSVDs of a
    non-rotating (NR), long-axis rotating (LR) and short-axis rotating
    (SR) Abel component are shown, with the corresponding DF
    parameters $w$, $u$ and $\delta$ given on the right. The height of
    each LOSVD is normalised to unity, and a (dashed) Gaussian
    distribution with zero mean and the same dispersion as the LOSVD
    is shown as a reference.}
  \label{fig:losvd}
\end{figure*}
%%%FIG

%%%FIG
\begin{figure*}
  \begin{center}
    \includegraphics[width=1.0\textwidth]{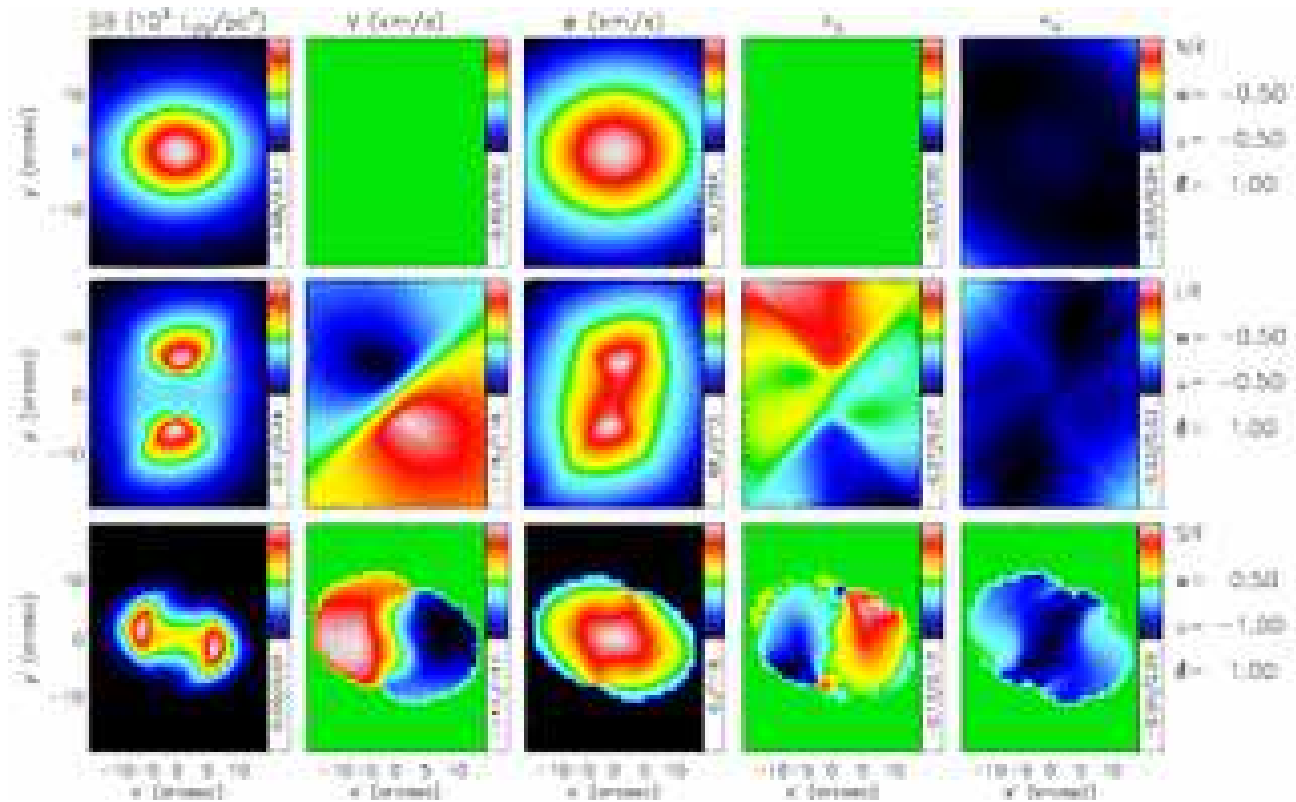}
  \end{center}
  \caption{Maps of the surface brightness (SB; in $10^3$ \Lsunpcsq),
    mean line-of-sight velocity $V$ and dispersion $\sigma$ (both in
    \kms), and higher order Gauss-Hermite moments $h_3$ and $h_4$, of
    the same three Abel DF components as in Fig.~\ref{fig:losvd},
    obtained by fitting a Gauss-Hermite series to the LOSVDs at each
    (pixel) position on the plane of the sky. The numerical artifacts
    at the edges of the compact SR component (third row) disappear
    when combined with components that extent over the full
    field-of-view (see e.g. the top row of
    Fig.~\ref{fig:obsmom_triax}).}
  \label{fig:ghmom}
\end{figure*}
%%%FIG

%%%FIG
\begin{figure*}
  \begin{center}
    \includegraphics[width=1.0\textwidth]{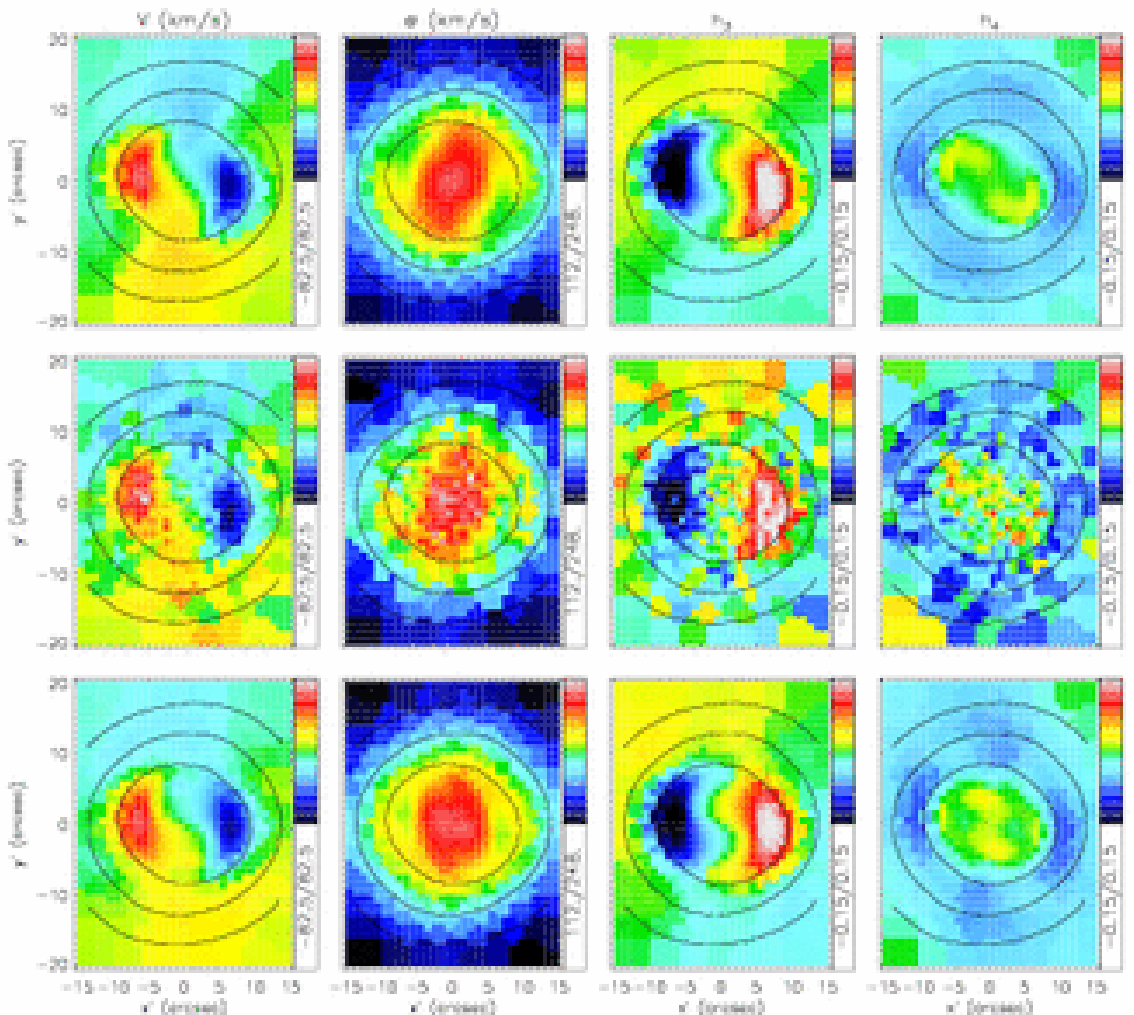}
  \end{center}
  \caption{Kinematic maps for a triaxial Abel model (top row) and
    converted to observables with realistic measurement errors added
    (middle row; see \S~\ref{sec:triaxabelmodel}), and for the
    best-fit triaxial Schwarzschild model (bottom row; see
    \S~\ref{sec:Schwmodelstriax}). From left to right: mean
    line-of-sight velocity $V$ and dispersion $\sigma$ (both in \kms),
    and Gauss-Hermite moments $h_3$ and $h_4$. Isophotes of the
    surface brightness of the Abel model are overplotted in each map.
    At the right side of each map, the (linear) scale of the
    corresponding kinematics is indicated by the colour bar, and the
    limits are given below.}
  \label{fig:obsmom_triax}
\end{figure*}
%%%FIG

As before, we choose the isochrone St\"ackel potential
\eqref{eq:isoV_S}, we take $\zeta=0.8$ and $\xi=0.64$ for the axis
ratios of the coordinate system \eqref{eq:defzetaxi}, resulting in a
triaxiality parameter \eqref{eq:triaxparT} of about $T=0.61$, and we
set the scale length $\sqrt{-\alpha}=10$\arcsec. Assuming a distance
of $D=20$ Mpc and a total mass of $10^{11}$\,\Msun\ results in a
central value for the potential $V_0 \sim 2.7\times10^5$
km$^2$\,s$^{-2}$, which also sets the unit of velocity. We restrict
the number of DF components to three, one of each type. For the first
component of type NR we set $w=u=-0.5/(-\alpha)$ and $\delta=1$, so
that the shape of the corresponding density is similar to that of
$\rho_S$, except in the outer parts where a steeper profile mimics the
presence of dark matter (see
Figs.~\ref{fig:flattening_wu}~and~\ref{fig:profile_delta}). For the
second and third component, respectively of type LR and SR, we adopt
the same parameters, expect that we take $w=0.5/(-\alpha)$ and
$u=-1.0/(-\alpha)$ for the SR component, which therefore is more
compact than the NR and LR component.

We set the line-of-sight by choosing $\vartheta=70$\dgr\ and
$\varphi=30$\dgr\ for the viewing angles. After rotation over the
misalignment angle $\psi=101$\dgr\ eq.~\eqref{eq:misalignment_psi}, we
compute for each DF component the LOSVD as a function of the positions
on a rectangular grid on the sky plane, illustrated in
Fig.~\ref{fig:losvd} for five sky positions. By fitting a
Gauss-Hermite series to each LOSVD, we obtain the maps of the mean
line-of-sight surface mass density $\Sigma$, velocity $V$, dispersion
$\sigma$ and higher-order Gauss-Hermite moments $h_3$ and $h_4$, shown
in Fig.~\ref{fig:ghmom}.  The parameters of each DF component are
given on the right.
% The colours indicate the range of the velocity moments in each panel,
% from small (blue) to large (red) values.
The NR component has zero (green) odd velocity moments.  For the LR
and SR component, the even velocity moments show a decrease in the
centre, because these components have zero density along respectively
the intrinsic long and short axis. We add the LOSVDs of the NR, LR and
SR components, weighted with mass fractions of respectively 80\%,
12.5\% and 7.5\%, and fit a Gauss-Hermite series to obtain maps of
$\Sigma$, $V$, $\sigma$, $h_3$ and $h_4$.  We convert $\Sigma$ to the
surface brightness by dividing by a constant stellar mass-to-light
ratio of $(M_\star/L)=4$ \MLsun.

To convert these `perfect' kinematics to `realistic' observations,
similar to those obtained with integral-field spectrographs such as
\sauron\ (Bacon et al.\ 2001\nocite{2001MNRAS.326...23B}), we finally
apply the following steps. We compute the kinematics on a rectangular
grid consisting of 30 by 40 square pixels of 1\arcsec in size. Using
the adaptive spatial two-dimensional binning scheme of Cappellari \&
Copin (2003\nocite{2003MNRAS.342..345C}), we bin the pixels according
to the criterion that each of the resulting (Voronoi) bins contains a
minimum in signal-to-noise (S/N), which we take proportional to the
square root of the surface brightness. For the mean errors in the
kinematics we adopt the typical values of $7.5$ \kms\ for $V$ and
$\sigma$ and $0.03$ for $h_3$ and $h_4$ in the kinematics of a
representative sample of early-type galaxies observed with \sauron\ 
(Emsellem et al.\ 2004\nocite{2004MNRAS.352..721E}).  We then weigh
these values with the S/N in each bin to mimic the observed variation
in measurement errors across the field. Finally, we use the computed
measurement errors to (Gaussian) randomise the kinematic maps. In this
way, we include the randomness that is always present in real
observations.
%% Moreover, without this randomization the goodness-of-fit parameter
%% $\chi^2$ \eqref{eq:chi2}, which we use to find the best-fit
%% Schwarzschild model, is not meaningful.
The resulting kinematic maps are shown in the top panels of
Fig.~\ref{fig:obsmom_triax}. Because of the eight-fold symmetry of the
triaxial model, the maps of the even (odd) velocity moments are always
point-(anti)-symmetric, apart from the noise added.

%============================= section 5 =============================
\section{Recovery of triaxial galaxy models}
\label{sec:recoverytriax}
%=====================================================================

We briefly describe our numerical implementation of Schwarz\-schild's
method in triaxial geometry (see vdB07 for a full description), which
we then use to fit the observables of the triaxial Abel model
constructed in \S~\ref{sec:triaxabelmodel}. We investigate the
recovery of the intrinsic velocity moments and, through the
distribution of the orbital mass weights, the recovery of the
three-integral DF.

%---------------------------------------------------------------------
\subsection{Triaxial Schwarzschild models}
\label{sec:Schwmodelstriax}
%---------------------------------------------------------------------

The first step is to infer the gravitational potential from the
observed surface brightness. We do this by means of the Multi-Gaussian
Expansion method (MGE; e.g., Cappellari
2002\nocite{2002MNRAS.333..400C}), which allows for possible position
angle twists and ellipticity variations in the surface brightness. For
a given set of viewing angles $(\vartheta,\varphi,\psi)$ (see
\S~\ref{sec:int2obscoordsystem}), we deproject the surface brightness
and we multiply it by a mass-to-light ratio $(M/L)$ to get the intrinsic
mass density, from which the gravitational potential then follows by
solving Poisson's equation.  We calculate orbits numerically in the
resulting gravitational potential.

To obtain a representative library of orbits, the integrals of motion
have to be sampled well. The energy can be sampled directly, but since
the other integrals of motion are generally not known, we start, at a
given energy, orbits from a polar grid in the $(x,z)$-plane, which is
crossed perpendicularly by all families of (regular) orbits. We
restrict ourselves to the region in the first quadrant that is
enclosed by the equipotential and the thin orbit curves to avoid
duplication of the tube orbits. To have enough box orbits to support
the triaxial shape, we also start orbits by dropping them from the
equipotential surface (Schwarzschild 1979\nocite{1979ApJ...232..236S},
1993\nocite{1993ApJ...409..563S}).

Assigning a mass weight $\gamma_j$ to each orbit $j$ from the library,
we compute their combined properties and find the weighted
superposition that best fits the observed surface brightness and
(two-dimensional) kinematics. However, the resulting orbital weight
distribution may vary rapidly, and hence probably corresponds to an
unrealistic DF. To obtain a smoothly varying DF, we both dither the
orbits by considering a bundle of integrated orbits that were started
close to each other, and we regularise when looking for the best-fit
set of orbital weights by requiring them to vary smoothly between
neighbouring orbits (in integral space). Finally, the best-fit
Schwarzschild model follows from the minimum in the (Chi-squared)
difference between (photometric and kinematic) observables and the
corresponding model predictions, weighted with the errors in the
observables.

%---------------------------------------------------------------------
\subsection{Fit to observables of a triaxial Abel model}
\label{sec:fit2obstriax}
%---------------------------------------------------------------------

In this case, the gravitational potential is known and given by the
isochrone St\"ackel potential $V_S$ eq.~\eqref{eq:isoUtau}. However,
to closely simulate the Schwarzschild modelling of real galaxies, we
infer the potential from a deprojection of an MGE fit of the surface
mass density $\Sigma_S$ generated by $V_S$. The resulting potential
reproduces $V_S$ to high precision, with relative differences less
than $10^{-3}$.  We compute a library of orbits by sampling 21
energies $E$ via a logarithmic grid in radius from 1\arcsec\ to
123\arcsec\ that contains $\ge$99.9 per cent of the total mass. At
each energy, we construct a uniform polar start space grid of 7 radii
by 8 angles within the first quadrant of the $(x,z)$-plane and drop
box orbits from a similar uniform polar grid on the equipotential
surface in the first octant.  This results in a total of
$21\times7\times8\times2=2352$ starting positions, from each of which
a bundle of $5^3$ orbits are started.  Taking into account the two
senses of rotation of the tube orbits, this results in a total
$441000$ orbits that are numerically integrated in the potential.

We sum the velocities of each bundle of orbits in histograms with 401
bins, at a velocity resolution of $10$ \kms. We fit the weighted sum
of the velocity histograms to the intrinsic mass density $\rho_\star$,
which we obtain from a deprojection of an MGE fit to the observed
surface brightness, multiplied with the (constant) $(M_\star/L)=4$
\MLsun.  Simultaneously, we fit the projected values of the velocity
histograms to the observed surface brightness and higher-order
velocity moments.  Finally, at the same time, we regularise the
orbital weights in $E$ and in the starting positions by minimising
their second order derivatives. The strenght of the regularisation is
given by the a smoothening parameter (e.g., Cretton et al.\ 
1999\nocite{1999ApJS..124..383C}), which we set to $\lambda=0.1$ (see
vdB07).

From Fig.~\ref{fig:obsmom_triax} it is clear that the (simulated)
observables of the triaxial Abel model (top panels) are very well
matched by the best-fit triaxial Schwarzschild model (bottom panels).
The signature of the kinematically decoupled component in the maps of
the mean line-of-sight velocity $V$ and Gauss-Hermite moment $h_3$ is
accurately fitted, as well as the kinematics of the main body up to
$h_4$ within the added noise (\S~\ref{sec:triaxabelmodel}). Below we
investigate how well the intrinsic velocity moments as well as the
three-integral DF --- which are not (directly) fitted --- are
recovered. Here, we keep the mass-to-light ratio and the viewing
angles fixed to the input values of the triaxial Abel model
(\S~\ref{sec:triaxabelmodel}), while in vdB07 we vary these global
parameters to study how well Schwarzschild's method is able to
determine them.

%---------------------------------------------------------------------
\subsection{Intrinsic velocity moments}
\label{sec:intmomtriax}
%---------------------------------------------------------------------

%%%FIG
\begin{figure*}
  \begin{center}
    \includegraphics[width=1.0\textwidth,trim=0 5mm 0 0]{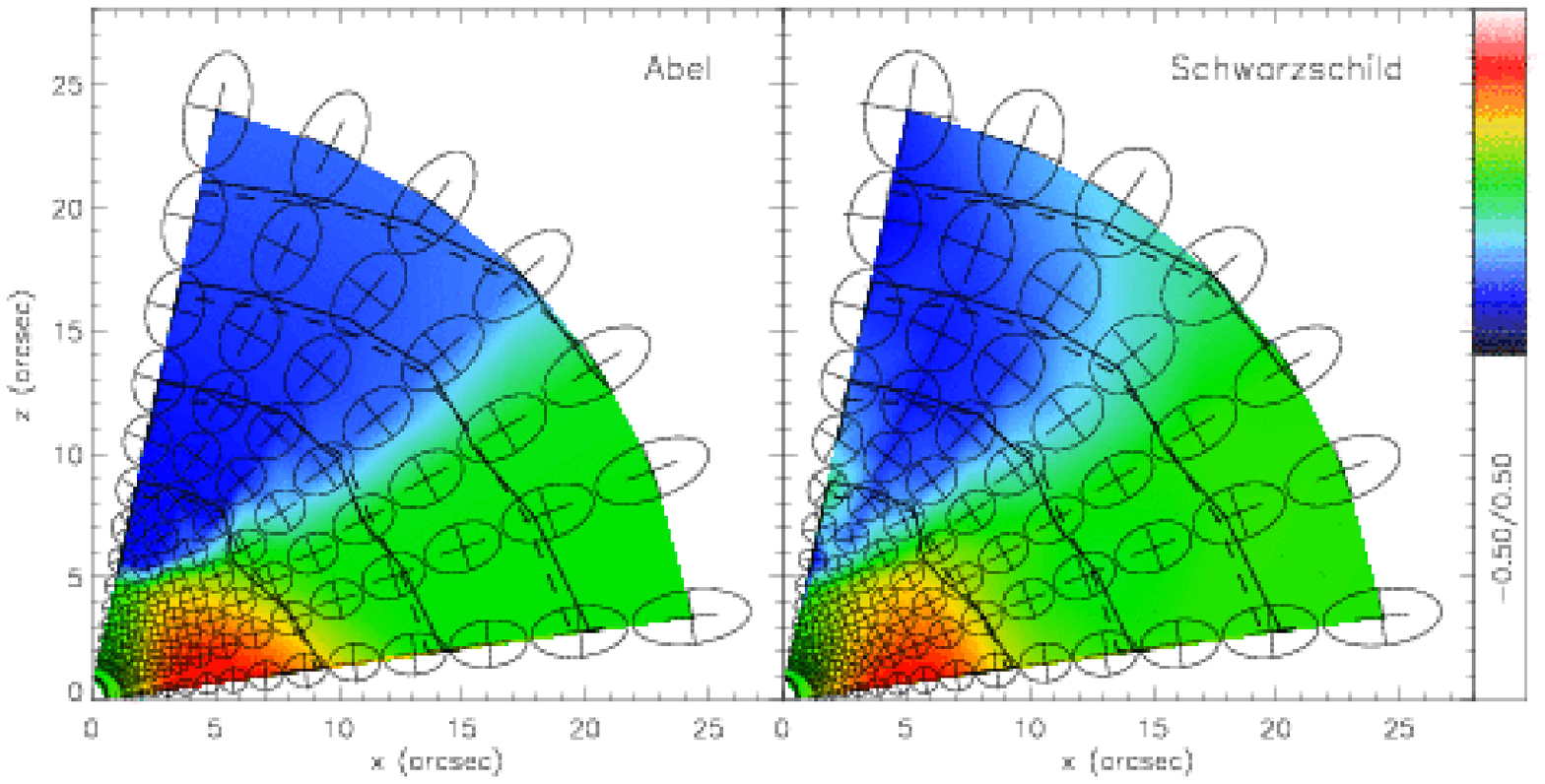}
  \end{center}
  \caption{The colours represent the mean motion $\langle v_y \rangle$
    perpendicular to the $(x,z)$-plane, normalised by
    $\sigma_\mathrm{RMS}$ (excluding the axes to avoid numerical
    problems), for the input triaxial Abel model (left) and for the
    best-fit triaxial Schwarzschild model (right). The ellipses are
    cross sections of the velocity ellipsoid with the $(x,z)$-plane
    and the crosses represent the (relative) size of the velocity
    ellipsoid in the perpendicular ($y$-axis) direction. The black
    curves are contours of constant mass density in steps of one
    magnitude, for the input Abel model (solid) and for the fitted
    Schwarzschild model (dashed). See \S~\ref{sec:intmomtriax} for
    details.}
  \label{fig:intVsig_triax}
\end{figure*}
%%%FIG

We calculate the intrinsic first and second order velocity moments of
the Schwarzschild model by combining the appropriate moments of the
orbits that receive weight in the superposition, and investigate how
well they compare with the intrinsic velocity moments of the Abel
model. In general, there are three first $\langle v_t \rangle$ and six
second order velocity moments $\langle v_s v_t \rangle$ ($s,t=x,y,z$).
Combining them yields the six dispersion components $\sigma_{st}$ of
the velocity dispersion tensor, where $\sigma_{st}^2 \equiv \langle
v_s v_t \rangle - \langle v_s \rangle \langle v_t \rangle$.

We first consider the $(x,z)$-plane, as it is crossed perpendicularly
by all four (major) orbit families. Because $\langle v_x \rangle =
\langle v_z \rangle = \sigma_{xy} = \sigma_{yz} = 0$, we are left with
$\langle v_y \rangle$ perpendicular to the $(x,z)$-plane as the only
non-vanishing mean motion and $\sigma_{zx}$ in the $(x,z)$-plane as
the only non-vanishing cross-term. The average root-mean-square
velocity dispersion $\sigma_\mathrm{RMS}$ is given by
$\sigma^2_\mathrm{RMS}= (\sigma^2_{xx} +\sigma^2_{yy} +
\sigma^2_{zz})/3$. The ratio $\langle v_y \rangle /
\sigma_\mathrm{RMS}$ of ordered-over-random motion is a measure of the
importance of rotation for the gravitational support of a galaxy. In
Fig.~\ref{fig:intVsig_triax}, the colours represent the values of this
ratio in the $(x,z)$-plane, for the input triaxial Abel model (left
panel) and for the best-fit triaxial Schwarzschild model (right
panel).

In a St\"ackel potential the axes of the velocity ellipsoid are
aligned with the confocal ellipsoidal coordinate system (e.g.,
Eddington 1915\nocite{1915MNRAS..76...37E}; van de Ven et al.
2003\nocite{2003MNRAS.342.1056V}). As a result, one of the axes of the
velocity ellipsoid is perpendicular to the $(x,z)$-plane, with
semi-axis length $\sigma_{yy}$. The other two axes lie in the
$(x,z)$-plane and have semi-axis lengths given by
\begin{equation}%
  \label{eq:semiaxislengthvelell}%
  \sigma^2_\pm = 
  \textstyle{\frac12}(\sigma_{xx}^2 +\sigma_{zz}^2) \pm
  \sqrt{\textstyle{\frac14}(\sigma_{xx}^2 -\sigma_{zz}^2)^2 +
    \sigma_{xz}^4}.
\end{equation}
The ellipses overplotted in Fig.~\ref{fig:intVsig_triax} show the
corresponding cross sections of the velocity ellipsoid with the
$(x,z)$-plane. The flattening of the ellipses is thus given by the
ratio $\sigma_-/\sigma_+$, while the angle $\theta_{xz}$ of the
major-axis with respect to the $x$-axis is given by\footnote{In case
  of alignment with the confocal ellipsoidal coordinate system, this
  angle is given by the tangent to the curves of constant $(\mu,\nu)$,
  i.e., $\tan\theta_{xz} = (z/x)(\lambda+\alpha)/(\lambda+\gamma)$,
  which indicates approaching alignment with the polar coordinate system
  at large radius.}
\begin{equation}
  \label{eq:orientationvelell}
  \tan(2\theta_{xz}) =
  2\sigma_{xz}^2/(\sigma_{xx}^2-\sigma_{zz}^2).
\end{equation}
In addition, the cross on top of each ellipse represents the ratio
$\sigma_{yy}/\sigma_+$, i.e., the (relative) size of the velocity
ellipsoid in the perpendicular direction. For an isotropic velocity
distribution the ellipses become circles and the crosses fill the
circles. Finally, the black curves are contours of constant mass
density in steps of one magnitude.

The density of the triaxial Abel model (solid curve) is well fitted by
the triaxial Schwarzschild model (dashed curve), with a
(biweight\footnote{The biweight mean (e.g., Andrews et al.
  1972\nocite{1972biweight}; Beers, Flynn \& Gebhardt
  1990\nocite{1990AJ....100...32B}) is robust estimators for a broad
  range of non-Gaussian underlying populations and is less sensitive
  to outliers than other moment estimators.})  mean fractional
difference below $1$\,\%. In both the Abel model and the fitted
Schwarzschild model the value of $\langle v_y \rangle /
\sigma_\mathrm{RMS}$ is relatively low, with a mean value $\sim 0.14$,
indicating that gravitational support is mainly due to random motion.
Still, the average rotation of the long-axis tube orbits (with
$\langle v_y \rangle < 0$) due to the maximum streaming LR component
in the input Abel model, as well as, the opposite maximum streaming of
the (compact) short-axis rotating component are clearly visible, and
well recovered by the best-fit Schwarzschild model. The average
absolute difference in both $\langle v_y \rangle$ and
$\sigma_\mathrm{RMS}$ is below $6$ \kms, and thus well within the
typical error of $7.5$ \kms\ assigned to the simulated mean
line-of-sight velocity $V$ and velocity dispersion $\sigma$ of the
Abel model (see \S~\ref{sec:triaxabelmodel}). The corresponding
uncertainty in $\langle v_y \rangle / \sigma_\mathrm{RMS}$ is $\sim
0.03$.

We see in Fig.~\ref{fig:intVsig_triax} that, at larger radii, the
ellipses become more radially elongated and the relative size of the
crosses decreases in the radial direction, but they stil fill the
ellipses in the angular direction.  This implies a velocity
distribution that becomes increasingly radially anisotropic outwards,
but remains close to isotropic in the tangenetial direction
everywhere. This shape and orientation of the velocity ellipsoid in
the input Abel model is well reproduced by the best-fit Schwarzschild
model, with only a (mild) underestimation of the radial anisotropy
towards the $z$-axis. This is likely the result of numerical
difficulties due to the small number of (sampled) long-axis tube
orbits that contribute in this region. The absolute difference in the
semi-axis lengths $\sigma_+$, $\sigma_-$ and $\sigma_{yy}$ of the
velocity ellipsoid is on average $\sim 8$\,\kms. This uncertainty
includes both deviations in shape and orientation of the velocity
ellipsoid, and is wihtin the expected range due to the errors in the
simulated kinematics. The corresponding axis ratios
$\sigma_-/\sigma_+$ and $\sigma_{yy}/\sigma_+$ of the velocity
ellipsoid are on average recovered within $\sim 5$\,\%.

%%%FIG
\begin{figure*}
  \begin{center}
    \includegraphics[width=1.0\textwidth,trim=0 5mm 0 0]{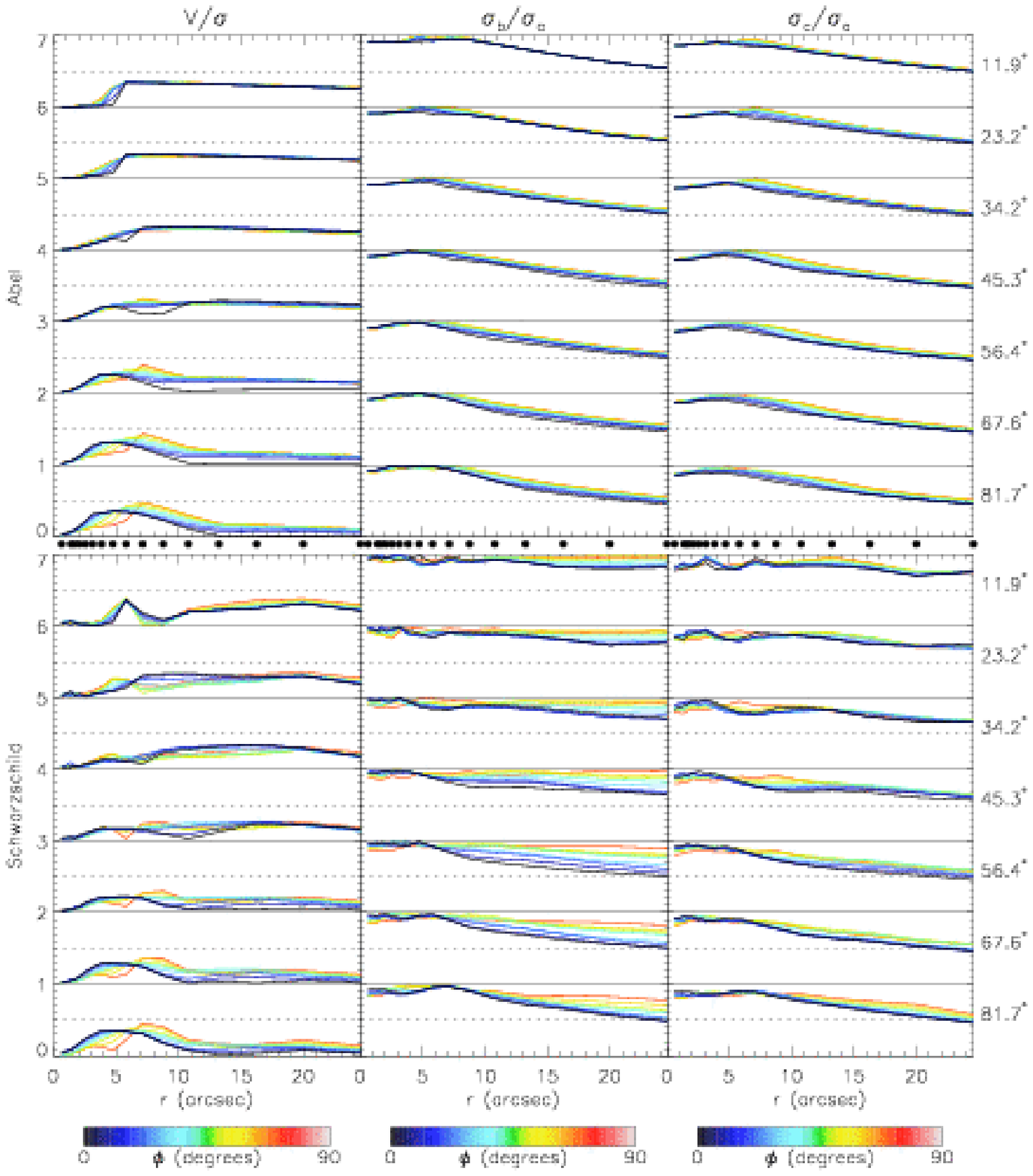}
  \end{center}
  \caption{Intrinsic velocity moments in three dimensions for the input
    triaxial Abel model (top) and for the best-fit triaxial
    Schwarzschild model (bottom). The first column shows the
    (amplitude) of the streaming motion $v_\mathrm{str}$, normalised
    by $\sigma_\mathrm{RMS}$. The second and third column show the
    axis ratios of the velocity ellipsoid, where $\sigma_a$,
    $\sigma_b$ and $\sigma_c$ are respectively the semi-lengths of the
    major, intermediate and minor principal axes. These quantities are
    computed on a polar grid $(r,\theta,\phi)$ in the first octant.
    The (logarithmic) sampling of the radius $r$ (in arcsec) is
    indicated by the black dots between the top and bottom panels.
    Each row is for a different polar angle $\theta$ (in degrees) as
    indicated on the right, with the top panel close to the $z$-axis
    and the bottom panel close to the $(x,y)$-plane. The colours
    represent the (linear) change in azimuthal angle $\phi$ (in
    degrees), with limits $0$\dgr\ and $90$\dgr\ corresponding to the
    $(x,z)$-plane and $(y,z)$-plane, respectively. See
    \S~\ref{sec:intmomtriax} for details.}
  \label{fig:intmom_triax}
\end{figure*}
%%%FIG

Away from the $(x,z)$-plane, the average fractional difference in the
density between the input Abel model and the best-fit Schwarzschild
model stays below $1$\,\%.  Fig.~\ref{fig:intmom_triax} compares the
intrinsic moments of the input triaxial Abel model (top) with those of
the best-fit triaxial Schwarzschild model (bottom) in three
dimensions. The first column shows the (amplitude) of the streaming
motion $v_\mathrm{str}$, given by $v_\mathrm{str}^2 = v_x^2 + v_y^2 +
v_z^2$, and normalised by $\sigma_\mathrm{RMS}$. These quantities are
computed on a polar grid $(r,\theta,\phi)$ in the first octant. The
(logarithmic) sampling of the radius $r$ is indicated by the black
dots between the top and bottom panels, while each row is for a
different polar angle $\theta$ as indicated on the right, and the
colours represent the (linear) change in azimuthal angle $\phi$.  The
limit $\phi=0$\dgr\ (black curves) corresponds to the $(x,z)$-plane
discussed above. The resulting ordered-over-random motion $V/\sigma$
is well recovered by the Schwarzschild model, apart from the upper
panel, which is likely the result of the above mentioned numerical
difficulties close to the $z$-axis. Overall, the average absolute
difference in both $v_\mathrm{str}$ and $\sigma_\mathrm{RMS}$ is below
$6$\,\kms\, and the uncertainty in
$v_\mathrm{str}/\sigma_\mathrm{RMS}$ is $\sim 0.03$.

The second and third column of Fig.~\ref{fig:intmom_triax} show
respectively the intermediate-over-major $\sigma_b/\sigma_a$ and
minor-over-major $\sigma_c/\sigma_a$ axis ratios of the velocity
ellipsoid. The velocity ellipsoid of the triaxial Abel model is
aligned with the confocal ellipsoidal coordinate system, so that its
semi-axis lenghts $\sigma_a \ge \sigma_b \ge \sigma_c$ follow directly
from $\sigma_\tau^2 = \langle v_\tau^2 \rangle - \langle v_\tau
\rangle^2$ with $\tau=\lambda,\mu,\nu$. In general, this is not the
case for the triaxial Schwarzschild model, and instead we diagonalize
the (symmetric) velocity dispersion tensor with components $\langle
\sigma_{st} \rangle$ ($s,t=x,y,z$). As before, the axis ratios of the
velocity ellipsoid are quite well recovered by the best-fit
Schwarzschild model, except towards the $z$-axis (upper panels) where
it underestimates the anisotropy in the velocity distribution of the
input Abel model. Similarly, away from the $(x,z)$-plane
($\phi=0$\dgr, black curves), the Schwarzschild model increasingly
overestimates the $\sigma_b/\sigma_a$ ratio, while the
$\sigma_c/\sigma_a$ remains well reproduced. It is plausible that the
recovery in the $(x,z)$-plane is better, because it is optimally
sampled as starting space for the numerical orbit calculations, and it
is crossed perpendicularly by all four major orbit families.
Nevertheless, the absolute difference in $\sigma_a$, $\sigma_b$ and
$\sigma_c$ between the input Abel model and the best-fit Schwarzschild
model is on average $\sim 9$\,\kms. The axis ratios
$\sigma_b/\sigma_a$ and $\sigma_c/\sigma_a$ are on average recovered
within $\sim 6$\,\%.

%---------------------------------------------------------------------
\subsection{Three-integral distribution function}
\label{sec:intdftriax}
%---------------------------------------------------------------------

%%%FIG
\begin{figure*}
  \begin{center}
    \includegraphics[width=1.0\textwidth]{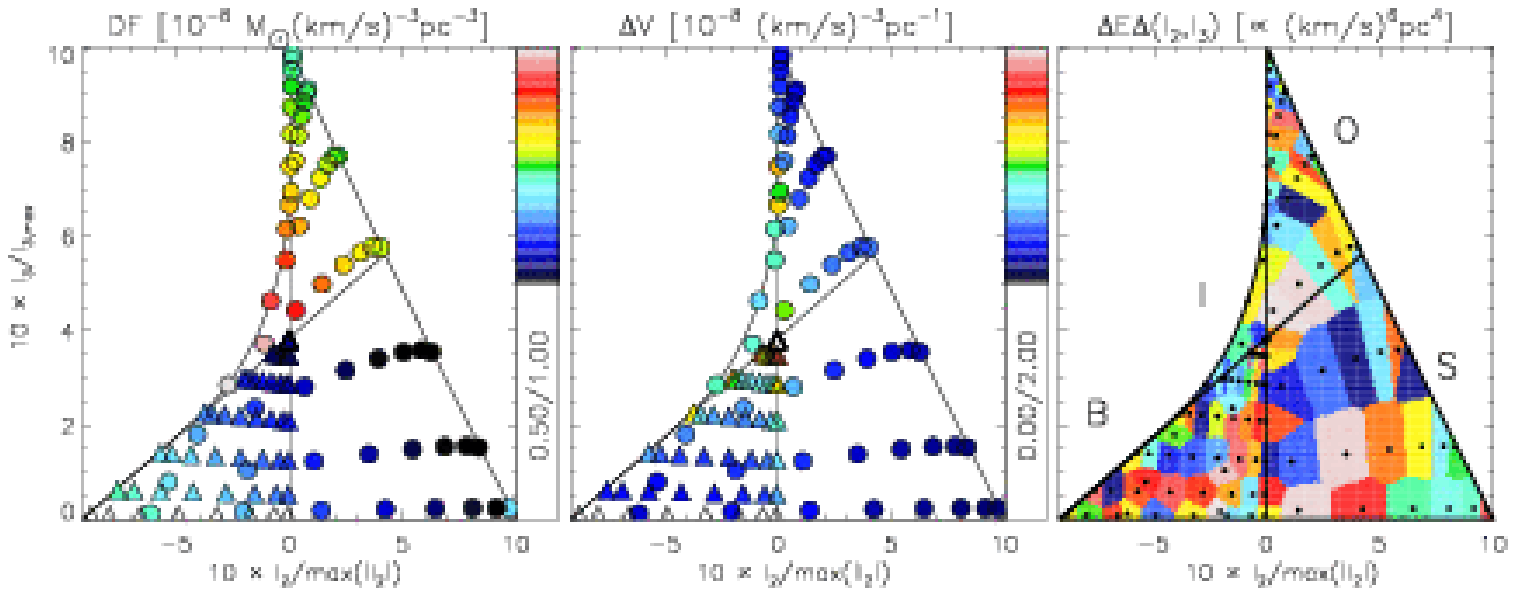}
  \end{center}
  \caption{Three quantities involved in the calculation of the orbital
    mass weights for a triaxial Abel model with an isochrone
    potential. For a given energy $E$, in each panel, the values of
    the second and third integral of motion, $I_2$ and $I_3$,
    indicated by the symbols, correspond to the orbital starting
    position and velocities in the triaxial Schwarzschild model that
    is fitted to the observables of this triaxial Abel model.  The
    solid curves, calculated with the expressions in
    \S~\ref{sec:integralspacecell}, bound and separate the regions of
    the box (B) orbits, inner (I) and outer (O) long-axis tube orbits
    and short-axis (S) tube orbits. The circles refer to orbits
    started in the $(x,z)$-plane and the triangles represent the
    additional set of orbits dropped from the equipotential surface
    (see \S~\ref{sec:Schwmodelstriax}). The latter box orbits may
    overlap with those started from the $(x,z)$-plane. The colours in
    the \textit{left panel} indicate the value of the DF
    $f(E,I_2,I_3)$ for each orbit in units of
    M$_\odot$\,(km/s)$^{-3}$\,pc$^{-3}$, with the (linear) scale given
    by the vertical bar on the right. The colours in the
    \textit{middle panel} represent the values of $\Delta
    V(E,I_2,I_3)$ defined in eq.~\eqref{eq:defdelV} in units of
    (km/s)$^{-3}$\,pc$^{-1}$. The area of each Voronoi bin in the
    \textit{right panel}, multiplied by the range in energy $E$,
    approximates the cell $\Delta E \Delta(I_2,I_3)$ in integral space
    for each orbit. The product of these three quantities yields an
    estimate of the mass weight $\gamma(E,I_2,I_3)$ for each orbit.}
  \label{fig:calcmassweight}
\end{figure*}
%%%FIG

The fitted triaxial Schwarzschild model results in a mass weight
$\gamma$ per orbit. These mass weights are a function of the three
integrals of motion $(E,I_2,I_3)$. In general, only the energy is
exact, but for a separable potential $I_2$ and $I_3$ are also known
explicitly and given by eq.~\eqref{eq:triaxEI2I3}. The orbital mass
weights follow from the DF by integrating $f(E,I_2,I_3)$ over the part
of phase-space $(\mathbf{x},\mathbf{v})$ that is accesible by the
orbit. Since each orbit is a (unique) delta-function in
integral-space, the resulting orbital mass weights are in principle
zero. However, as described in \S~\ref{sec:Schwmodelstriax} and
\S~\ref{sec:fit2obstriax}, final orbits consists each of a bundle of
125 orbits started closely to each other and their assigned mass
weights are required to vary smoothly between neighbouring orbits.

To estimate the orbital mass weights from the input triaxial Abel
model, we divide the integral-space in finite cells and link each cell
to the orbit that corresponds to its centroid.  The corresponding mass
weights then follow from
\begin{equation}%
\label{eq:massweight_triax}%
\gamma(E,I_2,I_3) \! = \!\! \iiint\limits_\mathrm{cell} \!\!
f(E,I_2,I_3) \; \Delta V(E,I_2,I_3) \; \du E \du I_2 \du I_3, \!\!
\end{equation}
where 
\begin{equation}
  \label{eq:defdelV}
  \Delta V(E,I_2,I_3) =  \iiint\limits_\Omega
  \left| \frac{\partial(v_x,v_y,v_z)}{\partial(E,I_2,I_3)}\right| 
  \, \du x \, \du y \, \du z,
\end{equation}
with $\Omega$ the volume in configuration space accessible by the
orbit. The multi-component DF of the input triaxial Abel model
consists of basis functions defined in eq.~\eqref{eq:simpleDF}, with
the DF parameters and weights per component given in
\S~\ref{sec:triaxabelmodel}. Below, we first calculate $\Delta V$ and
the cell in integral space, and then return to the comparison of the
orbital mass weights.

%---------------------------------------------------------------------

\subsubsection{Integral over configuration-space}
\label{sec:phasespacevolumetriax}

The expression for $\Delta V(E,I_2,I_3)$ of a single orbit in a
triaxial St\"ackel potential can be deduced from the relations in
\S~7.1 of de Zeeuw (1985a). It is given by
\begin{multline}
  \label{eq:deltaV}
  \Delta V(E,I_2,I_3) = (\gamma-\alpha) \iiint\limits_\Omega
  \frac{(\lambda-\mu)(\mu-\nu)(\nu-\lambda)}{a(\lambda)\,a(\mu)\,a(\nu)}
  \\ \times \sqrt{ \frac{8(\lambda+\beta)(\mu+\beta)(\nu+\beta)}
    {\left[E\!-\!V_\mathrm{eff}(\lambda)\right]
      \left[E\!-\!V_\mathrm{eff}(\mu)\right]
      \left[E\!-\!V_\mathrm{eff}(\nu)\right]} } \; \du \lambda \,\du
  \mu \,\du \nu,
\end{multline}
where $a(\tau)$, $\tau=\lambda,\mu,\nu$, is defined as
\begin{equation}
  \label{eq:defatau}
  a(\tau) = (\tau+\alpha)(\tau+\beta)(\tau+\gamma),
\end{equation}
the effective potential $V_\mathrm{eff}$ as
\begin{equation}
\label{eq:defVeff}
V_\mathrm{eff}(\tau) = \frac{I_2}{\tau+\alpha} +
\frac{I_3}{\tau+\gamma} +
U[\tau,-\alpha,-\gamma],
\end{equation}
and $\Omega$ is the volume in configuration space accessible by the orbit
in the triaxial separable potential that obeys $(E,I_2,I_3)$. The last
term in eq.~\eqref{eq:defVeff} is equal to the St\"ackel potential
\eqref{eq:triaxV_S} along the intermediate $y$-axis.

Because of the separability of the equations of motion, each orbit
in a triaxial separable potential can be considered as a sum of
three independent motions. Each of these one-dimensional motions
is either an oscillation or rotation in one of the three confocal
ellipsoidal coordinates $(\lambda,\mu,\nu)$, such that the
configuration space volume $\Omega$ is bounded by the
corresponding coordinate surfaces. The values of
$(\lambda,\mu,\nu)$ that correspond to these bounding surfaces can
be found from Table~\ref{tab:configspace} for the four families of
regular orbits: boxes (B), inner (I) and outer (O) long-axis
tubes, and short-axis (S) tubes.
Whereas $\alpha$, $\beta$ and $\gamma$ are the limits on
$(\lambda,\mu,\nu)$ set by the foci of the confocal ellipsoidal
coordinate system, the other limits are the solutions of
$E=V_\mathrm{eff}(\tau)$ (see Fig.~7 of de Zeeuw
1985a\nocite{1985MNRAS.216..273D}). In the case of the triaxial
isochrone St\"ackel potential \eqref{eq:isoV_S}, we can write this
equation as a fourth-order polynomial in $\sqrt{\tau}$. The solutions
are then the squares of three of the four roots of this polynomial
(the fourth root is always negative).

%%%TAB
\begin{table*}
\begin{center}
  \caption{Configuration space volume $\Omega$ accessible by the four
    families of regular orbits.}
  \begin{tabular}{lc*{4}{r@{\hspace{5pt}}c@{\hspace{5pt}}l}}
%  \begin{tabular}{cc rcl ccc}
    \hline
    \hline
    orbit & $I_2$ & &$E$& & &$\lambda$& & &$\mu$& & &$\nu$& \\
    \hline
    Box orbits & $<0$ &
    $V_\mathrm{eff}(-\beta)$ &$\ldots$& $0$ &
    $-\alpha$ &$\ldots$& $\lambda_\mathrm{max}$ &
    $-\beta $ &$\ldots$& $\mu_\mathrm{max}    $ &
    $-\gamma$ &$\ldots$& $\nu_\mathrm{max}    $ \\
    Inner long-axis tube orbits & $<0$ &
    $\min\left[V_\mathrm{eff}(\mu)\right]$ &$\dots$&
    $V_\mathrm{eff}(-\beta)$ &
    $-\alpha        $ &$\ldots$& $\lambda_\mathrm{max}$ &
    $\mu_\mathrm{min}$ &$\ldots$& $\mu_\mathrm{max}    $ &
    $-\gamma        $ &$\ldots$& $-\beta             $ \\
    Outer long-axis tube orbits & $>0$ &
    $\min\left[V_\mathrm{eff}(\lambda)\right]$ &$\ldots$&
    $V_\mathrm{eff}(-\beta)$ &
    $\lambda_\mathrm{min}$ &$\ldots$& $\lambda_\mathrm{max}$ &
    $\mu_\mathrm{min}    $ &$\ldots$& $-\alpha            $ &
    $-\gamma            $ &$\ldots$& $-\beta             $ \\
    Short-axis tube orbits & $>0$ &
    $\max\left\{V_\mathrm{eff}(-\beta) ,
      \min\left[V_\mathrm{eff}(\lambda)\right] \right\}$ &$\ldots$& $0$ &
    $\lambda_\mathrm{min}$ &$\ldots$& $\lambda_\mathrm{max}$ &
    $-\beta             $ &$\ldots$& $-\alpha            $ &
    $-\gamma            $ &$\ldots$& $\nu_\mathrm{max}    $ \\
    \hline
  \end{tabular}
  \label{tab:configspace}
\end{center}
\end{table*}
%%%TAB

For each orbit in our Schwarzschild model, we compute $(E,I_2,I_3)$ by
substituting the starting position and velocities of the orbit into
the expressions \eqref{eq:triaxEI2I3}. From the value of $E$ and the
sign of $I_2$ (while always $I_3\ge0$), we determine to which orbit
family it belongs. The corresponding configuration space volume
$\Omega$ is then given by the boundaries for $\lambda$, $\mu$ and
$\nu$ in the last three columns of Table~\ref{tab:configspace}. The
value of $\Delta V(E,I_2,I_3)$ follows by numerical evaluation of the
right-hand side of eq.~\eqref{eq:deltaV}.

The integrand in eq.~\eqref{eq:deltaV} contains singularities at the
integration limits, which can be easily removed for a triaxial
isochrone potential. We write the integrand completely in terms of
$(\sqrt{\sigma}\pm\sqrt{\tau})^{1/2}$, where
$\sigma,\tau=\lambda,\mu,\nu$ or a constant value. Suppose now that
the integral over $\lambda$ ranges from $\lambda_0$ to $\lambda_1$ and
the terms $(\sqrt{\lambda}-\sqrt{\lambda_0})^{1/2}$ and
$(\sqrt{\lambda_1}-\sqrt{\lambda})^{1/2}$ appear in the denominator.
The substitution $\sqrt{\lambda} = \sqrt{\lambda_0} +
(\sqrt{\lambda_1} - \sqrt{\lambda_0}) \sin^2\eta$ then removes both
singularities since $d\lambda / [(\sqrt{\lambda}-\sqrt{\lambda_0})
(\sqrt{\lambda_1}-\sqrt{\lambda})]^{1/2} = 4\sqrt{\lambda}\,d\eta$.

%---------------------------------------------------------------------

\subsubsection{Cell in integral space}
\label{sec:integralspacecell}

We approximate the triple integration over the cell in integral space
in eq.~\eqref{eq:massweight_triax} by the volume $\Delta E
\Delta(I_2,I_3)$. Here $\Delta E$ is the (logarithmic) range in $E$
between subsequent sets of orbits at different energies (see
\S~\ref{sec:Schwmodelstriax}), with limits given by the central
potential and $E=0$. Because we do not directly sample $I_2$ and $I_3$
in our implementation of Schwarzschild's method, as their expressions
are in general unknown, we cannot directly calculate the area
$\Delta(I_2,I_3)$. Instead, we compute the Voronoi diagram of the
points in the $(I_2,I_3$)-plane that correspond to the starting
position and velocities of each orbit, at a given energy $E$. An
example is given in the right panel of Fig.~\ref{fig:calcmassweight}.
The area of the Voronoi bins approximates the area $\Delta(I_2,I_3)$
for each orbit.

The four families of regular orbits are separated by two lines that
follow from $I_2=0$ and $E=V_\mathrm{eff}(-\beta)$. The latter
provides the part of the boundary on $I_2$ and $I_3$ for the box
orbits. The remainder of this boundary is given by the positivity
constraint on $I_3$ and by the solution of (cf.\ eqs 64 and 65 of de
Zeeuw 1985a\nocite{1985MNRAS.216..273D})
\begin{equation}
  \label{eq:boundaryI2I3}
  E=V_\mathrm{eff}(\kappa_0)
  \quad \mathrm{and} \quad
  \left[ \frac{\du V_\mathrm{eff}(\kappa)}{\du \kappa}
  \right]_{\kappa_0}
  \hspace{-10pt}=0, \qquad \kappa_0\ge-\beta.
\end{equation}
Substituting $V_\mathrm{eff}$ from eq.~\eqref{eq:defVeff} and
using $\du U[\tau,-\alpha,-\gamma]/\du \tau =
U[\tau,\tau,-\alpha,-\gamma]$, we find the solution
\begin{equation}
  \label{eq:solboundI2I3}
  I_2 = \frac{(\kappa_0+\alpha)^2}{(\alpha-\gamma)} 
  \Bigl\{ E - U[-\alpha,\kappa_0,\kappa_0] \Bigr\},
\end{equation}
and similarly for $I_3$ by interchanging $\alpha \leftrightarrow
\gamma$. For $-\beta\le\kappa_0\le-\alpha$, the solution describes the
boundary curve for which $I_2\le0$ and corresponds to the thin I tube
orbits.  For $\kappa_0\ge-\alpha$, we find the boundary curve for
which $I_2\ge0$, corresponding to the thin O and S tube orbits.

There are limits on the values of $\kappa_0$ depending on the value of
$E$, and sometimes there are no valid solutions for $\kappa_0$, which
implies that only box orbits contribute at that energy. These limits
can be obtained from the thin orbit curves in the $(x,z)$-plane. With
$y=v_x=v_z=0$, the expressions \eqref{eq:triaxEI2I3} for the integrals
of motion reduce in this plane to
\begin{eqnarray}
  \label{eq:EI2I3xzplane}
  E & = & \textstyle{\frac12} v_y^2 + U[\lambda,\kappa,-\beta],
  \nonumber \\
  I_2 & = & x^2 \, \Bigl\{ 
  \textstyle{\frac12} v_y^2 + (\alpha-\beta)
  U[\lambda,\kappa,-\beta,-\alpha] \Bigr\},
  \\
  I_3 & = & z^2 \, \Bigl\{ 
  \textstyle{\frac12} v_y^2 + (\gamma-\beta)
  U[\lambda,\kappa,-\beta,-\gamma] \Bigr\},
  \nonumber
\end{eqnarray}
with $-\gamma\le\kappa\le-\alpha$ replacing $\mu$ and $\nu$
respectively above and below the focal curve given by
$z^2/(\gamma-\beta)-x^2/(\beta-\alpha)=1$. Next, we substitute the
expression for $E$ in those for $I_2$ and $I_3$ and we use that
$(\tau+\beta)U[\lambda,\kappa,\tau,-\beta] = U[\lambda,\kappa,\tau] -
U[\lambda,\kappa,-\beta]$, respectively for $\tau=-\alpha$ and
$\tau=-\gamma$. We find that the thin orbit curves follow by solving
$I_2=0$ and thus $E=U[\lambda,\kappa_0,\kappa_0]$ for I tubes, and
$I_3=0$ and thus $E=U[\kappa_0,\kappa_0,\kappa]$, with $\kappa=\mu$
for O tubes and $\kappa=\nu$ for S tubes. In general these equations
have to be solved numerically, but in the case of the triaxial
isochrone potential \eqref{eq:isoV_S}, they reduce to a second order
polynomial in $\sqrt{\kappa_0}$ and the solutions simply follow from
the roots of the polynomial.

%---------------------------------------------------------------------

\subsubsection{Orbital mass weight distribution}
\label{sec:orbmassweight}

%%%FIG
\begin{figure*}
  \begin{center}
    \includegraphics[width=1.0\textwidth]{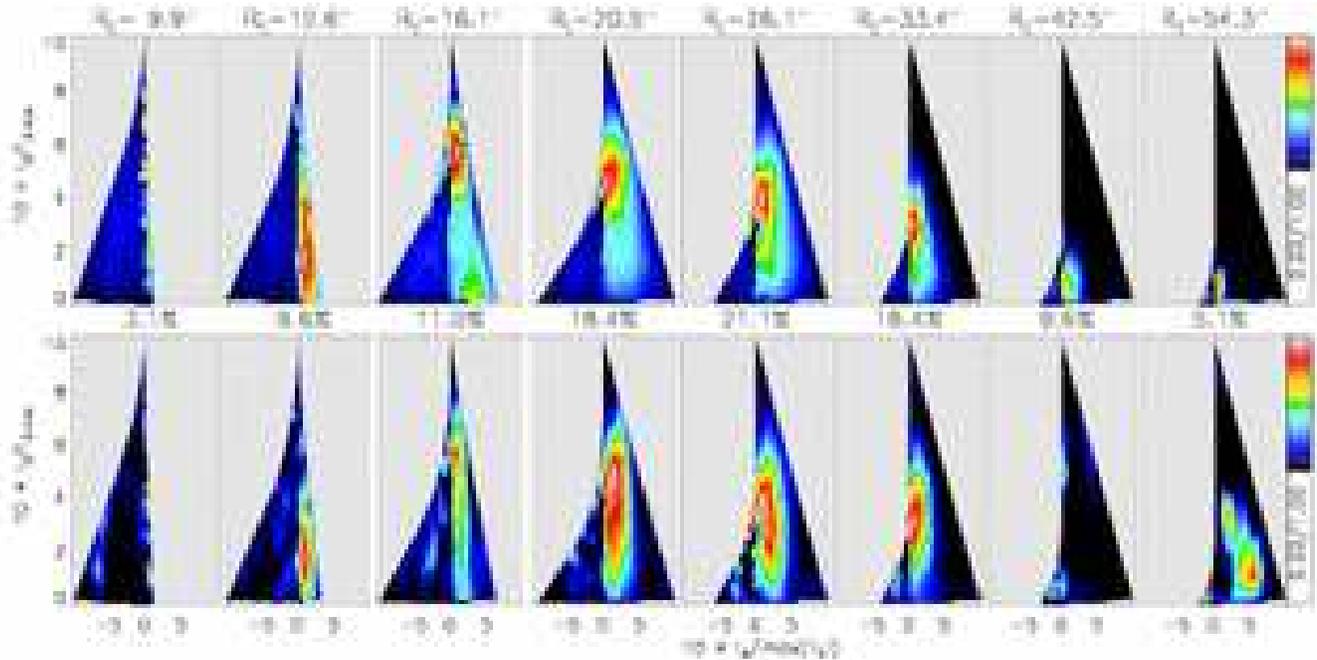}
  \end{center}
  \caption{ The orbital mass weight distribution for the input
    triaxial Abel model (top) and for the best-fit triaxial
    Schwarzschild model (bottom). From left to right the energy
    increases, corresponding to increasing distance from the centre,
    indicated by the radius $R_E$ (in arcsec) of the thin short-axis
    tube orbit on the $x$-axis.  The vertical and horizontal axes
    represent respectively the second and third integral of motion,
    $I_2$ and $I_3$, normalised by their maximum amplitude (for given
    $E$).  In each panel, the colours represent the mass weights,
    normalised with respect to the maximum in that panel, and with the
    (linear) scale given by the vertical bars on the right. Between
    the two rows of panels, the fraction (in \%) of the included mass
    with respect to the total mass is indicated.}
  \label{fig:df_triax}
\end{figure*}
%%%FIG

Once we have computed for each orbit the DF $f(E,I_2,I_3)$, $\Delta
V(E,I_2,I_3)$ and the cell $\Delta E \Delta(I_2,I_3)$ in integral
space (Fig.~\ref{fig:calcmassweight}), its (approximate) mass weight
$\gamma(E,I_2,I_3)$ follows by multiplication of these three
quantities. As before, the choice of maximum streaming for the (LR and
SR) rotating components reduces the accessible integral space, and
thus also the corresponding orbital mass weights, by a factor two. 

The resulting orbital mass weight distribution of the input triaxial
Abel model is shown in the top panels of Fig.~\ref{fig:df_triax}, and
that of the fitted triaxial Schwarzschild model in the bottom panels.
The energy $E$ increases from left to right, which corresponds to
increasing distance from the centre as is indicated by the radius
$R_E$ (in arcsec) at the top of each panel. For this representative
radius we use the radius of the corresponding thin (S) tube orbit on
the $x$-axis. The values of $I_2$ and $I_3$, on the horizontal and
vertical axes respectively, are both normalised with respect to their
maximum amplitude at the given energy. In each panel, the mass weight
values are normalised with respect to the maximum in that panel.
Between the two rows of panels, the fraction of the summed values in
each panel with respect to the total mass weight in all panels is
given (in \%).

The panels with $R_E \la 40$\arcsec\ are best constrained by the
kinematic observables. This takes into account that even orbits that
extend beyond the maximum radius covered by the observables can
contribute significantly at smaller radii. In these panels, the main
features of the orbital mass weight distribution of the triaxial Abel
model are recovered. In the outer parts the Schwarzschild model is
still constrained by the mass model, which extends to a radius of
about 100\arcsec, but the orbital mass weight distribution deviates
from that of the input Abel model due to the lack of kinematic
constraints. A point-by-point comparison yields an average fractional
error of $\sim 50$\,\%, and if we consider in each panel the mass
weights above the mean value, which together contribute more than half
of the total mass, the fractional error decreases to $\sim 30$\,\%.
However, this way of quantifying the recovery is (somewhat) misleading
since the relatively large fractional errors are at least partially
caused by the strong peaks in the orbital mass weight distribution.
For example, if in the input Abel model a certain orbit gets a
significant weight, but in the Schwarzschild model, due to numerical
uncertainties, this weight is assigned to a neighboring orbit with a
(slightly) different value of $I_3$, the relative error at each of the
corresponding points in the integral space can be very large.

%%%FIG
\begin{figure*}
  \begin{center}
    \includegraphics[width=1.0\textwidth]{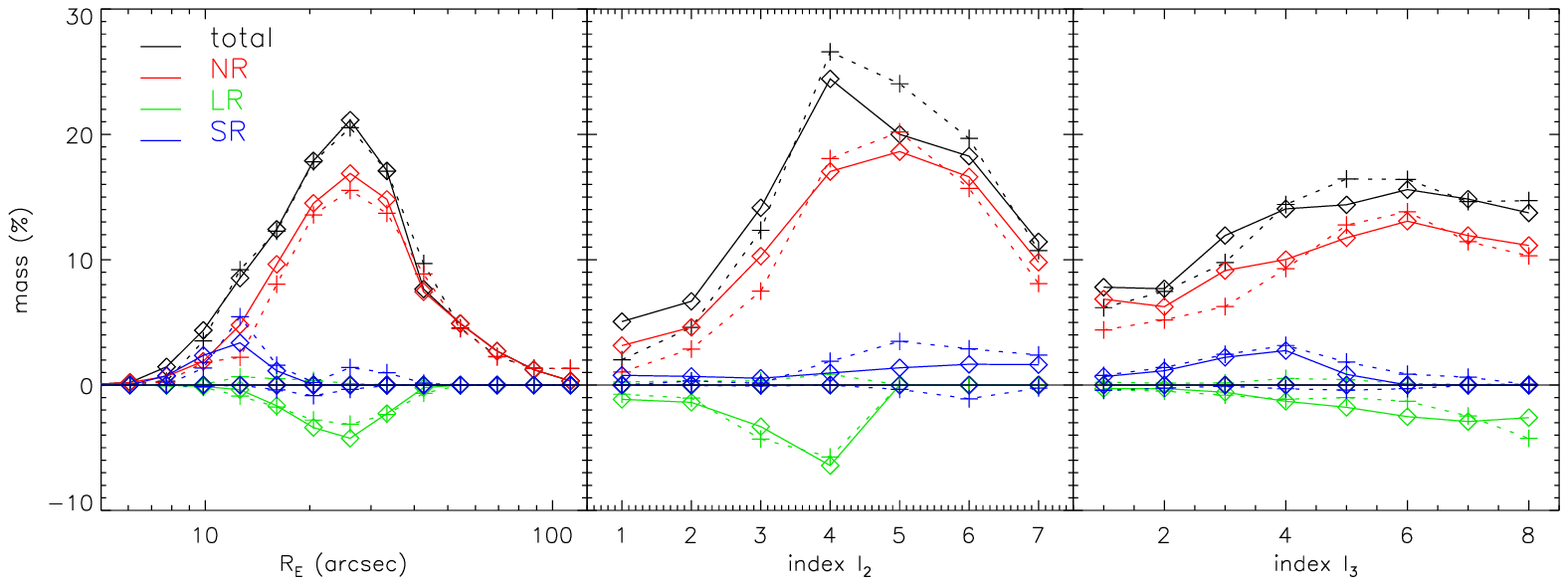}
  \end{center}
  \caption{The orbital mass weights (in \% of the total mass) for the input
    triaxial Abel model (diamonds connected by solid curves) and for
    the best-fit triaxial Schwarzschild model (crosses connected by
    dotted curves), as function of each of the three integrals of
    motion. These 'projections' of the three-dimensional orbital mass
    weight distribution shown in Fig.~\ref{fig:df_triax} are obtained
    by collapsing the cube in $(E,I_2,I_3)$ in two dimensions. As
    before, we represent the energy $E$ in the first panel by the
    radius $R_E$ (in arcsec) of the thin short-axis tube orbit on the
    $x$-axis.  For the second and third integral of motion, $I_2$ and
    $I_3$, we use the index in the cube, since the (range of) their
    values changes with $E$. The total distribution (black colour) is
    split into contributions from the non-rotating (NR; red),
    long-axis rotating (LR; green) and short-axis rotating (SR; blue)
    components. Moreover, for each rotating component the
    contributions from the two directions of rotation are separated by
    making the mass weights for one of the directions negative.}
  \label{fig:orbmw_triax}
\end{figure*}
%%%FIG

Henceforth, we show in Fig.~\ref{fig:orbmw_triax} the orbital mass
weights as function of each of the three integrals of motion
separately by collapsing the cube in $(E,I_2,I_3)$ in the remaining
two dimensions. We again use $R_E$ as a representative radius for $E$
(first panel), but since the (range of) values for $I_2$ and $I_3$
change with $E$ (see also Fig.~\ref{fig:df_triax}), we use their index
in the cube instead. In addition to the total distribution, we also
show the contribution of the three different NR, LR and SR components
separately, as well as for the latter two rotating components the
contributions from the two directions of rotation by making the mass
weights for one of the directions negative. Since the input triaxial
Abel model (diamonds connected by solid curves) is constructed with
maximum streaming in one of the two directions for both the LR and SR
component (see \S~\ref{sec:triaxabelmodel}), the opposite direction in
both cases has zero mass weight. This is nicely reproduced by the
best-fit triaxial Schwarzschild model (crosses connected by dotted
curves) in which $\sim 2$\,\% of the total mass, or $\sim 10$\,\% of
the mass of the LR and SR components, is wrongly assigned to the
opposite direction.  Keeping in mind that the orbital mass weights
itself are not directly fitted and that the typical velocity error of
7.5 \kms\ is more than $10$\,\% of the maximum in the simulated
velocity field (see \S~\ref{sec:triaxabelmodel}), these per centages
are (well) within the expected uncertainties.

From the first panel of Fig.~\ref{fig:orbmw_triax}, we see that mass
as function of $E$ is well recovered, even in the outer parts where
(nearly) all the constraints come only from the mass model. The
average absolute difference is $\sim 0.7$\,\%.  Whereas for $E$ the
constraints provided by the mass model already seem sufficient, for
$I_2$ and $I_3$ the kinematic constraints are essential. Not
suprisingly, we then also see that the recovery is less good with an
average absolute difference of $\sim 1.9$\,\% in $I_2$ and $\sim
1.0$\,\% in $I_2$. The main contribution is from the NR component,
while the two rotating components seem to better recovered.

%============================= section 6 =============================
\section{Axisymmetric three-integral galaxy models}
\label{sec:axigalmodels}
%=====================================================================

We now consider three-integral galaxy models in the axisymmetric
limit. As we have seen in the Introduction
(Section~\ref{sec:abelintroduction}), various groups have successfully
developed independent axisymmetric implementations of Schwarzschild's
method and verified their codes in a number of ways.  The published
tests to recover a known (analytical) input model have been limited to
spherical geometry or to an axisymmetric DF that is a function of the
two integrals of motion $E$ and $L_z$ only.

Here, we present the velocity moments of the three-integral Abel DF in
the axisymmetric limit and we choose again the isochrone form in
eq.~\eqref{eq:isoUtau} for the St\"ackel potential. The properties of
the resulting three-integral Kuzmin-Kutuzov models can be expressed
explicitly in cylindrical coordinates. We construct an axisymmetric
oblate Abel model and fit Schwarzschild models to the resulting
observables to test how well the axisymmetric implementation of
Schwarzschild's method, as presented in Cappellari et al.\
(2006\nocite{2006MNRAS.366.1126C}), recovers the intrinsic velocity
moments as well as the three-integral DF.

%---------------------------------------------------------------------
\subsection{Velocity moments and line-of-sight velocity distribution}
\label{sec:axivelmom}
%---------------------------------------------------------------------

When two of the three constants $\alpha$, $\beta$ or $\gamma$ are
equal, the confocal ellipsoidal coordinates $(\lambda,\mu,\nu)$ reduce
to spheroidal coordinates and the triaxial St\"ackel potential
\eqref{eq:triaxV_S} becomes axisymmetric.

%---------------------------------------------------------------------

\subsubsection{Oblate axisymmetric model}
\label{sec:oblateaxi}

When $\beta=\alpha\neq\gamma$ (triaxiality parameter $T=0$), we cannot
use $\mu$ as a coordinate and replace it by the azimuthal angle
$\phi$, defined as $\tan\phi = y/x$.  The relation between
$(\lambda,\phi,\nu)$ and the usual cylindrical coordinates
$(R,\phi,z)$ is given by
\begin{equation}
  \label{eq:oblate_sph2car}
  R^2 = \frac{(\lambda+\alpha)(\nu+\alpha)}{\alpha-\gamma},
  \quad
  z^2 = \frac{(\lambda+\gamma)(\nu+\gamma)}{\gamma-\alpha}.
\end{equation}
The St\"ackel potential $V_S(\lambda,\nu)=U[\lambda,-\alpha,\nu]$ is
\textit{oblate axisymmetric}. The corresponding integrals of motion
follow by substitution of $\mu=-\beta=-\alpha$ in the expressions
\eqref{eq:triaxEI2I3}, so that the second integral of motion reduces
to $I_2=\frac12L_z^2$.

With the choice \eqref{eq:dfabel} for the DF, the expression for
velocity moments $\mu_{lmn}(\lambda,\nu)$ is that of the triaxial case
given in eq.~\eqref{eq:mugeneral}, but with $\mu=-\beta=-\alpha$. From
Fig.~\ref{fig:Slim}, we see that the lower limit on $w$ vanishes.  For
the NR type of components, $S_\mathrm{max} =
S_\mathrm{top}(\lambda,\mu,-\gamma)$ and the corresponding velocity
moments $\mu^\mathrm{NR}_{lmb}(\lambda,\nu)$ vanish when either $l$,
$m$ or $n$ is odd. Because the only family of orbits that exists are
the short-axis tube orbits, we can introduce net rotation (around the
short $z$-axis) by setting the DF to zero for $L_z<0$, so that
$\mu^\mathrm{SR}_{lmn}(\lambda,\nu) =
\frac12\mu^\mathrm{NR}_{lmn}(\lambda,\nu)$. These SR velocity moments
vanish when either $l$ or $n$ is odd, but are non-zero if $m$ is odd.
They should be multiplied with $(-1)^m$ for maximum streaming in the
opposite direction. By choosing different weights for both senses of
rotation, we can control the direction and the amount of streaming
motion.

In the conversion to observables described in
\S~\ref{sec:observables}, the matrix $\mathbf{Q}$, which transforms
the velocity components $(v_\lambda,v_\phi,v_\nu)$ to $(v_x,v_y,v_z)$,
reduces to
\begin{equation}
  \label{eq:obl_convN}
  \mathbf{Q}
  =
  \begin{pmatrix}
    A\cos\phi & -\sin\phi & -B\cos\phi \\
    A\sin\phi &  \cos\phi & -B\sin\phi \\
    B         &         0 &  A
  \end{pmatrix},
\end{equation}
where $A$ and $B$ are defined as
\begin{equation}
  \label{eq:obl_defAB}
  A^2 = \frac{(\lambda+\gamma)(\nu+\alpha)}
  {(\lambda-\nu)(\alpha-\gamma)},
  \quad
  B^2 = \frac{(\lambda+\alpha)(\nu+\gamma)}
  {(\lambda-\nu)(\gamma-\alpha)}.
\end{equation}
Because of the symmetry around the short-axis, the azimuthal viewing
angle $\varphi$ looses its meaning and the misalignment angle
$\psi=0$\dgr. We are left with only the polar viewing angle
$\vartheta$, which is commonly referred to as the inclination $i$,
with $i=0$\dgr\ face-on and $i=90$\dgr\ edge-on viewing. As a consequence,
the projection matrix $\mathbf{P}$ is a function of $i$ only and
follows by substituting $\vartheta=i$ and $\varphi=0$ in
eq.~\eqref{eq:projmatP}. The rotation matrix $\mathbf{R}$ in
eq.~\eqref{eq:rotmatR} reduces to the identity matrix, so that
$\mathbf{M}=\mathbf{P}\mathbf{S}\mathbf{Q}$.

The expression for the LOSVD follows from that of the triaxial case in
eq.~\eqref{eq:triaxAbelLOSVD} by substituting $\mu=-\beta=-\alpha$.
For the NR components, again $\Delta\xi'_\mathrm{NR} = 2\pi$ and the
simplified expression \eqref{eq:triaxLOSVD_NRsimpleDf} holds in case
of a DF basis function as defined in eq.~\eqref{eq:simpleDF}. To
introduce net rotation, we require that $(v_\mu=)\,v_\phi\ge0$ as in
\S~\ref{sec:triaxLOSVD_SR}, which yields SR components with maximum
streaming.  As illustrated in the right panel of
Fig.~\ref{fig:partintspace}, $\Delta\xi'_\mathrm{SR}$ is the length of
the part of the circle between the intersections
$\xi_\pm=2\arctan(u_\pm)$ with the line (with $u_\pm$ given in
eq.~\ref{eq:triaxline2angles_SR}), and which is on the correct side of
the line in eq.~\eqref{eq:triaxstreaming_SR}. This is again similar to
SR components in the triaxial case, but without the restriction to
stay within the ellipses.

%---------------------------------------------------------------------

\subsubsection{Prolate axisymmetric model}
\label{sec:prolateaxi}

When $\beta=\gamma\neq\alpha$ ($T=1$), we replace the coordinate $\nu$
by the angle $\chi$, defined as $\tan\chi = z/y$. The resulting
coordinates $(\lambda,\mu,\chi)$ follow from the above coordinates
$(\lambda,\phi,\nu)$ by taking $\nu\to\mu$, $\phi\to\chi$, and
$\gamma\to\alpha\to\beta$. The St\"ackel potential
$V_S(\lambda,\mu)=U[\lambda,\mu,-\gamma]$ is now \textit{prolate
  axisymmetric}. By substituting $\nu=-\beta=-\gamma$ in
eqs \eqref{eq:triaxEI2I3} and \eqref{eq:mugeneral}, we obtain the
expressions respectively for the integrals of motion (with
$I_3=\frac12L_x^2$) and for the intrinsic velocity moments
$\mu_{lmn}(\lambda,\mu)$. From Fig.~\ref{fig:Slim}, we see that now
the upper limit on $u$ vanishes. For the NR components,
$S_\mathrm{max} = S_\mathrm{top}(\lambda,\mu,-\gamma)$, and since we
only have the long-axis tube orbits, we can introduce net rotation
(around the $x$-axis) by setting the DF to zero for $L_x<0$, so that
$\mu^\mathrm{LR}_{lmn}(\lambda,\mu) =
\frac12\mu^\mathrm{NR}_{lmn}(\lambda,\mu)$. These LR velocity moments
vanish if either $l$ or $m$ is odd and multiplication with $(-1)^n$
yields net rotation in the opposite direction.

The matrix $\mathbf{Q}$, which transforms $(v_\lambda,v_\mu,v_\chi)$
to $(v_x,v_y,v_z)$, in this case reduces to
\begin{equation}
  \label{eq:pro_convN}
  \mathbf{Q}
  =
  \begin{pmatrix}
    C         &        -D &  0        \\
    D\cos\chi & C\cos\chi & -\sin\chi \\
    D\sin\chi & C\sin\chi &  \cos\chi
  \end{pmatrix},
\end{equation}
where $C$ and $D$ are given by
\begin{equation}
  \label{eq:pro_defCD}
  C^2 = \frac{(\lambda+\beta)(\mu+\alpha)}
  {(\lambda-\mu)(\alpha-\beta)},
  \quad
  D^2 = \frac{(\lambda+\alpha)(\mu+\beta)}
  {(\lambda-\mu)(\beta-\alpha)}.
\end{equation}
%
%where $C$ and $D$ follow from respectively $A$ and $B$ in
%\eqref{eq:obl_defAB} by replacing $\nu$ by $\mu$, and $\gamma$ by
%$\beta$. 
In the projection matrix $\mathbf{P}$ in eq.~\eqref{eq:projmatP}, we
substitute $\vartheta=\pi/2-i$ and $\varphi=0$, so that for
inclination $i=0$\dgr\ and $i=90$\dgr, we are respectively viewing the
prolate mass model end-on and side-on. In the rotation matrix
$\mathbf{R}$ we take $\psi=90$\dgr\ to align the projected major axis
horizontally.  The expression for the LOSVD follows from
eq.~\eqref{eq:triaxAbelLOSVD} by substituting $\nu=-\beta=-\gamma$,
and by requiring $(v_\nu=)\,v_\chi\ge0$ we obtain LR components with
maximum streaming. As for the oblate case and illustrated in the left
panel of Fig.~\ref{fig:partintspace}, $\Delta\xi'_\mathrm{SR}$ is the
length of the circle part between the angles $\xi_\pm=2\arctan(u_\pm)$
(with $u_\pm$ given in eq.~\ref{eq:triaxline2angles_LR}) which is on
the correct side of the line in eq.~\eqref{eq:triaxstreaming_LR}.

%---------------------------------------------------------------------
\subsection{Kuzmin-Kutuzov potential}
\label{sec:KKmodels}
%---------------------------------------------------------------------

In the axisymmetric limit, the form \eqref{eq:isoUtau} for $U(\tau)$
results in the Kuzmin-Kutuzov (1962\nocite{1962KK}) potential. We
give the properties relevant for our analysis, while further details
can be found in Dejonghe \& de Zeeuw
(1988\nocite{1988ApJ...333...90D}), including expressions and
plots of the mass density $\rho_S$, its axis ratios, and the
two-integral DF $f(E,L_z^2)$ consistent with $\rho_S$ [see also
Batsleer \& Dejonghe (1993\nocite{1993A&A...271..104B}), who also
corrected a typographical error in $f(E,L_z^2)$].

When $\beta=\alpha$, the oblate axisymmetric potential
$V_S(\lambda,\nu)=U[\lambda,-\alpha,\nu]$ and the third order
divided difference $U[\lambda,-\alpha,\nu,\sigma]$, which both
appear in the expressions for the integral of motions
\eqref{eq:triaxEI2I3}, have the simple forms
\begin{eqnarray}%
   \label{eq:KKpot}%
   V_S(\lambda,\nu) \hspace{-5pt} & = & \hspace{-5pt}
   \frac{-GM}{\sqrt{\lambda}+\sqrt{\nu}},
   \\
   \label{eq:KKdd3thorder}%
   U[\lambda,-\alpha,\nu,\sigma] \hspace{-5pt} & = & \hspace{-5pt}
   \frac{GM}{ (\sqrt{\lambda}+\sqrt{\nu}) (\sqrt{\lambda}+\sqrt{\sigma})
   (\sqrt{\nu}+\sqrt{\sigma}) },
\end{eqnarray}
where again $GM=\sqrt{-\gamma}+\sqrt{-\alpha}$, so that $V_S=-1$ in
the centre. By means of the relations
\begin{equation}
  \label{eq:oblate_car2sph}
  \lambda + \nu = R^2 + z^2 - \alpha - \gamma,
  \quad
  \lambda\nu = \alpha\gamma - \gamma R^2 - \alpha z^2,
\end{equation}
and $(\sqrt{\lambda}+\sqrt{\nu})^2 = \lambda + \nu +
2\sqrt{\lambda\nu}$ and $(\sqrt{\lambda}+\sqrt{\sigma})
(\sqrt{\nu}+\sqrt{\sigma}) = \sqrt{\lambda\nu} +
\sqrt{\sigma}(\sqrt{\lambda}+\sqrt{\nu}) + \sigma$, we can write
the potential and integrals of motion explicitly as elementary
expressions in the usual cylindrical coordinates.

When $\beta=\gamma$, the prolate potential
$V_S(\lambda,\mu)=U[\lambda,\mu,-\gamma]$ and the third order divided
difference $U[\lambda,\mu,-\gamma,\sigma]$ follow respectively from
\eqref{eq:KKpot} and \eqref{eq:KKdd3thorder} by replacing $\nu$ by
$\mu$.

%---------------------------------------------------------------------
\subsection{An axisymmetric Abel model}
\label{sec:axiabelmodel}
%---------------------------------------------------------------------

%%%FIG
\begin{figure*}
  \begin{center}
    \includegraphics[width=1.0\textwidth]{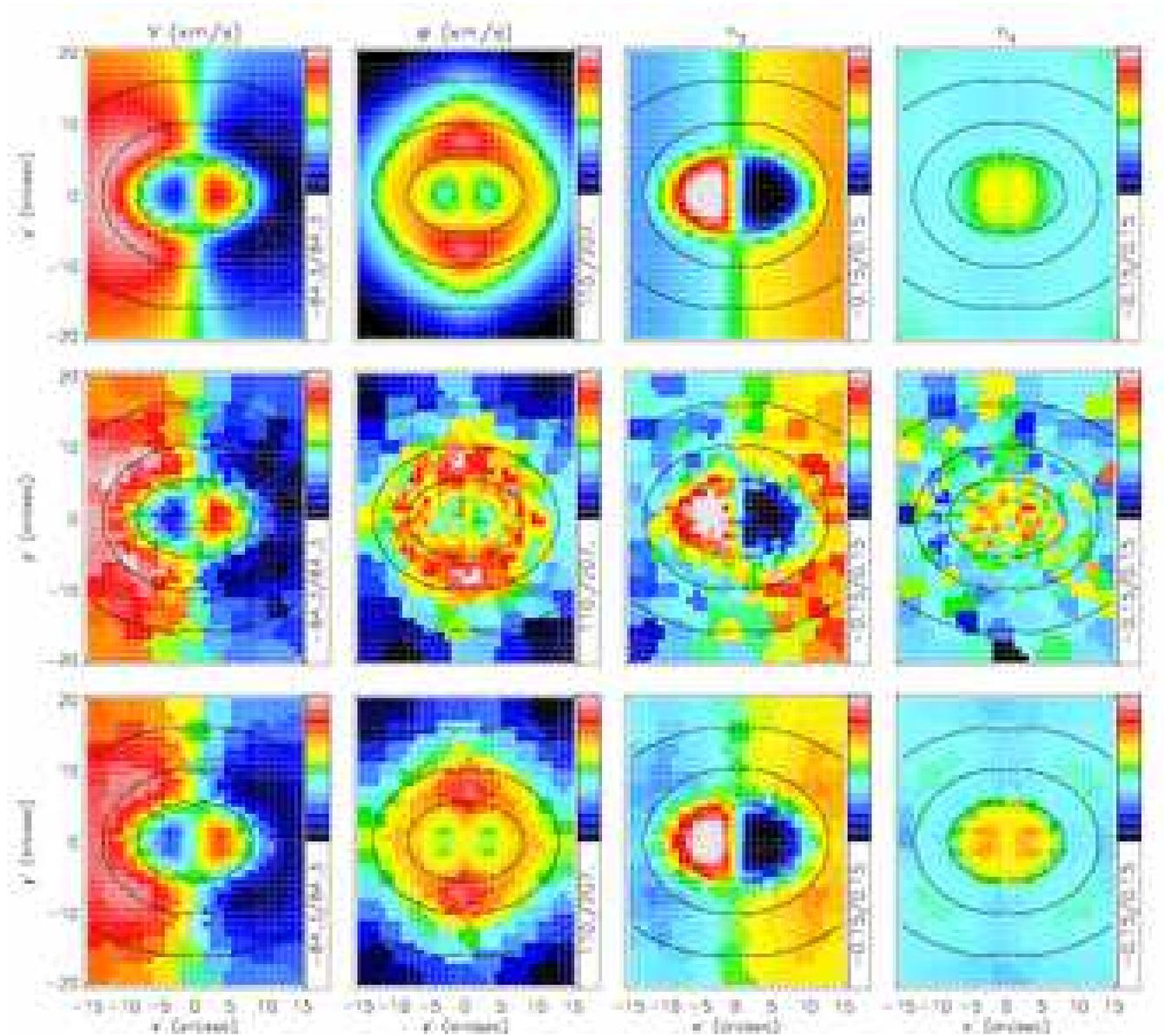}
  \end{center}
  \caption{Kinematic maps for an oblate axisymmetric Abel model (top
    row) and converted to observables with realistic measurement
    errors added (middle row; see \S~\ref{sec:axiabelmodel}) and for
    the fitted axisymmetric Schwarzschild model (bottom row; see
    \S~\ref{sec:recoveryaxi}). Parameters and colour scale are as in
    Fig.~\ref{fig:obsmom_triax}.}
  \label{fig:obsmom_axi}
\end{figure*}
%%%FIG

The above constructed triaxial Abel model
(\S~\ref{sec:triaxabelmodel}) transforms into an oblate axisymmetric
Abel model if we let $\zeta$ approach unity, while keeping $\xi=0.64$
fixed. Similar to the triaxial case, the DF contains a NR component
with the same parameters, $u=w=-0.5/(-\alpha)$ and $\delta=1$, but we
exclude the LR component since long-axis tube orbits do not exist in
an oblate axisymmetric galaxy. We include two SR components, one with
the same parameters as the NR component, and for the other we set
$w=0.5/(-\alpha)$ and $u=-1.0/(-\alpha)$, and we choose the sense of
rotation in the opposite direction. The latter implies a compact
counter-rotating component, which is clearly visible in the kinematic
maps shown in the top panels of Fig.~\ref{fig:obsmom_axi}. The
inclination is the same value as the polar angle $\vartheta$ for the
triaxial Abel model, i.e.  $i=70$\dgr, and the mass fractions of the
three DF components are respectively 20\%, 60\% and 20\%. Due to
axisymmetry, all maps of the even (odd) velocity moments are
bi-(anti)-symmetric and the velocity field shows a straight
zero-velocity curve. The signatures of the counter-rotation are
similar in the velocity field and $h_3$ (but anti-correlated), and
result in a decrease of $\sigma$ and an increase of $h_4$ in the
centre.

%---------------------------------------------------------------------
\subsection{Recovery of axisymmetric three-integral models}
\label{sec:recoveryaxi}
%---------------------------------------------------------------------

We now describe the application of our axisymmetric implementation of
Schwarzschild's method to the observables of the oblate axisymmetric
Abel model of \S~\ref{sec:axiabelmodel}, while highlighting the
differences with the application in triaxial geometry described in
Section~\ref{sec:recoverytriax}.

%---------------------------------------------------------------------

\subsubsection{Axisymmetric Schwarzschild model fit to observables of
  an oblate axisymmetric Abel model }
\label{sec:Schwmodelsaxi}

We use the implementation of Schwarzschild's method in axisymmetric
geometry that is described in detail in Cappellari et al.\
(2006\nocite{2006MNRAS.366.1126C}).  The main differences with respect
to our triaxial implementation are certain simplifications due to the
extra symmetry. There are no twists in the surface brightness and of
the four families of regular orbits only the short-axis tube orbits
are supported. We use the same set-up as in the triaxial case, but
since there are no box orbits, the additional dropping of orbits from
the equipotential surface is not needed.

Fig.~\ref{fig:obsmom_axi} shows that the (simulated) observables of
the oblate axisymmetric Abel model (top panels) are very well matched
by the best-fit axisymmetric Schwarzschild model (bottom panels). The
kinematics of the main body as well as the signatures of the
counter-rotating core are accurately fitted within the (added) noise.

%---------------------------------------------------------------------

\subsubsection{Intrinsic velocity moments}
\label{sec:intmomaxi}

%%%FIG
\begin{figure*}
  \begin{center}
    \includegraphics[width=1.0\textwidth]{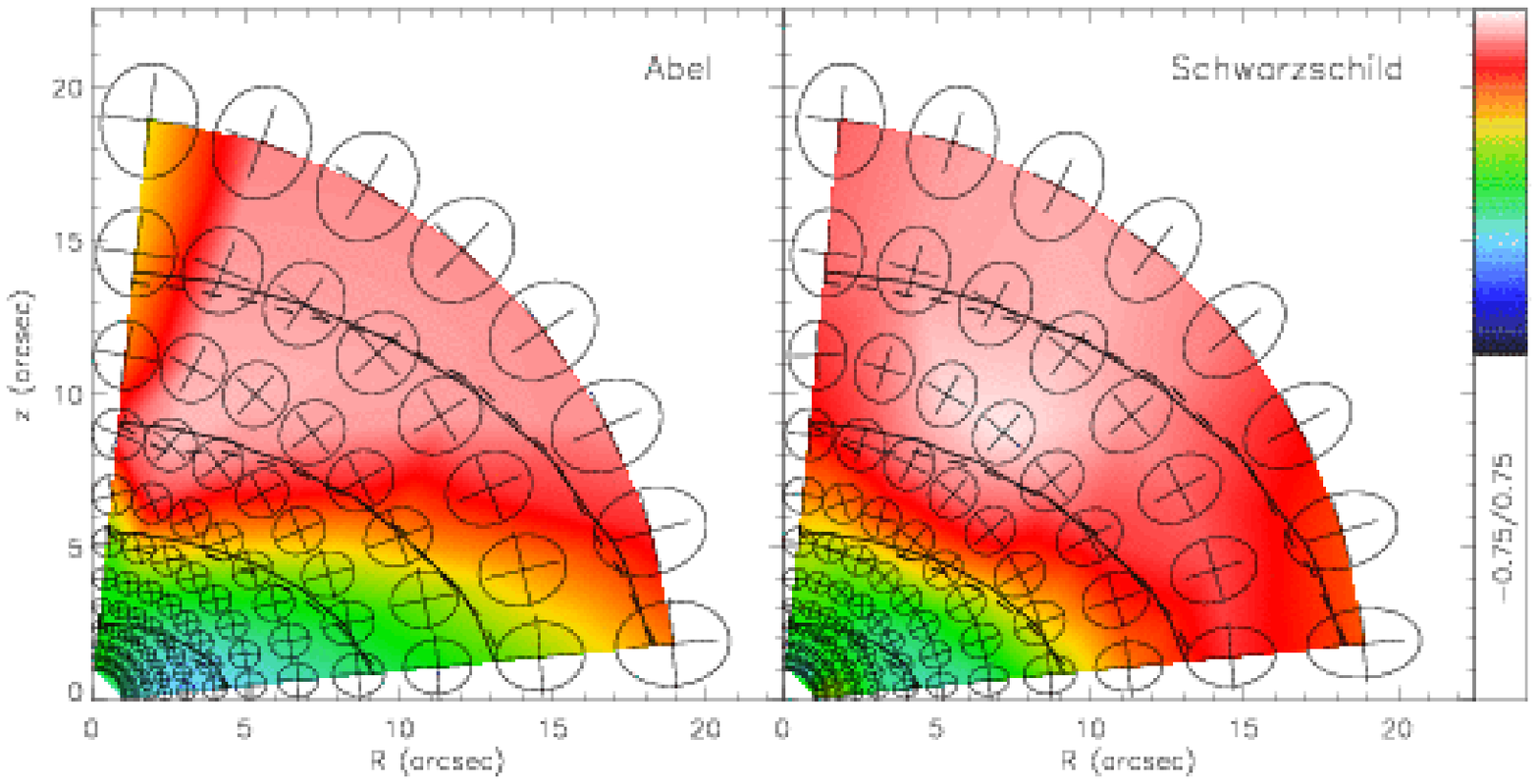}
  \end{center}
  \caption{ The mean azimuthal motion $\langle v_\phi\rangle$
    perpendicular to the meridional plane, normalised by
    $\sigma_\mathrm{RMS}$, for an oblate axisymmetric Abel model
    (left) and for the best-fit axisymmetric Schwarzschild model
    (right).  Parameters as in Fig.~\ref{fig:intVsig_triax}.}
  \label{fig:intVsig_axi}
\end{figure*}
%%%FIG

It is convenient to analyse the intrinsic velocity moments of (oblate)
axisymmetric models in cylindrical coordinates $(R,\phi,z)$.  Because
of axisymmetry the models are independent of the azimuthal angle
$\phi$, and it is sufficient to consider the meridional $(R,z)$-plane.
The analysis of the intrinsic velocity moments in the $(R,z)$-plane is
similar to that for the triaxial case in the $(x,z)$-plane
(\S~\ref{sec:intmomtriax}). In this case, the mean azimuthal rotation
$\langle v_\phi \rangle$, perpendicular to the meridional plane, is
the only non-vanishing first-order velocity moment. In
Fig.~\ref{fig:intVsig_axi}, we compare the values of $\langle v_\phi
\rangle / \sigma_\mathrm{RMS}$, indicated by the colours, for the Abel
model (left panel) with those for the fitted Schwarzschild model
(right panel). The root-mean-square velocity dispersion
$\sigma_\mathrm{RMS}$ is defined as $\sigma^2_\mathrm{RMS}=
(\sigma_R^2 +\sigma_\phi^2 + \sigma_z^2)/3$. The azimuthal axis of the
velocity ellipsoid, with semi-axis length $\sigma_\phi$ defined as
$\sigma_\phi^2=\langle v_\phi^2 \rangle - \langle v_\phi \rangle^2$,
is perpendicular to the meridional plane. The cross sections with the
meridional plane are indicated by the ellipses in
Fig.~\ref{fig:intVsig_axi}, where the semi-axis lengths follow from
\eqref{eq:semiaxislengthvelell} by replacing $(x,z)$ with $(R,z)$.

As in the triaxial case the density (solid curve) is well fitted by
the axisymmetric Schwarzschild model (dashed curve). The Abel model
shows a strong gradient in $\langle v_\phi \rangle /
\sigma_\mathrm{RMS}$, which is correctly recovered by the axisymmetric
Schwarzschild model. The absolute difference is on average less than
$0.06$, except near the symmetry $z$-axis. This is likely the result
of numerical difficulties due to the small number of (sampled)
short-axis tube orbits that contribute in this region.  The shape and
orientation of the ellipses are nearly identical, indicating that the
anisotropic velocity distribution of the Abel model is reproduced
within the expected uncertainties due to the errors in the simulated
kinematics. The axis ratios $\sigma_-/\sigma_+$ and
$\sigma_\phi/\sigma_+$ of the velocity ellipsoid are on average
recovered within $\sim 5$\,\%.

%---------------------------------------------------------------------

\subsubsection{Three-integral distribution function}
\label{sec:intdfaxi}

%%%FIG
\begin{figure*}
  \begin{center}
    \includegraphics[width=1.0\textwidth]{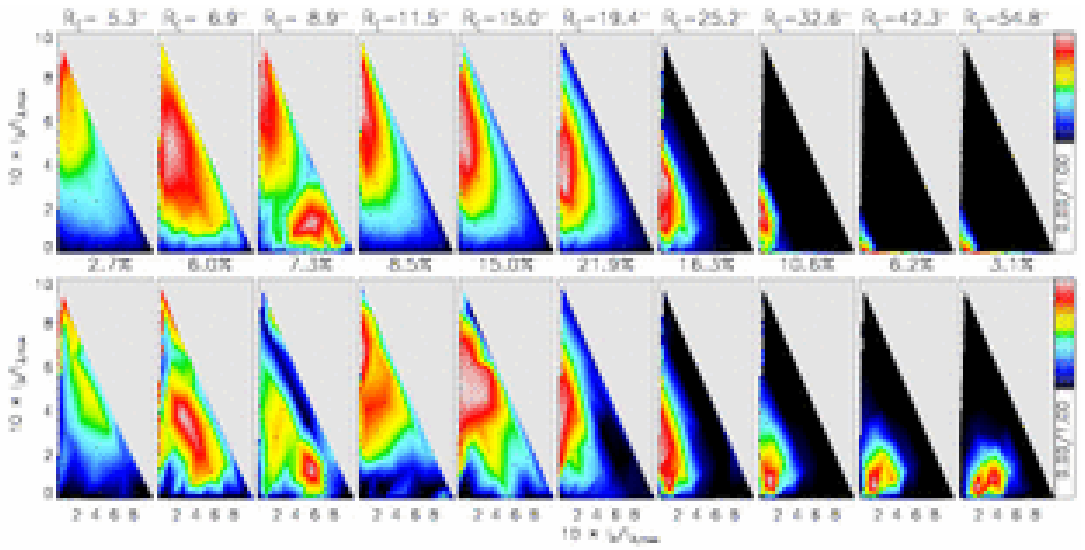}
  \end{center}
  \caption{The mass weight distribution for the input oblate Abel
    model (top) and for the best-fit axisymmetric Schwarzschild 
    model (bottom). Parameters as in Fig.~\ref{fig:df_triax}. The
    second integral of motion $I_2=\frac12L_z^2\ge0$, where $L_z$ is
    the component of the angular momentum parallel to the symmetry
    $z$-axis.}
  \label{fig:df_axi}
\end{figure*}
%%%FIG

In the oblate axisymmetric case, all (regular) orbits are short-axis
tube orbits with $I_2=\frac12L_z^2$ and energy $E$ ranging from
$\min\left[V_\mathrm{eff}(\lambda)\right]$ to zero.  The expression
for $\Delta V$ in eq.~\eqref{eq:deltaV} reduces to
\begin{multline}%
  \label{eq:phasespacevolumeaxi}%
  \Delta V(E,L_z,I_3) =
  \frac{4\pi}{\,|L_z|} \hspace{-3pt}
  \int\limits_{-\gamma}^{\nu_\mathrm{max}}
  \int\limits_{\lambda_\mathrm{min}}^{\lambda_\mathrm{max}}
  \frac{(\nu-\lambda)}{
  (\lambda\!+\!\alpha)(\lambda\!+\!\gamma)(\nu\!+\!\alpha)(\nu\!+\!\gamma)}
\\ \times
  \sqrt{
    \frac{(\lambda\!+\!\alpha)(\nu\!+\!\alpha)}
    {\left[E\!-\!V_\mathrm{eff}(\lambda)\right]
      \left[E\!-\!V_\mathrm{eff}(\nu)\right]}}
  \; \du \lambda \, \du \nu
\end{multline}
where as before $\nu_\mathrm{max}$, $\lambda_\mathrm{min}$ and
$\lambda_\mathrm{max}$ are the solutions of $E=V_\mathrm{eff}(\tau)$
(see Fig.~23 of de Zeeuw 1985a\nocite{1985MNRAS.216..273D}).  The
factor in front of the double integral includes the factor $2\pi$ from
the integration over the azimuthal angle $\phi$. 

%%%FIG
\begin{figure*}
  \begin{center}
    \includegraphics[width=1.0\textwidth]{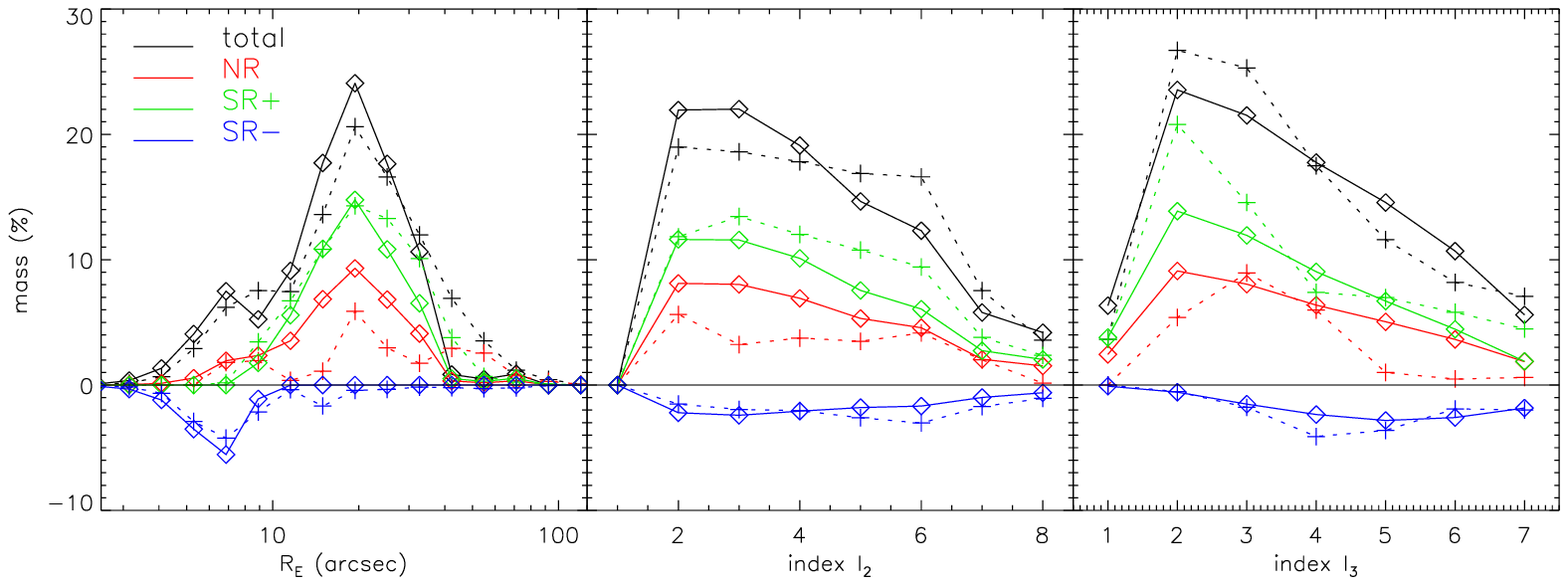}
  \end{center}
  \caption{The orbital mass weights for the input oblate Abel model (diamonds
    connected by solid curves) and for the best-fit axisymmetric
    Schwarzschild model (crosses connected by dotted curves). The
    parameters are as in in Fig.~\ref{fig:orbmw_triax}, except that
    rotation can only come from short-axis rotating (SR) components,
    for which the two directions of rotation are indicated
    separately.}
  \label{fig:orbmw_axi}
\end{figure*}
%%%FIG

In Fig.~\ref{fig:df_axi}, we compare the orbital mass weight
distribution of the input oblate Abel model (top panels), with that of
the best-fit axisymmetric Schwarzschild model (bottom panels). The
three-integral mass weight distributions are quite similar, even in
the panels with a relatively low mass content. The average fractional
error is $\sim 30$\,\%, and if we consider in each panel the mass
weights above the mean value, which together contribute more than half
of the total mass, the fractional error decreases to around $\sim
20$\,\%.  Because of possible strong point-to-point fluctuations as
discussed in \S~\ref{sec:orbmassweight}, we also show in
Fig.~\ref{fig:orbmw_axi} the orbital mass distribution as function of
each of the three integrals of motion separately by collapsing the
cube in $(E,I_2,I_3)$ in the remaining two dimensions. Besides the
total distribution, we show separately the contributions from the NR
component and the two opposite rotating SR components in the input
oblate Abel model (see \S~\ref{sec:axiabelmodel}). While the compact
counter-rotating SR component (blue) is nicely reproduced by the
best-fit axisymmetric Schwarzschild model, the mass assigned to the
main SR component is too high ($\sim 10$\,\% of the total mass), which
also results in an underestimation of the NR component. This is
reflected in the average absolute difference in mass as function of
$E$, which is $\sim 1.3$\,\%. As for the triaxial case, the recovery
for $I_2$ and $I_3$ is less good with average uncertainties of $\sim
2.1$\,\% and $\sim 2.4$\,\%, respectively.

A similar good recovery was found by Krajnovi{\'c} et al.\ 
(2005\nocite{2005MNRAS.357.1113K}) for the case of a two-integral DF
$f(E,L_z)$, which implies an isotropic velocity distribution in the
meridional plane. Thomas et al.\ (2004\nocite{2004MNRAS.353..391T})
showed that their independent axisymmetric numerical implementation of
Schwarzschild's method is similarly able to recover an analytical
$f(E,L_z)$. Our results show that the orbital mass weight distribution
that follows from a fully three-integral DF $f(E,L_z,I_3)$ can be
recovered as well.

%============================= section 7 =============================
\section{Discussion and conclusions}
\label{sec:abeldiscconcl}
%=====================================================================

We have extended the Abel models introduced by DL91 and generalised by
MD99, and shown that, in addition to the intrinsic velocity moments,
the full LOSVD of these models can be calculated in a straightforward
way. We have then used the Abel models to construct realistic
axisymmetric and triaxial galaxy models to test the accuracy of
Schwarzschild's orbit superposition method.

Although Abel models have separable potentials with a central core and
assume a specific functional form for the (three-integral) DF, they
display a large range of shapes and their observables, which can be
calculated easily, include many of the features seen in the kinematic
maps of early-type galaxies. We have used an isochrone St\"ackel
potential that in the axisymmetric limit reduces to the Kuzmin-Kutuzov
model and becomes H\'enon's isochrone in the spherical limit. Because
of the simple form of the isochrone potential, the resulting Abel
models are ideally suited to test numerical implementations of the
Schwarzschild orbit superposition method. The calculation of $\Delta
V$, needed when comparing the orbital mass weight distribution of the
Schwarzschild models with the three-integral DF of the Abel models,
simplifies significantly for this case.

Integral-field observations in principle provide the LOSVD as a
function of position on the sky, so that it is a function
$\mathcal{L}(x',y',v_{z'})$ that depends on three variables. The
oblate axisymmetric and triaxial galaxy models we have constructed,
have a DF which is a sum of Abel components $f(S)=f(-E+wI_2+uI_3)$
with different values of the parameters $w$ and $u$, so that the DF is
a function of three variables as well, namely the integrals of motion
$E$, $I_2$ and $I_3$. We have shown that the simulated integral-field
observables of these models are matched in detail by the best-fit
Schwarzschild model. This does not automatically imply that the
intrinsic velocity moments and the three-integral DF --- which are not
directly fitted --- are also correctly recovered.

First consider three-integral oblate models, i.e., with a DF that is a
function $f(E,L_z,I_3)$. In the special case that a galaxy happens to
be well approximated by a two-integral DF $f(E, L_z)$, the density
$\rho(R,z)$ uniquely determines the even part of $f(E,L_z)$ and the
mean streaming $\rho\langle v_\phi \rangle$ in the meridional plane
fixes the part of $f(E,L_z)$ that is odd in $L_z$ (Dejonghe
1986\nocite{1986PhR...133..217D}). Ignoring non-uniqueness in the
deprojection of the surface density $\Sigma$ (Rybicki
1987\nocite{1987IAUS..127..397R}) and the mean streaming motion $V$ on
the plane of the sky, these two quantities define a two-integral DF
completely.  The observed velocity dispersion and higher-order
velocity moments of the LOSVD then provide additional information,
which for example can be used to constrain the inclination (e.g.,
Cappellari et al.\ 2006\nocite{2006MNRAS.366.1126C}).  However, the
reliability of the derived inclination, of course, depends on the
correctness of the assumption of a two-integral DF.  In the more
realistic case of a three-integral DF $f(E,L_z,I_3)$, such a
one-to-one relation with (the velocity moments of) the observed LOSVD
$\mathcal{L}(x',y',v_{z'})$ has not been established.  Nevertheless,
we showed that, given integral-field observations of the velocity
moments of the LOSVD (up to $h_4$), recovery of the full
three-integral DF is possible with Schwarzschild's method, for the
correct inclination and mass-to-light ratio.

In the triaxial case, the DF is again a function of three integrals of
motion, but the orbital structure in these models is substantially
richer than in the oblate axisymmetric models, with four major orbit
families, instead of only one.  This introduces a fundamental
non-uniqueness in the recovery of the DF.  Whereas in the oblate
axisymmetric case $\rho(R,z)$ uniquely defines the even part of
$f(E,L_z)$, in the (separable) triaxial case the density $\rho(x,y,z)$
does \emph{not} uniquely determine the even part of $f(E,I_2,I_3)$,
although both of these are functions of three variables (Hunter \& de
Zeeuw 1992\nocite{1992ApJ...389...79H}). It is (yet) unknown how much
specification of $\mathcal{L}(x',y',v_{z'})$ can narrow down the range
of possible DFs further, even ignoring the non-uniqueness caused by
the required deprojection of the surface brightness. Our results show
that Schwarzschild's method recovers the correct orbital mass weight
distribution associated to $f(E,I_2,I_3)$. Given the very large
freedom in the orbit choice for this case, the modest resolution of
our orbit library, and the resulting approximations in the evaluation
of the phase space volume, the agreement between the orbital mass
weights found in \S~\ref{sec:intdftriax} is in fact remarkable. It may
be possible to improve the DF recovery further by refining the
sampling of the orbits and the regularisation of the orbital mass
weights.

Our analysis shows that it is clear that Abel models are useful for
testing orbit-based modelling methods such as Schwarzschild's method.
In particular the oblate limiting case with a Kuzmin-Kutuzov potential
(\S~\ref{sec:recoveryaxi}) provides a new and convenient test for
existing axisymmetric Schwarzschild codes. Furthermore, because Abel
models with a few DF components can already provide quite a good
representation of observed early-type galaxies, they can be used as a
way to (numerically) build three-integral dynamical models of these
galaxies (see e.g. MD96 for an application to Centaurus~A).

We conclude that Schwarzschild's method is able to recover the
internal dynamical structure of realistic models of early-type
galaxies. We show in vdB07 that Schwarzschild's method also allows for
an accurate determination of the mass-to-light ratio and provides
significant constraints on the viewing direction and intrinsic shape.
The axisymmetric Schwarzschild method has already been successfully
applied by us and other groups to determine the black hole mass,
mass-to-light ratio (profile), dark matter profile as well as the
(three-integral) DF of early-type galaxies.  With our extension to
triaxial geometry, described in detail in vdB07, we are now able to
model early-type galaxies --- in particular the giant ellipticals ---
which show clear signatures of non-axisymmetry, including isophote
twist, kinematic misalignment and kinematic decoupled components.
Moreover, since triaxial galaxies may appear axisymmetric (or even
spherical) in projection, we can investigate the effect of intrinsic
triaxiality on the measurements of e.g.\ black hole masses based on
axisymmetric model fits to observations of galaxies. Work along these
lines is in progress.

%=====================================================================
% ACKNOWLEDGMENTS
%=====================================================================

\section*{acknowledgements}
\label{sec:acknowledgments}

We sincerely thank the referee for constructive comments and
suggestions that improved the presentation, and Michele Cappellari and
Anne-Marie Weijmans for a careful reading of an earlier version of the
manuscript. GvdV acknowledges support provided by NASA through grant
NNG04GL47G and through Hubble Fellowship grant HST-HF-01202.01-A
awarded by the Space Telescope Science Institute, which is operated by
the Association of Universities for Research in Astronomy, Inc., for
NASA, under contract NAS 5-26555. RvdB acknowledges support by the
Netherlands Organization for Scientific Research (NWO) through grant
614.000.301.

%=====================================================================
% REFERENCES
%=====================================================================

%\bibliographystyle{mn2e}
%\bibliography{vdven_abel}

%=====================================================================
% APPENDICES
%=====================================================================

\appendix

%============================= appendix A ============================
\section{Limiting cases}
\label{sec:limitingcases}
%=====================================================================

When two or all three of the constants $\alpha$, $\beta$ or $\gamma$
that define the confocal ellipsoidal coordinate system are equal, the
triaxial Abel models reduce to limiting cases with more symmetry and
thus with fewer degrees of freedom. The oblate and prolate
axisymmetric limits are described in \S~\ref{sec:axivelmom}. DL91
derived the velocity moments for the non-rotating Abel models for
elliptic discs and in the spherical limit. We summarise their results
and give the rotating Abel models as well as the expressions for the
LOSVD for these limiting cases. At the same time, we also derive the
properties of the non-rotating and rotating Abel models in the limit
of large radii.

%---------------------------------------------------------------------
\subsection{Elliptic disc potential}
\label{sec:limelldisc}
%---------------------------------------------------------------------

The two-dimensional analogues of the triaxial Abel models are the
elliptic Abel discs with St\"ackel potential
$V_S(\lambda,\mu)=U[\lambda,\mu]$ in confocal elliptic coordinates
$(\lambda,\mu)$. The relations with $(x,y)$ follow from those in
\S~\ref{sec:triaxpot} by setting $z=0$ or equivalently $\nu=-\gamma$.
The two integrals of motion $E$ and $I_2$ are given by
\begin{eqnarray}
  \label{eq:discEI2}
  E   & = & \textstyle{\frac12} \left( v_x^2 + v_y^2 \right)
  + U[\lambda,\mu],
  \nonumber\\[-5pt]\\[-5pt]\nonumber
  I_2 & = & \textstyle{\frac12} L_z^2
  + \textstyle{\frac12} (\alpha-\beta) v_x^2
  + (\alpha-\beta) x^2 U[\lambda,\mu,-\alpha].
\end{eqnarray}
%

%---------------------------------------------------------------------

\subsubsection{Velocity moments}
\label{sec:disc_velmom}

Choosing the DF as $f(E,I_2) = f(S)$, with $S = -E+w\,I_2$, the
velocity moments can be evaluated as
\begin{multline}
  \label{eq:discmulm}
  \mu_{lm}(\lambda,\mu) =
  \sqrt{\frac{2^{l+m+2}}{h_\mu^{l+1} h_\lambda^{m+1}}}
  \\ \times
   \int\limits_{S_\mathrm{min}}^{S_\mathrm{max}} T_{lm} \, \left[
    S_\mathrm{top}(\lambda,\mu) - S \right]^{(l+m)/2}
    f(S) \, \du S,
\end{multline}
with the terms $h_\mu$ and $h_\lambda$ defined as
\begin{equation}
  \label{eq:discdefhtau}
  h_\tau = 1 - (\tau+\alpha)\,w,
  \qquad \tau=\lambda,\mu.
\end{equation}
As in the general triaxial case, $S_\mathrm{min} \ge S_\mathrm{lim}$,
where the expression of the latter is given along the $w$-axis ($u=0$)
in Fig.~\ref{fig:Slim}. The accessible part of the $(E,I_2)$-integral
space is now a triangle, the top of which is
$S_\mathrm{top}(\lambda,\mu)= -U[\lambda,\mu] + w (\lambda+\alpha)
(\mu+\alpha) U[\lambda,\mu,-\alpha]$.

For the NR components we have that
$S_\mathrm{max}=S_\mathrm{top}(\lambda,\mu)$ and $T^\mathrm{NR}_{lm} =
B(\frac{l+1}{2},\frac{m+1}{2})$. Of the two possible orbit families,
the box orbits have no net rotation and the tube orbits rotate around
the axis perpendicular to the disc (the $z$-axis). Since this is
similar to the short-axis tube orbits in the general triaxial case, we
refer to the rotating type as the SR type.  This SR type reaches the
region of the accessible integral space (the triangle) for which
$v_\mu^2\ge0$ at $\mu=-\alpha$ (or $I_2\ge0$).  Therefore,
$S_\mathrm{max}= S_\mathrm{top}(\lambda,-\alpha)$ and
\begin{equation}
  \label{eq:conTlm_SR}
  T^\mathrm{SR}_{lm} =
  2\int_0^{\arcsin(\sqrt{a_2})} \hspace{-10pt}
  \sin^l\theta \cos^m\theta \, d\theta,
\end{equation}
where $a_2$ is defined as
\begin{equation}
  \label{eq:disca2}
  a_2 =
  \frac{(\lambda+\alpha)\,h_\mu\,
    \left[S_\mathrm{top}(\lambda,-\alpha)-S\right]}
  {(\lambda-\mu)\,h_{(-\alpha)}\,
    \left[S_\mathrm{top}(\lambda,\mu)-S\right]}.
\end{equation}
The integral \eqref{eq:conTlm_SR} can be evaluated in terms of
elementary functions (e.g., Gradshteyn \& Ryzhik
1994\nocite{GR..tables}, relations 2.513 on p.160--162). The NR
velocity moments $\mu^\mathrm{NR}_{lm}(\lambda,\mu)$ vanish when
either $l$ or $m$ is odd, and the SR velocity moments
$\mu^\mathrm{SR}_{lm}(\lambda,\mu)$ vanish when $l$ is odd. The latter
should be multiplied with $(-1)^m$ for net rotation in the opposite
direction.

The matrix $\mathbf{Q}$, which transforms the velocity components
$(v_\lambda,v_\mu,v_\nu)$ to $(v_x,v_y,v_z)$, is that for the prolate
case given in eq.~\eqref{eq:pro_convN}, but with $\chi=0$ substituted.
The sign matrix $\mathbf{S}$, projection matrix $\mathbf{P}$ and
rotation matrix $\mathbf{R}$ are the same as for the triaxial case
given in respectively eqs \eqref{eq:signmatS}--\eqref{eq:rotmatR}.
The polar angle is the inclination, $\vartheta=i$, and the azimuthal
angle $\varphi$ the orientation of the infinitesimally thin disc
($\gamma=0$) in the plane $z=0$. In the expression
\eqref{eq:misalignment_psi} for the misalignment angle $\psi$, the
triaxiality parameter thus reduces to $T=1-\beta/\alpha$, with
$0<\beta<\alpha$, bracketing the limiting cases of a needle and a
circular disc. 

%---------------------------------------------------------------------

\subsubsection{Line-of-sight velocity distribution}
\label{sec:disc_losvd}

Starting with the expression for the stellar (surface) mass density
$\Sigma_\star(x',y')=\mu_{00}(\lambda,\mu)$ from
eq.~\eqref{eq:discmulm}, we derive the Abel LOSVD for the elliptic
disc in a similar way as in \S~\ref{sec:triaxabelLOSVD} for the
triaxial case\footnote{Alternatively, one can invert the relations
  $S=S_\mathrm{top}(\lambda,\mu) - \frac12 ( h_\mu v_\lambda^2 +
  h_\lambda v_\mu^2 )$ and $v_{z'} = M_{31} \, v_\lambda + M_{32} \,
  v_\mu$ to find the Jacobian to transform from the coordinates
  $(v_\lambda,v_\mu)$ to $(S,v_{z'})$. Leaving out the integral over
  $v_{z'}$ yields the same expression for the LOSVD as in
  eq.~\eqref{eq:discAbelLOSVD}.}. The cross section of the unit sphere
with a plane, reduces to the cross section of the unit circle
$X^2+Y^2=1$ with a line $e_1 X + e_2 Y = v_{z'}/g(S)$. Here, the
variable $X$ is defined as
\begin{equation}
  \label{eq:discdefXY}
  X = \frac{h\,v_\lambda}{g(S)\sqrt{h_\lambda}}, 
  \quad \mathrm{with} \quad
  g(S) = h \sqrt{\frac{2[S_\mathrm{top}(\lambda,\mu)-S]}{h_\lambda h_\mu}},
\end{equation}
and the variable $Y$ follows by interchanging $\lambda \leftrightarrow
\mu$. The coefficients of the line are $e_1 = \sqrt{h_\lambda}
M_{31}/h$ and $e_2 = \sqrt{h_\mu} M_{32}/h$. We can write the
normalisation $h^2 = h_\lambda M_{31}^2 + h_\mu M_{32}^2$, explicitly
as
\begin{multline}
  \label{eq:discexprh}
  h = \sin i \left\{ 
    [1 - (\lambda+\alpha)\,w] \, (C\cos\varphi+D\sin\varphi)^2 
    \right. \\ \left.
    + [1 - (\mu+\alpha)\,w] \, (C\sin\varphi-D\cos\varphi)^2 
  \right\}^\frac12,
\end{multline}
with $C$ and $D$ defined in eq.~\eqref{eq:pro_defCD}. In this way we
can write $\Sigma_\star$ as a double integral over $S$ and $v_{z'}$,
so that at a given line-of-sight velocity the LOSVD becomes
\begin{multline}
  \label{eq:discAbelLOSVD}
  \mathcal{L}(x',y',v_{z'}) 
  =
  \frac1h 
  \int\limits_{S_\mathrm{min}}^{S_\mathrm{up}(v_{z'})} \!\!
  \frac{f(S)}{\sqrt{2\left[G(v_{z'})-S\right]}} 
  \, \Delta(v_{z'},S) \, \du S,
  \\
  = \frac{\sqrt{2\left[G(v_{z'})-S_\mathrm{min}\right]}}{h}
  \int\limits_0^{\eta_\mathrm{up}} \!\!
  f(S) \, \Delta(v_{z'},S) \, \sin\eta \, \du \eta,
\end{multline}
which vanishes when $|v_{z'}|$ exceeds the 'terminal velocity' $v_t =
g(S_\mathrm{min})$. The second expression removes the possible
singularity at the upper limit of $S$, given by $S_\mathrm{up} =
\min[G(v_{z'}),S_\mathrm{max}]$, with
\begin{equation}
  \label{eq:discdefGvzp}
  G(v_{z'}) = S_\mathrm{top}(\lambda,\mu)
  - h_\lambda h_\mu v_{z'}^2/(2h^2).
\end{equation}
Hence, $\eta_\mathrm{up}$ is given by $\sin^2\eta_\mathrm{up} =
[S_\mathrm{up}-S_\mathrm{min}]/[G(v_{z'})-S_\mathrm{min}]$.  Depending
on the integral space accessible by the orbits, the value of
$\Delta(v_{z'},S)$ is either zero, one or two.

For the NR components, $S_\mathrm{max} = S_\mathrm{top}(\lambda,\mu)$,
and, since the full integral space is accessible, $\Delta_\mathrm{NR}
= 2$, independent of $S$ and $v_{z'}$. In the case of a basis
function $f_\delta(S)$ as defined in eq.~\eqref{eq:simpleDF}, the
integral over $S$ can be evaluated explicitly resulting in
\begin{equation}
  \label{eq:discLOSVD_NRsimpleDF}
  \mathcal{L}_\delta^\mathrm{NR} =
  \frac{4^{\delta+1}B(\delta+1,\delta+1)}
  {\sqrt{2}\,(1-S_\mathrm{min})^\delta} \; \frac1h
  \left[ G(v_{z'}) - S_\mathrm{min} \right]^{\delta+\frac12}.
\end{equation}
For SR components with maximum streaming, both $v_\mu^2\ge0$ at
$\mu=-\alpha$ and $v_\mu \ge 0$, which is equivalently to $0 \le Y \le
\sqrt{a_2}$, where $a_2$ is defined in eq.~\eqref{eq:disca2}. The
intersection of the above unit circle and line provides the following
two solutions
\begin{equation}
  \label{eq:disctwosolY}
  Y_\pm = e_2 \, [v_{z'}/g(S)] \pm e_1 \sqrt{1-[v_{z'}/g(S)]^2}.
\end{equation}
Given the values of $v_{z'}$ and $S$, $\Delta_\mathrm{SR}$ is thus
equal to 0, 1 or 2 if for respectively none, one or both of the
solutions $0 \le Y_\pm \le \sqrt{a_2}$.

The expression for $h$ in eq.~\eqref{eq:discexprh} shows that the
LOSVD in eq.~\eqref{eq:discAbelLOSVD} is inversely proportional to
$\sin i$. For face-on viewing at inclination $i=0$\dgr\, the LOSVD
reduces to $\Sigma_\star(x',y')\delta(v_{z'})$. Because the velocity
perpendicular to the disc $v_z=0$, the face-on LOSVD is zero at all
line-of-sight velocities, except at $v_{z'}=0$ when it equals the
surface mass density. For edge-on viewing at inclination $i=90$\dgr,
the LOSVD follows upon substituting $y'=0$ in
eq.~\eqref{eq:discAbelLOSVD} and integrating over the line-of-sight
$z'$. For $i<90$\dgr, the latter integration is not needed, since at
each position $(x',y')$ on the plane of the sky there is only a single
(unique) point along $z'$ where it intersects the infinitesimally thin
disc. The edge-on LOSVD and also $\Sigma_\star$ are thus spatially
only one-dimensional functions of $x'$, and vanish for non-zero
$y'$-values.

Further information on elliptic St\"ackel discs can be found in Teuben
(1987\nocite{1987MNRAS.227..815T}), de Zeeuw, Hunter \& Schwarzschild
(1987\nocite{1987ApJ...317..607D}), and Evans \& de Zeeuw
(1992\nocite{1992MNRAS.257..152E}).

%---------------------------------------------------------------------
\subsection{Large distance limit}
\label{sec:limlargedist}
%---------------------------------------------------------------------

At large radii, $\lambda \to r^2 \gg -\alpha$, so that the confocal
ellipsoidal coordinates of \S~\ref{sec:triaxpot} reduce to conical
coordinates $(r,\mu,\nu)$, with $r$ the usual distance to the origin,
i.e., $r^2=x^2+y^2+z^2$, and $\mu$ and $\nu$ angular coordinates on the
sphere. In these coordinates the St\"ackel potential is of the form
$V_S(r,\mu,\nu) = V(r) + U[\mu,\nu]/r^2$, where $V(r)$ is an arbitrary
smooth function of $r$. The corresponding integrals of motion are
given by
\begin{eqnarray}
  \label{eq:conEI2I3}
  E   & = & \textstyle{\frac12} \left( v_x^2 + v_y^2 + v_z^2 \right)
  + V_S(r,\mu,\nu),
  \nonumber \\
  I_2 & = & \textstyle{\frac12} T L_y^2
  + \textstyle{\frac12} L_z^2
  + (\alpha-\beta) \frac{x^2}{r^2} U[\mu,\nu,-\alpha],
  \\ \nonumber
  I_3 & = & \textstyle{\frac12} L_x^2
  + \textstyle{\frac12} (1-T) L_y^2
  + (\gamma-\beta) \frac{z^2}{r^2} U[\mu,\nu,-\gamma].
\end{eqnarray}
With the choice \eqref{eq:dfabel} for the DF, the expression for the
velocity moments becomes
\begin{multline}
  \label{eq:con_mulmn}
  \mu_{lmn}(r,\mu,\nu) = \frac{1}{r^{m+n+2}} \,
  \sqrt{ \frac{2^{l+m+n+3}}
    {F_{\nu}^{m+1} F_{\mu}^{n+1}}}
  \\ \times
  \int\limits_{S_\mathrm{min}}^{S_\mathrm{max}} T_{lmn} \, \left[
  S_\mathrm{top}(r,\mu,\nu) - S \right]^{(l+m+n+1)/2}
  f(S) \, \du S,
\end{multline}
where $F_\nu$ and $F_\mu$ are defined as
\begin{equation}
  \label{eq:conFtau}
  F_\tau = \frac{1}{r^2} +
  \frac{(\tau+\alpha)\,w-(\tau+\gamma)\,u}{\gamma-\alpha},
  \quad \tau=\mu,\nu.
\end{equation}
As in the general triaxial case, $S_\mathrm{min} \ge S_\mathrm{lim}$,
where $S_\mathrm{lim}$ can be obtained from Fig.~\ref{fig:Slim}. The
expressions of $S_\mathrm{max}$ and $T_{lmn}$ for the NR, LR and SR
types are those given in \S\S~\ref{sec:triaxNR}--\ref{sec:triaxSR}
respectively, but with $S_\mathrm{top}(\lambda,\mu,\nu)$
(eq.~\ref{eq:S_top}) replaced by
\begin{eqnarray}
  \label{eq:con_S_top}
  S_\mathrm{top}(r,\mu,\nu)
  &=&
  - V_S(r,\mu,\nu) \nonumber \\
  && - w\,\frac{(\mu\!+\!\alpha)(\nu\!+\!\alpha)}{\gamma\!-\!\alpha}
  \; U[\mu,\nu,-\alpha] \nonumber \\
  && - u\,\frac{(\mu\!+\!\gamma)(\nu\!+\!\gamma)}{\alpha\!-\!\gamma}
  \; U[\mu,\nu,-\gamma],
\end{eqnarray}
and the parameters $a_0$ and $b_0$ (\ref{eq:defa0b0}) reduce to
\begin{eqnarray}
  \label{eq:con_defa0b0}
  a_0 & = & \frac{S_\mathrm{top}(r,\mu,-\beta) - S}
  {S_\mathrm{top}(r,\mu,\nu) - S},
  \nonumber\\[-5pt]\\[-5pt]\nonumber
  b_0 & = & \frac{(\mu+\beta)\,F_{\nu}\,
    \left[ S_\mathrm{top}(r,\mu,-\beta) - S \right]}
  {(\mu-\nu)\,F_{(-\beta)}\,
    \left[ S_\mathrm{top}(r,\mu,\nu) - S \right]},
\end{eqnarray}
which by interchanging $\nu \leftrightarrow \mu$ become $a_1$ and
$b_1$, and in turn $a_2$ and $b_2$ follow by $\beta \leftrightarrow
\alpha$. 

In the conversion to observables described in
\S~\ref{sec:observables}, in the matrix $\mathbf{Q}$, which transforms
the velocity components $(v_r,v_\mu,v_\nu)$ to $(v_x,v_y,v_z)$, all
terms $\lambda+\sigma$ ($\sigma=-\alpha,-\beta,-\gamma,\mu,\nu$)
cancel out (cf.\ eq.~25 of Statler 1994\nocite{1994ApJ...425..458S}).
The expression for the LOSVD follows from that of the triaxial case in
eq.~\eqref{eq:triaxAbelLOSVD} by substituting $H_{\mu\nu}=1$,
$H_{\nu\lambda}=r^2 F_\nu$, $F_{\lambda\mu}=r^2 F_\mu$ and
$S_\mathrm{top}(\lambda,\mu,\nu) = S_\mathrm{top}(r,\mu,\nu)$.

Suppose now that at large radii $r$, the function $V(r)$ in the
St\"ackel potential vanishes and we keep in the above expressions only
the dominant terms. In this case, $F_\mu$, $F_\nu$ and
$S_\mathrm{top}$ reduce to functions of $\mu$ and $\nu$ only. As a
result, the velocity moments \eqref{eq:con_mulmn} are independent of
$r$, except for the prefactor $1/r^{m+n+2}$, and therefore are
scale-free. Once we have calculated the velocity moments at a radius
$r$, those at radius $r'=qr$, with $q$ a constant, follow by a simple
scaling, $\mu_{lmn}(r',\mu,\nu)=\mu_{lmn}(r,\mu,\nu)/q^{m+n+2}$. The
same holds true for the line-of-sight velocity moments
$\mu_k(r,\mu,\nu)$, but not for the LOSVD.

%---------------------------------------------------------------------
\subsection{Spherical potential}
\label{sec:limspherical}
%---------------------------------------------------------------------

When $\alpha=\beta=\gamma$, both $\mu$ and $\nu$ loose their
meaning and we replace them by the customary polar angle $\theta$
and azimuthal angle $\phi$. The expressions for the Abel models in
these spherical coordinates $(r,\theta,\phi)$ follow in a
straightforward way from those in \S~\ref{sec:limlargedist} for
the large distance limit in conical coordinates $(r,\mu,\nu)$.

The St\"ackel potential $V_S=V(r)$ is spherically symmetric. The
expressions for the integrals of motion follow from
(\ref{eq:conEI2I3}), where for $I_2$ and $I_3$ the right-most terms
vanish. The triaxiality parameter $T$ is now a free parameter, so
that, together with the parameters $w$ and $u$, we can rewrite
$S=-E+w\,I_2+u\,I_3$ as
\begin{equation}
  \label{eq:sphS}
  S = -E + \textstyle{\frac12} u L_x^2 +
  \textstyle{\frac12} [(1-T) u + T w] L_y^2 +
  \textstyle{\frac12} w L_z^2.
\end{equation}
This means that with the choice \eqref{eq:dfabel} for the DF, we
cover the most general homogeneous quadratic form in the
velocities that is allowed by the integrals of motion in a
spherical symmetric potential, i.e., the energy $E$ and all three
components of the angular momentum vector $\mathbf{L}$ (cf.\
DL91).  These include the models considered by Osipkov
(1979\nocite{1979PAZh....5...77O}) and Merritt
(1985\nocite{1985AJ.....90.1027M}) with the DF of the from
$f(-E \pm L^2/r_a^2)$ and those studied by Arnold
(1990\nocite{1990MNRAS.244..465A}) with a more general DF of
the form $f(-E \pm L^2/r_a^2 \pm L_z^2/r_b^2)$. These models follow
by setting $u=w=\pm2/r_a^2$, and by taking $u=\pm2/r_a^2$,
$w=u\pm2/r_b^2$ and $T=0$, respectively.

%---------------------------------------------------------------------

\subsubsection{Velocity moments}
\label{sec:sph_velmom}

The velocity moments follow from eq.~\eqref{eq:con_mulmn}, with
\begin{multline}
  \label{eq:sphFtau}
  F_\tau = \frac{1}{r^2} - \textstyle{\frac12}(w+u) \\
  + \textstyle{\frac12}(w-u)\left[\cos^2\theta +
    T\,(\sin^2\theta\sin^2\phi-1) \pm \sqrt{\Lambda} \right],
\end{multline}
where the positive and negative sign are for $F_\mu$ and $F_\nu$, and
\begin{multline}
  \label{eq:sphdefLambda}
  \Lambda = \left[ \sin^2\!\theta
    + T\,(\sin^2\!\theta\sin^2\!\phi-1) \right]^2 \\
  + 4T\,\sin^2\theta\cos^2\theta\sin^2\phi.
\end{multline}
Taking $\alpha=\beta=\gamma$ in Fig.~\ref{fig:Slim}, we see that the
boundaries on $w$ and $u$ both vanish. The separatrices $L_1$ and
$L_2$, defined in eq.~\eqref{eq:sepL1andL2}, reduce to the negative
$w$-axis and the line $w=u$, respectively. Furthermore,
$S_\mathrm{max} = S_\mathrm{top} = -V(r)$, and for $T_{lmn}$ we use
the expression \eqref{eq:Tlmn_NR}.  The resulting velocity moments
$\mu_{lmn}(r,\theta,\phi)$, which are in general \emph{not}
spherically symmetric, vanish when either $l$, $m$ or $n$ is odd.

The latter implies no net rotation, which is the case when the
(conserved) angular momentum vectors $\mathbf{L}$ for the orbits are
randomly oriented. We may introduce net rotation by assuming that (a
fraction of) the orbits have a preferred sense of rotation around an
angular momentum vector $\mathbf{L}_0$ that points in a specific
direction given by $\theta_0$ and $\phi_0$. Using the projection
matrix $P$ in eq.~\eqref{eq:projmatP} with $\vartheta=\theta_0$ and
$\varphi=\phi_0$, we transform to the coordinate system
$(r'=r,\theta',\phi')$, in which $\mathbf{L}_0$ is aligned with the
$z'$-axis. If we next set the DF to zero for $L_{z'}<0$, we find
$\mu'_{lmn}(r,\theta',\phi') =
\textstyle{\frac12}\mu_{lmn}(r,\theta',\phi')$, which does still
vanish when $l$ or $m$ is odd, but is non-zero when $n$ is odd,
resulting in maximum streaming around the $z'$-axis, and
multiplication with $(-1)^n$ for opposite direction of rotation. With
the inverse of the projection matrix, we can then transform these
velocity moments to the original coordinates system $(r,\theta,\phi)$.
In this way, we can build spherical Abel models, which in addition to
a non-rotating part consist of a component or several components with
a preferred rotation axis. Mathieu, Dejonghe \& Hui
(1996\nocite{1996A&A...309...30M}) used this approach to construct a
dynamical model of Centaurus~A, with a spherical potential, but
triaxial luminosity density and DF components with rotation around the
apparent long and short axis.

From the customary definition of the spherical coordinate system,
$x=r\sin\theta\cos\phi$, $y=r\sin\theta\sin\phi$ and $z=r\cos\theta$,
it follows directly that the matrix $\mathbf{Q}$, which transforms the
velocity components $(v_r,v_\theta,v_\phi)$ to $(v_x,v_y,v_z)$, is
given by
\begin{equation}
  \label{eq:sph_convN}
  \mathbf{Q}
  =
  \begin{pmatrix}
    \sin\theta\cos\phi &  \cos\theta\cos\phi & -\sin\phi \\
    \sin\theta\sin\phi &  \cos\theta\sin\phi &  \cos\phi \\
    \cos\theta         & -\sin\theta         &  0
  \end{pmatrix}.
\end{equation}
In case the orbits have no preferred sense of rotation, we may set the
viewing angles $\vartheta=\varphi=0$ without loss of generality, so
that with $\psi=0$ from eq.~\eqref{eq:misalignment_psi},
$(x',y')=(y,-x)$ on the plane of the sky and $z'=z$ along the
line-of-sight, and similarly for the Cartesian velocity components.

%---------------------------------------------------------------------

\subsubsection{Line-of-sight velocity distribution}
\label{sec:sph_losvd}

There is no obvious further simplification of the LOSVD for rotating
components. For the NR components, the LOSVD follows from
eq.~\eqref{eq:triaxAbelLOSVD} with
$S_\mathrm{max}=S_\mathrm{top}=-V(r)$ and $\Delta\xi'=2\pi$, or from
eq.~\eqref{eq:triaxLOSVD_NRsimpleDf} after substituting the basis
function $f_\delta(S)$ from eq.~\eqref{eq:simpleDF}. Since the
line-of-sight velocity $v_{z'}= \mathrm{sgn}(z) [ \cos\theta\,v_r -
\sin\theta\,v_\theta ]$, it follows that $M_{31}^2=\cos^2\theta$,
$M_{32}=\sin^2\theta$ and $M_{33}^2=0$ in eq~\eqref{eq:defnormh} for
$h$. Moreover, $H_{\mu\nu}=1$ and $H_{\tau\lambda}=r^2 F_\tau$
($\tau=\mu,\nu$), with $F_\tau$ given in eq.~\eqref{eq:sphFtau}.

Carollo et al.\ (1995\nocite{1995MNRAS.276.1131C}) compute the LOSVD
for Osipkov-Merritt models with $f(-E - L^2/(2r_a^2))$, which is
a special case of the Abel DF $f(S)$, that follows from
eq.~\eqref{eq:sphS} by setting $u=w=-1/r_a^2$. Substituting the latter
in eq.~\eqref{eq:sphFtau}, we find that
$H_{\lambda\mu}=H_{\lambda\nu}=(r_a^2+r^2)/r_a^2$, so that 
%the expressions \eqref{eq:defnormh} and \eqref{eq:triaxdefGvzp} for
%$h$ and $G(v_{z'})$ reduce to
%
\begin{eqnarray}
  \label{eq:sphnormh_OM}
  h^2 & = & (r_a^2+r^2) (r_a^2+r^2 - R'^2) / r_a^4, \\
  \label{eq:sphGvzp_OM}
  G(v_{z'}) & = & -V(r) - \frac{r_a^2+r^2}{r_a^2+r^2 - R'^2} \frac{v_{z'}^2}{2},
\end{eqnarray}
with radius $R'=r\sin\theta$ on the plane of the sky. After
substitution in eq.~\eqref{eq:triaxAbelLOSVD}, and transforming the
integral over $\du z'$ to $\du r$, we find the following LOSVD
\begin{equation}
  \label{eq:sphLOSVD_OM}
  \mathcal{L}(R',v_{z'}) =
  4\pi \!\!
  \int\limits_{R'}^{\infty} \!\!
  \frac{r}{h\sqrt{r^2-R'^2}} \!\!
  \int\limits_{S_\mathrm{min}}^{G(v_{z'})} \!\!
  f(S) \, \du S \, \du r.
\end{equation}
This is the same as the (unnormalised) velocity profile in eq.~(27) of
Carollo et al.\ (1995\nocite{1995MNRAS.276.1131C}), with
$\Phi_\infty$, the lower limit of their (relative) potential
$\Phi(r)=-V(r)$, equal to $S_\mathrm{min}$, which from
Fig.~\ref{fig:Slim} in this case has $S_\mathrm{lim}=0$ as lower
limit. Their function $g(r,R')$ and upper limit $Q_\mathrm{max}$ in
eqs (25) and (26), are equivalent to respectively the inverse of $h$
and $G(v_{z'})$ in
eqs \eqref{eq:sphnormh_OM} and \eqref{eq:sphGvzp_OM} above. The
well-known isotropic case follows upon taken the limit $r_a\to\infty$,
so that $f(S) \to f(-E)$, $h \to 1$ and $G(v_{z'}) \to -V(r) -
v_{z'}^2/2$.

%============================= appendix B ============================
\section{The function $\mathcal{M}$}
\label{sec:funcM}
%=====================================================================

The function $\mathcal{M}$ that appears in the velocity moments of the
rotating Abel components is defined as
\begin{equation}
  \label{eq:defM}
  \mathcal{M}(s,i,j;a,b,\phi) =
  \int\limits_0^\phi
  \left(\frac{\partial}{\partial a}\right)^i
  \left(\frac{\partial}{\partial b}\right)^j
  \frac{1\!-\!\sqrt{\left[1\!-\!p(\theta)\right]^{s+1}}}{p(\theta)}
  \, \du \theta,
\end{equation}
with $p(\theta) \equiv a\cos^2\theta + b\sin^2\theta$. For odd $s$,
corresponding to odd velocity moments, the integral can be evaluated
in a straightforward way in terms of elementary functions. In
Table~\ref{tab:Msodd}, we give the resulting expressions for
$s=1,3,5$.

%%%TAB
\begin{table}
\begin{center}
  \caption{The function $\mathcal{M}$ for odd $s$.}
  \begin{tabular}{ll}
    \hline
    \hline
    $s\,i\,j$ & $\mathcal{M}(s,i,j;a,b,\phi)$ \\
    \hline
    100 & $\phi$ \\
    300 & $\frac12(4-a-b)\,\phi + \frac14(b-a)\sin2\phi$ \\
    310 & $-\frac12\phi - \frac14\sin2\phi$ \\
    301 & $-\frac12\phi + \frac14\sin2\phi$ \\
    500 & $\frac18(24-12a-12b+3a^2+3b^2+2ab)\,\phi$ \\
        & $+ \frac14(b-a)(3-a-b)\sin2\phi + \frac{1}{32}(b-a)^2\sin4\phi$ \\
    510 & $-\frac14(6-3a-b)\,\phi - \frac14(3-2a)\sin2\phi - \frac{1}{16}(b-a)\sin4\phi$ \\
    501 & $-\frac14(6-3b-a)\,\phi + \frac14(3-2b)\sin2\phi + \frac{1}{16}(b-a)\sin4\phi$ \\
    520 & $\frac34\phi + \frac12\sin2\phi + \frac{1}{16}\sin4\phi$ \\
    511 & $\frac14\phi - \frac{1}{16}\sin4\phi$ \\
    502 & $\frac34\phi - \frac12\sin2\phi + \frac{1}{16}\sin4\phi$ \\[+5pt]
    \hline
  \end{tabular}
  \label{tab:Msodd}
\end{center}
\end{table}
%%%TAB

%%%TAB
\begin{table*}
\begin{center}
  \caption{The function $\mathcal{M}$ for even $s$.}
  \begin{tabular}{ll}
    \hline
    \hline
    $s\,i\,j$ & $\mathcal{M}(s,i,j;a,b,\phi)$ \\
    \hline
    000 & $A - F + J$ \\
    200 & $A - (1-a)\,F - (b-a)\,D + J$ \\
    210 & $-\frac{1}{2a}\left[ A + Q - (1+a)\,F + (1-a)\,D + J \right]$ \\
    201 & $-\frac{1}{2b}\left[ A - Q - F - (1-b)\,D + J \right]$ \\
    400 & $A + \frac13(b-a)\,P - \frac13(2a^2+ab-6a+3)\,F + \frac13(2a+2b-7)(b-a)\,D + J$ \\
    410 & $-\frac{1}{2a}\left[ A + a\,P + Q - (1+2a)(1-a)\,F + (2a^2-2a-ab+1)\,D + J \right]$ \\
    401 & $-\frac{1}{2b}\left[ A - b\,P - Q - (1-ab)\,F - (2b^2-2b-ab+1)\,D + J \right]$ \\
    420 & $\frac{3}{4a^2}\left\{ A +
    \frac{a^2\,p(\phi)-ab}{3(b-a)p(\phi)}\,P +
    \frac{5a\cos^2\phi+3b\sin^2\phi}{3p(\phi)}\,Q +
    \frac{2a^3-3a^2b+4a^2+3a-3ab-3b}{3(b-a)}\,F -
    \frac{(2a^2+5a-4ab-3b)(1-a)}{3(b-a)}\,D + J \right\}$ \\
    411 & $\frac{1}{4ab}\left\{ A
      + \frac{ab-ab\,p(\phi)}{(b-a)p(\phi)}\,P
      + \frac{b\sin^2\phi-a\cos^2\phi}{p(\phi)}\,Q %\right.\\&\left.
      + \frac{a^2b-ab+a-b}{b-a}\,F
      + \frac{a^2b+ab^2-4ab+a+b}{b-a}\,D
      + J \right\}$ \\
    402 & $\frac{3}{4b^2}\left\{ A
      + \frac{b^2\,p(\phi)-ab}{3(b-a)p(\phi)}\,P
      - \frac{3a\cos^2\phi+5b\sin^2\phi}{3p(\phi)}\,Q
      - \frac{3b-3a-ab+ab^2}{3(b-a)}\,F
      - \frac{(2b^2+5b-4ab-3a)(1-b)}{3(b-a)}\,D + J \right\}$ \\[+5pt]
    \hline
  \end{tabular}
  \label{tab:Mseven}
\end{center}
\end{table*}
%%%TAB

\begin{figure*}
  \begin{center}
    \includegraphics[width=1.0\textwidth]{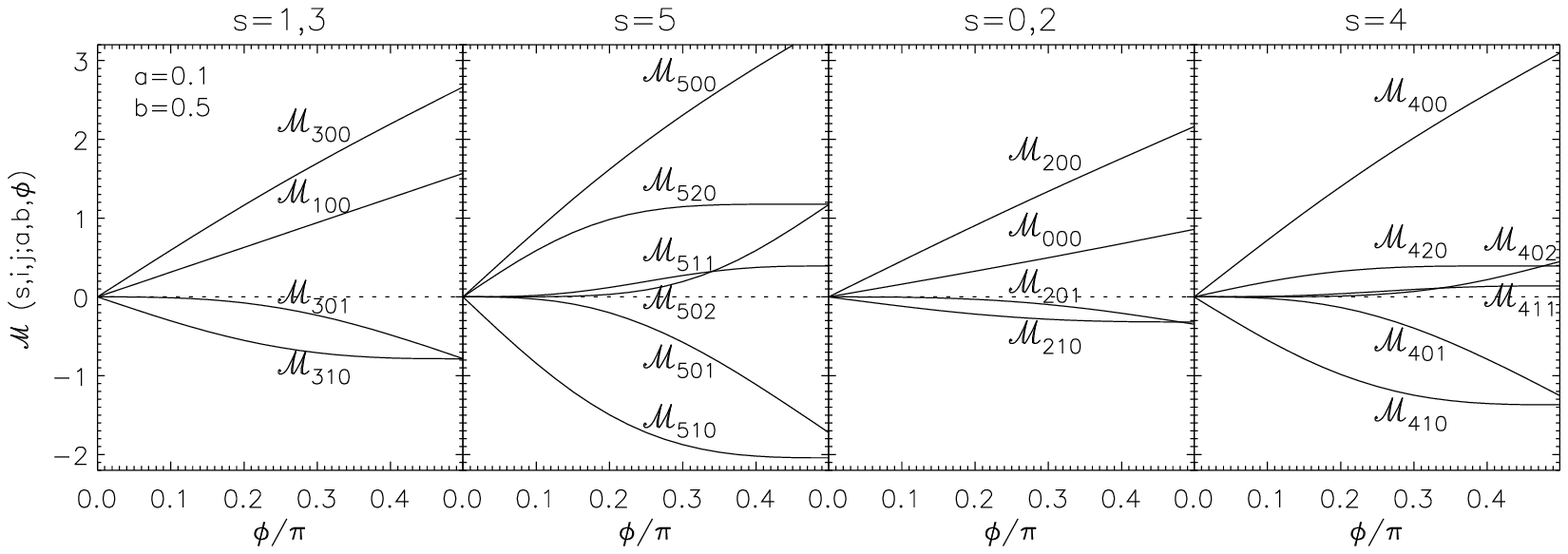}
  \end{center}
  \caption{The function $\mathcal{M}(s,i,j;a,b,\phi)$ defined in
    eq.~\eqref{eq:defM} plotted against $\phi$, for $a=0.1$ and
    $b=0.5$, up to order $s=5$. The curves in the left two panels are
    for odd values of $s$ corresponding to the odd velocity moments,
    whereas the curves in the right two panels are for even values of
    $s$. The indices of the labels $\mathcal{M}_{sij}$ refer to the
    first three parameters of the function $\mathcal{M}$.}
  \label{fig:funcM}
\end{figure*}

For even $s$, the integral can be evaluated in terms of the
(incomplete) elliptic integrals. To simplify the numerical evaluation
we use Carlson's (1977\nocite{C77..carlsonfunc}) symmetrical
forms $R_F$, $R_D$ and $R_J$ (for the relations between both forms see
e.g.\ de Zeeuw \& Pfenniger 1988\nocite{1988MNRAS.235..949D}). In
Table~\ref{tab:Mseven}, we give the expressions for $s=0,2,4$, where
we have introduced the following quantities based on these symmetric
elliptic integrals
\begin{eqnarray}
  F & = & \frac{\sqrt{1-a}\,\sin\phi}{a} \,
  R_F(\cos^2\phi,\Delta^2,1),
  \nonumber \\
  \label{eq:defMcarlsonFDJ}
  D & = & \frac{\sin^3\phi}{3\sqrt{1-a}} \,
  R_D(\cos^2\phi,\Delta^2,1),
  \\ \nonumber
  J & = & \frac{(b-a)\sin^3\phi}{3a^2\sqrt{1-a}} \,
  R_J(\cos^2\phi,\Delta^2,1,\frac{p(\phi)}{a}),
\end{eqnarray}
with $\Delta^2=[1-p(\phi)]/(1-a)$, and we have defined the terms
\begin{eqnarray}
  A & = & \frac{1}{\sqrt{ab}}\,
  \arctan\left(\sqrt{\frac{b}{a}}\tan\phi\right),
  \nonumber \\
  \label{eq:defAPQ}
  P & = & \sin\phi\cos\phi\sqrt{1-p(\phi)},
  \\ \nonumber
  Q & = & \sin\phi\cos\phi\,\frac{1-\sqrt{1-p(\phi)}}{p(\phi)}.
\end{eqnarray}
In Fig.~\ref{fig:funcM}, we show the $\mathcal{M}(s,i,j;a,b,\phi)$ as
function of $\phi$ for the case that $a=0.1$ and $b=0.5$, up to order $s=5$.

%%%FIG
%%%FIG

We now consider some special cases. When either $a$ or $b$ is zero,
the corresponding velocity moments vanish
(eqs \ref{eq:Tlmn_LR} and \ref{eq:Tlmn_SR}), and when $a_i > b_i$ the
arguments of the function $\mathcal{M}$ are interchanged
(eqs \ref{eq:C_LR}, \ref{eq:C_SR_I} and \ref{eq:C_SR_II}). This means
we only have to consider the range $0 < a \le b$, together with
$0<\phi\le\pi/2$, since $\mathcal{M}$ vanishes when $\phi=0$.

When $a=b$, it follows that $p(\theta)=a$ in eq.~\eqref{eq:defM}, so
we can separate $\mathcal{M}(s,i,j;a,a,\phi) = \mathcal{M}_1(s,i,j;a)
\, \mathcal{M}_2(i,j;\phi)$, where
\begin{eqnarray}
    \label{eq:defM1M2}
    \mathcal{M}_1(s,i,j;a)
    \hspace{-5pt}&=&\hspace{-5pt}
        \frac{\du^{i+j}}{\du a^{i+j}} \frac{1-\sqrt{(1-a)^{s+1}}}{a}
    \nonumber\\[-5pt]\\[-5pt]\nonumber
    \mathcal{M}_2(i,j;\phi)
    \hspace{-5pt}&=&\hspace{-5pt}
        \int_0^\phi \cos^{2i}\theta \sin^{2j}\theta
        \int\limits_0^\phi \cos^{2i}\theta \sin^{2j}\theta
    \du \theta.
\end{eqnarray}
For $a=1$, the expression for $\mathcal{M}_1$ simplifies to
$(-1)^{i+j}(i+j)!$. The integral in the expression for $\mathcal{M}_2$
can be evaluated explicitly using e.g.\ the relations 2.513 of
Gradshteyn \& Ryzhik (1994\nocite{GR..tables}). For $\phi=\pi/2$, it
reduces to the beta function $B(i+1/2,j+1/2)$.

When $a<b=1$, the elliptic integrals become elementary, so
that the quantities $F$, $D$ and $J$ in
eq.~\eqref{eq:defMcarlsonFDJ} reduce to
\begin{eqnarray}
  \label{eq:defMcarlsonFDJ_b=1}
  F & = & \frac{\sqrt{1-a}}{a}\,
    \ln\left[\tan\left(\frac\pi4+\frac\phi2\right)\right],
    %\label{eq:defMcarlsonb1F}
    \nonumber \\
  D & = & \frac{a}{1-a}\,F - \frac{\sin\phi}{\sqrt{1-a}}, \quad
    %\label{eq:defMcarlsonb1D}
    \\ \nonumber
  J & = & F - \frac{1}{\sqrt{a}}\,
  \arctan\left(\sqrt{\frac{1-a}{a}}\sin\phi\right).
    %\label{eq:defMcarlsonb1J}
\end{eqnarray}
Although $F$ diverges when $\phi\to\pi/2$, substitution of
these reduced quantities in the expressions of $\mathcal{M}$ for even
$s$ (Table~\ref{tab:Mseven}), shows that all terms with $F$ cancel.
For $\phi=\pi/2$, the function $\mathcal{M}$ is thus everywhere
finite, with $A=\pi/(2\sqrt{ab})$ and $P=Q=0$.

% limiting/special cases at focal curves not included

%============================= appendix C ============================
\section{Edgeworth expansion}
\label{sec:edgeworth}
%=====================================================================

For the (re)construction of the LOSVD from its true line-of-sight
velocity moments, one can use the well-known Gram-Charlier series, the
terms of which are simple functions of the true moments (see e.g.\ 
Appendix B2 of van der Marel \& Franx
1993\nocite{1993ApJ...407..525V}), but it has poor convergence
properties. The terms in the Edgeworth (1905\nocite{Edgeworth1905})
expansion are also directly related to the true moments, but since it
is a true asymptotic expansion its accuracy is controlled, so that,
unlike the Gauss-Hermite and Gram-Charlier expansions, convergence
plays no role (see Blinnikov \& Moessner
1998\nocite{1998A&AS..130..193B} for a comparison between the
expansions and for further references).

The Edgeworth expansion of the LOSVD up to order $N$ is given by
\begin{equation}
  \label{eq:EDexpansion}
  \mathcal{L}_N^\mathrm{ED}(v) =
  \Sigma\frac{e^{-\frac12 w^2}}{\sqrt{2\pi}\sigma}
  \left[ 1 + \sum_{n=3}^N D_n \right],
\end{equation}
with $w = (v-V)/\sigma$ and
\begin{equation}\label{eq:EDtermD_n}
  D_n = \sum_{\{l_{i-2}\}} \mathcal{H}_{n+2(l-1)}(w)
  \prod_{i=3}^{n} \frac{1}{l_{i-2}!}
  \left(\frac{d_i}{i!}\right)^{l_{i-2}}.
\end{equation}
The Hermite polynomials $\mathcal{H}_m$ are related to those
defined by van der Marel \& Franx
(1993\nocite{1993ApJ...407..525V}) as $\mathcal{H}_m(w) =
\sqrt{m!}\,H_m(w/\sqrt{2})$. We have defined
$l=\sum_{j=1}^{n-2}l_j$, where the sets $\{l_j\}$ are the
non-negative integer solutions of the Diophantine equation
\begin{equation}
  \label{eq:diophantineeq}
  l_j+2l_j+\dots+(n-2)l_{n-2}=n-2, \qquad n\ge3,
\end{equation}
Substituting these solutions, we find up to order $N=5$
\begin{eqnarray}
  \label{eq:EDexpansion_N=5}
  \mathcal{L}_5^\mathrm{ED}(v)
  \hspace{-5pt}&=&\hspace{-5pt}
  \Sigma\frac{e^{-\frac12 w^2}}{\sqrt{2\pi}\sigma}
  \Biggl[ 1
    + \mathcal{H}_3(w)\frac{d_3}{3!}
    + \mathcal{H}_4(w)\frac{d_4}{4!}
    \Biggr. \nonumber \\ \Biggl.
    \hspace{-5pt}&&\hspace{-5pt}
    + \mathcal{H}_6(w)\frac12\left(\frac{d_3}{3!}\right)^2
    + \mathcal{H}_5(w)\frac{d_5}{5!}
    \Biggr. \nonumber \\ \Biggl.
    \hspace{-5pt}&&\hspace{-5pt}
    + \mathcal{H}_7(w)\frac{d_3}{3!}\frac{d_4}{4!}
    + \mathcal{H}_9(w)\frac16\left(\frac{d_3}{3!}\right)^3
    \Biggr].
\end{eqnarray}
The lower-order moments $\Sigma$, $V$ and $\sigma$ are equivalent
to those in eq.~\eqref{eq:def_velmomlos}, while the higher-order
moments $d_i$ ($i\ge3$) are cumulants of the true moments
\begin{equation}
  \label{eq:defdiED}
  d_i = \frac{i!}{\sigma^n} \sum_{\{l_k\}} (-1)^{l-1} (l\!-\!1)!
  \prod_{k=1}^{i} \frac{1}{l_k!} \left(\frac{\mu_k}{k!}\right)^{l_k},
\end{equation}
so that
\begin{equation}
  d_3 = \xi_1 , \quad
  d_4 = \xi_2-3 , \quad \mathrm{and} \quad
  d_5 = \xi_3 - 10\xi_1.
\end{equation}
The central moments $\xi_1$ (skewness), $\xi_2$ (kurtosis) and
$\xi_3$ are related to the true moments respectively as
\begin{eqnarray}
  \label{eq:centralmom}
  (\mu_0\sigma)^3\,\xi_1
  \hspace{-5pt}&=&\hspace{-5pt}
  \mu_0^2\,\mu_3 - 3\,\mu_0\,\mu_1\,\mu_2 + 2\,\mu_1^3,\\
  (\mu_0\sigma)^4\,\xi_2
  \hspace{-5pt}&=&\hspace{-5pt}
  \mu_0^3\,\mu_4 - 4\,\mu_0^2\,\mu_1\,\mu_3 +
  6\,\mu_0\,\mu_1^2\,\mu_2 - 3\,\mu_1^4,\\
  (\mu_0\sigma)^5\,\xi_3
  \hspace{-5pt}&=&\hspace{-5pt}
  \mu_0^4\,\mu_5 - 5\,\mu_0^3\,\mu_1\,\mu_4 +
  10\,\mu_0^2\,\mu_1^2\,\mu_3 %- 10\,\mu_0\,\mu_1^3\,\mu_2 + 4\,\mu_1^5.
  \nonumber \\ \hspace{-5pt}&&\hspace{-5pt}
  - 10\,\mu_0\,\mu_1^3\,\mu_2 + 4\,\mu_1^5.
\end{eqnarray}
Substituting the line-of-sight true moments $\mu_k$ for $k=0,\dots,K$,
we can compute $\mathcal{L}_K^\mathrm{ED}(v)$ at each position on the
plane of the sky.

%%%FIG
\begin{figure}
  \begin{center}
    \includegraphics[width=1.0\columnwidth]{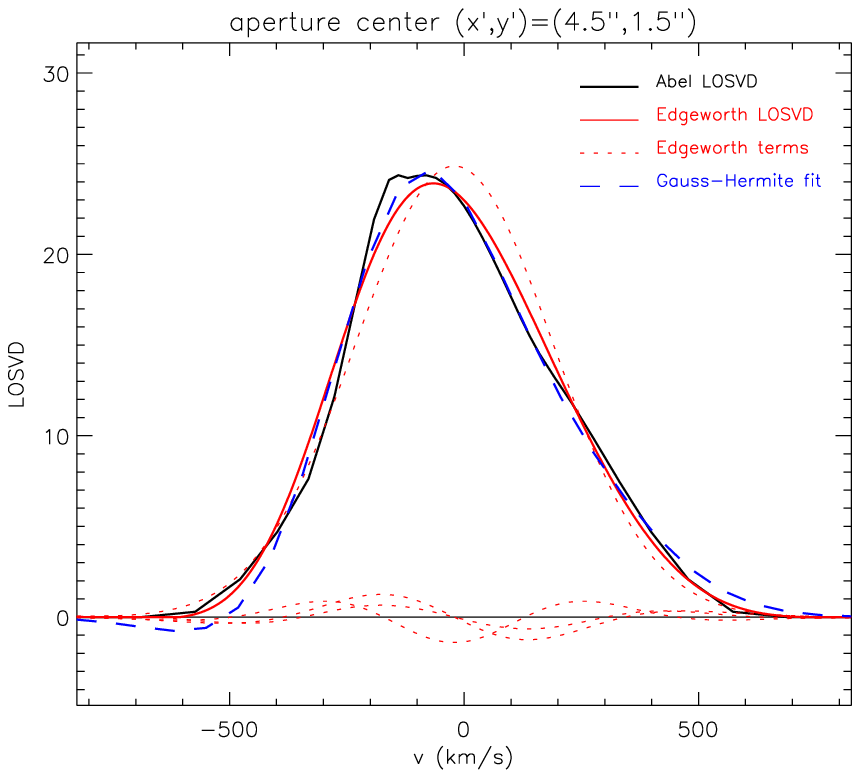}
  \end{center}
  \caption{Line-of-sight velocity distribution (LOSVD) of the triaxial
    Abel model presented in \S~\ref{sec:triaxabelmodel}, at the
    (aperture) position on the sky-plane given at the top of the
    figure. The black solid curve is the LOSVD computed directly via
    eq.~\eqref{eq:triaxAbelLOSVD}. The red solid curve show the
    Edgeworth LOSVD constructed from the true line-of-sight velocity
    moments, based on the intrinsic velocity moments computed via
    eq.~\eqref{eq:mugeneral}. The Gaussian and the higher order terms
    of the Edgeworth expansion \eqref{eq:EDexpansion} are shown by the
    red dotted curves. The blue dashed curve shows the best-fit
    Gauss-Hermite LOSVD.}
    \label{fig:edgeworth}
\end{figure}
%%%FIG

\begin{figure*}
  \begin{center}
    \includegraphics[width=1.0\textwidth]{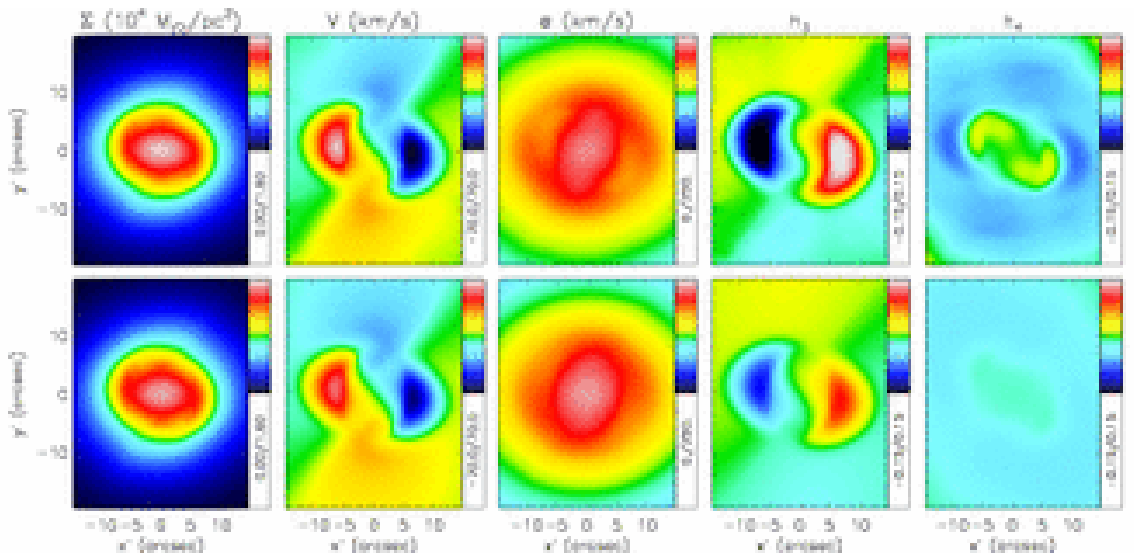}
  \end{center}
  \caption{
    Maps of the surface mass density ($\Sigma$; in $10^4$ \Msunpcsq),
    mean line-of-sight velocity $V$ and dispersion $\sigma$ (both in
    \kms), and higher order Gauss-Hermite moments $h_3$ and $h_4$, of
    the triaxial Abel model constructed in
    \S~\ref{sec:triaxabelmodel}. The top panels follow from fitting,
    at each (aperture) position on the sky-plane, Gauss-Hermite series
    to the LOSVD computed directly via eq.~\eqref{eq:triaxAbelLOSVD}.
    For the bottom panels the fit is applied to the Edgeworth LOSVD
    constructed from the true line-of-sight velocity moments, based on
    the intrinsic velocity moments computed via
    eq.~\eqref{eq:mugeneral}.}
    \label{fig:comp_ghmom_triax}
\end{figure*}

In Fig.~\ref{fig:edgeworth}, we show an example of a LOSVD (black
solid line) computed directly via eq.~\eqref{eq:triaxAbelLOSVD} for
the triaxial Abel model constructed in \S~\ref{sec:triaxabelmodel}.
The Edgeworth LOSVD (red solid line) is constructed from the true
line-of-sight velocity moments, based on the intrinsic velocity
moments computed via eq.~\eqref{eq:mugeneral}. The Edgeworth
reconstruction approximates (very) well the directly-computed LOSVD,
as well as the corresponding best-fit Gauss-Hermite series (blue
dashed line). In Fig.~\ref{fig:comp_ghmom_triax}, we show the
Gauss-Hermite moments after fitting at each (aperture) position on the
sky-plane the directly-computed LOSVD (top panels) as well as the
reconstructed Edgeworth LOSVD (bottom panels). The resulting maps are
very similar, except for a suppression of the higher-order
Gauss-Hermite moments in case of the Edgeworth reconstruction. This is
expected, since intrinsic velocity moments of order higher than $N=5$
are needed to accurately determine the wings of the LOSVD.
Nevertheless, this comparison is important to show the correctness of
both (independent) approaches, and that the Edgeworth expansion
provides a reliable and efficient way to reconstruct the LOSVD (and
obtain Gauss-Hermite moments) from true moments that are in general
(numerically) easier to compute than the full LOSVD.

%%%FIG
%%%FIG

%=====================================================================
% END DOCUMENT
%=====================================================================

\bsp % ``This paper has been produced using the ...''

\label{lastpage}

\end{document}